\documentclass[aps,prd,twocolumn,superscriptaddress,nofootinbib,preprintnumbers]{revtex4-1}
\usepackage[utf8]{inputenc}
\usepackage[english]{babel}
\usepackage{natbib}
\usepackage{graphicx}
\usepackage{amsmath,amssymb}
\usepackage{hyperref}
\usepackage{braket}
\usepackage{float}
\usepackage[dvipsnames]{xcolor}
\usepackage{soul}
\usepackage{rotating}
\usepackage{multirow}
\usepackage{mathtools}
\usepackage{makecell}

\usepackage[caption = false]{subfig}
\usepackage{txfonts}
\usepackage[margin=0.75in]{geometry}
\usepackage{tabularx}
\let\oldequation\equation
\let\oldendequation\endequation
\renewenvironment{equation}
  {\linenomathNonumbers\oldequation}
  {\oldendequation\endlinenomath}

\usepackage{hyperref}
\usepackage[T1]{fontenc}
\usepackage{ae,aecompl}

\usepackage{graphicx}	
\usepackage{amsmath}	
\usepackage{adjustbox}
\usepackage{multirow}
\usepackage{newtxmath}  
\usepackage{booktabs}

\definecolor{purple}{RGB}{150,0,200}

\usepackage{lineno}

\newcommand{\maglim}{\textsc{MagLim}\xspace}
\newcommand{\maglimpp}{\textsc{MagLim++}\xspace}

\newcommand{\nz}{\textit{n(z)}\xspace}

\usepackage{xcolor}
\definecolor{Gray}{gray}{0.9}
\definecolor{myorange}{RGB}{255, 104, 51}

\usepackage{ulem}

\usepackage{xspace} 

\defcitealias{y6-gold}{\textsc{Y6Gold}}
\defcitealias{y6-piff}{\textsc{Y6PSF}}
\defcitealias{y6-mask}{\textsc{Y6Mask}}
\defcitealias{y6-balrog}{\textsc{Y6Balrog}}
\defcitealias{y6-metadetect}{\textsc{Y6Shear}}
\defcitealias{y6-maglim}{\textsc{Y6Pos}}
\defcitealias{y6-sourcepz}{\textsc{Y6ShearPZ}}
\defcitealias{y6-lenspz}{\textsc{Y6PosPZ}}
\defcitealias{y6-wz}{\textsc{Y6WZ}}
\defcitealias{y6-nzmodes}{\textsc{Y6Modes}}
\defcitealias{y6-magnification}{\textsc{Y6Mag}}
\defcitealias{y6-gglens}{\textsc{Y6GGL}}
\defcitealias{y6-methods}{\textsc{Y6Model}}
\defcitealias{y6-ppd}{\textsc{Y6PPD}}
\defcitealias{y6-imagesims}{\textsc{Y6Imsim}}
\defcitealias{y6-cardinal}{\textsc{Y6Mock}}
\defcitealias{y6-1x2pt}{\textsc{Y6-1x2pt}}
\defcitealias{y6-2x2pt}{\textsc{Y6-2x2pt}}
\defcitealias{y6-3x2pt}{\textsc{Y6-3x2pt}}
\defcitealias{y6-extensions}{\textsc{Y6Extend}}

\usepackage{xcolor}
\definecolor{Gray}{gray}{0.9}
\definecolor{myorange}{RGB}{255, 104, 51}
\definecolor{mygreen}{RGB}{3, 170, 60}

\definecolor{Darkred}{RGB}{180,0,0}

\usepackage{soul} 

\usepackage{ulem}

\usepackage{xspace} 

\begin{document}
\title[DES Y6 Lenses Redshift Calibration]{Dark Energy Survey Year 6 Results: Redshift Calibration of the \maglim++ Lens Sample}

\author{G.~Giannini}\email{giulia.giannini15@gmail.com}
\affiliation{Institute of Space Sciences (ICE, CSIC),  Campus UAB, Carrer de Can Magrans, s/n,  08193 Barcelona, Spain}
\affiliation{Kavli Institute for Cosmological Physics, University of Chicago, Chicago, IL 60637, USA}

\author{A.~Alarcon}
\affiliation{Institute of Space Sciences (ICE, CSIC),  Campus UAB, Carrer de Can Magrans, s/n,  08193 Barcelona, Spain}

\author{W.~d'Assignies}
\affiliation{Institut de F\'{\i}sica d'Altes Energies (IFAE), The Barcelona Institute of Science and Technology, Campus UAB, 08193 Bellaterra (Barcelona) Spain}

\author{G.~M.~Bernstein}
\affiliation{Department of Physics and Astronomy, University of Pennsylvania, Philadelphia, PA 19104, USA}

\author{M.~A.~Troxel}
\affiliation{Department of Physics, Duke University Durham, NC 27708, USA}

\author{C.~Chang}
\affiliation{Department of Astronomy and Astrophysics, University of Chicago, Chicago, IL 60637, USA}
\affiliation{Kavli Institute for Cosmological Physics, University of Chicago, Chicago, IL 60637, USA}

\author{B.~Yin}
\affiliation{Department of Physics, Duke University Durham, NC 27708, USA}

\author{A.~Amon}
\affiliation{Department of Astrophysical Sciences, Princeton University, Peyton Hall, Princeton, NJ 08544, USA}

\author{J.~Myles}
\affiliation{Department of Astrophysical Sciences, Princeton University, Peyton Hall, Princeton, NJ 08544, USA}

\author{N.~Weaverdyck}
\affiliation{Berkeley Center for Cosmological Physics, Department of Physics, University of California, Berkeley, CA 94720, US}
\affiliation{Lawrence Berkeley National Laboratory, 1 Cyclotron Road, Berkeley, CA 94720, USA}

\author{A.~Porredon}
\affiliation{Ruhr University Bochum, Faculty of Physics and Astronomy, Astronomical Institute, German Centre for Cosmological Lensing, 44780 Bochum, Germany}
\affiliation{Centro de Investigaciones Energ\'eticas, Medioambientales y Tecnol\'ogicas (CIEMAT), Madrid, Spain}

\author{D.~Anbajagane}
\affiliation{Kavli Institute for Cosmological Physics, University of Chicago, Chicago, IL 60637, USA}

\author{S.~Avila}
\affiliation{Centro de Investigaciones Energ\'eticas, Medioambientales y Tecnol\'ogicas (CIEMAT), Madrid, Spain}

\author{K.~Bechtol}
\affiliation{Physics Department, 2320 Chamberlin Hall, University of Wisconsin-Madison, 1150 University Avenue Madison, WI  53706-1390}

\author{M.~R.~Becker}
\affiliation{Argonne National Laboratory, 9700 South Cass Avenue, Lemont, IL 60439, USA}

\author{J.~Blazek}
\affiliation{Department of Physics, Northeastern University, Boston, MA 02115, USA}

\author{M.~Crocce}
\affiliation{Institut d'Estudis Espacials de Catalunya (IEEC), 08034 Barcelona, Spain}
\affiliation{Institute of Space Sciences (ICE, CSIC),  Campus UAB, Carrer de Can Magrans, s/n,  08193 Barcelona, Spain}

\author{D.~Gruen}
\affiliation{University Observatory, LMU Faculty of Physics, Scheinerstr. 1, 81679 Munich, Germany}

\author{M.~Rodriguez-Monroy}
\affiliation{Instituto de Física Teórica UAM/CSIC, Universidad Autónoma de Madrid, 28049 Madrid, Spain}
\affiliation{Laboratoire de physique des 2 infinis Irène Joliot-Curie, CNRS Université Paris-Saclay, Bât. 100, F-91405 Orsay Cedex, France}

\author{C.~S{\'a}nchez}
\affiliation{Institut de F\'{\i}sica d'Altes Energies (IFAE), The Barcelona Institute of Science and Technology, Campus UAB, 08193 Bellaterra (Barcelona) Spain}
\affiliation{Departament de F\'{\i}sica, Universitat Aut\`{o}noma de Barcelona (UAB), 08193 Bellaterra, Barcelona, Spain}

\author{I.~Sevilla-Noarbe}
\affiliation{Centro de Investigaciones Energ\'eticas, Medioambientales y Tecnol\'ogicas (CIEMAT), Madrid, Spain}

\author{M.~Aguena}
\affiliation{INAF-Osservatorio Astronomico di Trieste, via G. B. Tiepolo 11, I-34143 Trieste, Italy}
\affiliation{Laborat\'orio Interinstitucional de e-Astronomia - LIneA, Av. Pastor Martin Luther King Jr, 126 Del Castilho, Nova Am\'erica Offices, Torre 3000/sala 817 CEP: 20765-000, Brazil}

\author{S.~Allam}
\affiliation{Fermi National Accelerator Laboratory, P. O. Box 500, Batavia, IL 60510, USA}

\author{O.~Alves}
\affiliation{Department of Physics, University of Michigan, Ann Arbor, MI 48109, USA}

\author{F.~Andrade-Oliveira}
\affiliation{Physik-Institut, University of Zürich, Winterthurerstrasse 190, CH-8057 Zürich, Switzerland}

\author{D.~Bacon}
\affiliation{Institute of Cosmology and Gravitation, University of Portsmouth, Portsmouth, PO1 3FX, UK}

\author{S.~Bocquet}
\affiliation{University Observatory, LMU Faculty of Physics, Scheinerstr. 1, 81679 Munich, Germany}

\author{D.~Brooks}
\affiliation{Department of Physics \& Astronomy, University College London, Gower Street, London, WC1E 6BT, UK}

\author{R.~Camilleri}
\affiliation{School of Mathematics and Physics, University of Queensland,  Brisbane, QLD 4072, Australia}

\author{A.~Carnero~Rosell}
\affiliation{Universidad de La Laguna, Dpto. Astrofísica, E-38206 La Laguna, Tenerife, Spain}
\affiliation{Laborat\'orio Interinstitucional de e-Astronomia - LIneA, Av. Pastor Martin Luther King Jr, 126 Del Castilho, Nova Am\'erica Offices, Torre 3000/sala 817 CEP: 20765-000, Brazil}
\affiliation{Instituto de Astrofisica de Canarias, E-38205 La Laguna, Tenerife, Spain}

\author{J.~Carretero}
\affiliation{Institut de F\'{\i}sica d'Altes Energies (IFAE), The Barcelona Institute of Science and Technology, Campus UAB, 08193 Bellaterra (Barcelona) Spain}

\author{R.~Cawthon}
\affiliation{Oxford College of Emory University, Oxford, GA 30054, USA}

\author{L.~N.~da Costa}
\affiliation{Laborat\'orio Interinstitucional de e-Astronomia - LIneA, Av. Pastor Martin Luther King Jr, 126 Del Castilho, Nova Am\'erica Offices, Torre 3000/sala 817 CEP: 20765-000, Brazil}

\author{M.~E.~da Silva Pereira}
\affiliation{Hamburger Sternwarte, Universit\"{a}t Hamburg, Gojenbergsweg 112, 21029 Hamburg, Germany}

\author{T.~M.~Davis}
\affiliation{School of Mathematics and Physics, University of Queensland,  Brisbane, QLD 4072, Australia}

\author{J.~De~Vicente}
\affiliation{Centro de Investigaciones Energ\'eticas, Medioambientales y Tecnol\'ogicas (CIEMAT), Madrid, Spain}

\author{D.~L.~DePoy}
\affiliation{George P. and Cynthia Woods Mitchell Institute for Fundamental Physics and Astronomy, and Department of Physics and Astronomy, Texas A\&M University, College Station, TX 77843,  USA}

\author{S.~Desai}
\affiliation{Department of Physics, IIT Hyderabad, Kandi, Telangana 502285, India}

\author{H.~T.~Diehl}
\affiliation{Fermi National Accelerator Laboratory, P. O. Box 500, Batavia, IL 60510, USA}

\author{S.~Dodelson}
\affiliation{Fermi National Accelerator Laboratory, P. O. Box 500, Batavia, IL 60510, USA}
\affiliation{Department of Astronomy and Astrophysics, University of Chicago, Chicago, IL 60637, USA}
\affiliation{Kavli Institute for Cosmological Physics, University of Chicago, Chicago, IL 60637, USA}

\author{P.~Doel}
\affiliation{Department of Physics \& Astronomy, University College London, Gower Street, London, WC1E 6BT, UK}

\author{C.~Doux}
\affiliation{Department of Physics and Astronomy, University of Pennsylvania, Philadelphia, PA 19104, USA}
\affiliation{Universit\'e Grenoble Alpes, CNRS, LPSC-IN2P3, 38000 Grenoble, France}

\author{A.~Drlica-Wagner}
\affiliation{Kavli Institute for Cosmological Physics, University of Chicago, Chicago, IL 60637, USA}
\affiliation{Fermi National Accelerator Laboratory, P. O. Box 500, Batavia, IL 60510, USA}
\affiliation{Department of Astronomy and Astrophysics, University of Chicago, Chicago, IL 60637, USA}

\author{J.~Elvin-Poole}
\affiliation{Department of Physics and Astronomy, University of Waterloo, 200 University Ave W, Waterloo, ON N2L 3G1, Canada}

\author{S.~Everett}
\affiliation{California Institute of Technology, 1200 East California Blvd, MC 249-17, Pasadena, CA 91125, USA}

\author{A.~E.~Evrard}
\affiliation{Department of Astronomy, University of Michigan, Ann Arbor, MI 48109, USA}
\affiliation{Department of Physics, University of Michigan, Ann Arbor, MI 48109, USA}

\author{B.~Flaugher}
\affiliation{Fermi National Accelerator Laboratory, P. O. Box 500, Batavia, IL 60510, USA}

\author{J.~Frieman}
\affiliation{Department of Astronomy and Astrophysics, University of Chicago, Chicago, IL 60637, USA}
\affiliation{Fermi National Accelerator Laboratory, P. O. Box 500, Batavia, IL 60510, USA}
\affiliation{Kavli Institute for Cosmological Physics, University of Chicago, Chicago, IL 60637, USA}

\author{J.~Garc\'ia-Bellido}
\affiliation{Instituto de Fisica Teorica UAM/CSIC, Universidad Autonoma de Madrid, 28049 Madrid, Spain}

\author{M.~Gatti}
\affiliation{Kavli Institute for Cosmological Physics, University of Chicago, Chicago, IL 60637, USA}
\affiliation{Institute of Space Sciences (ICE, CSIC),  Campus UAB, Carrer de Can Magrans, s/n,  08193 Barcelona, Spain}

\author{E.~Gaztanaga}
\affiliation{Institute of Cosmology and Gravitation, University of Portsmouth, Portsmouth, PO1 3FX, UK}
\affiliation{Institute of Space Sciences (ICE, CSIC),  Campus UAB, Carrer de Can Magrans, s/n,  08193 Barcelona, Spain}
\affiliation{Institut d'Estudis Espacials de Catalunya (IEEC), 08034 Barcelona, Spain}

\author{P.~Giles}
\affiliation{Department of Physics and Astronomy, Pevensey Building, University of Sussex, Brighton, BN1 9QH, UK}

\author{R.~A.~Gruendl}
\affiliation{Department of Astronomy, University of Illinois at Urbana-Champaign, 1002 W. Green Street, Urbana, IL 61801, USA}
\affiliation{Center for Astrophysical Surveys, National Center for Supercomputing Applications, 1205 West Clark St., Urbana, IL 61801, USA}

\author{G.~Gutierrez}
\affiliation{Fermi National Accelerator Laboratory, P. O. Box 500, Batavia, IL 60510, USA}

\author{K.~Herner}
\affiliation{Fermi National Accelerator Laboratory, P. O. Box 500, Batavia, IL 60510, USA}

\author{S.~R.~Hinton}
\affiliation{School of Mathematics and Physics, University of Queensland,  Brisbane, QLD 4072, Australia}

\author{D.~L.~Hollowood}
\affiliation{Santa Cruz Institute for Particle Physics, Santa Cruz, CA 95064, USA}

\author{K.~Honscheid}
\affiliation{Department of Physics, The Ohio State University, Columbus, OH 43210, USA}
\affiliation{Center for Cosmology and Astro-Particle Physics, The Ohio State University, Columbus, OH 43210, USA}

\author{D.~Huterer}
\affiliation{Department of Physics, University of Michigan, Ann Arbor, MI 48109, USA}

\author{D.~J.~James}
\affiliation{Center for Astrophysics $\vert$ Harvard \& Smithsonian, 60 Garden Street, Cambridge, MA 02138, USA}

\author{K.~Kuehn}
\affiliation{Australian Astronomical Optics, Macquarie University, North Ryde, NSW 2113, Australia}
\affiliation{Lowell Observatory, 1400 Mars Hill Rd, Flagstaff, AZ 86001, USA}

\author{O.~Lahav}
\affiliation{Department of Physics \& Astronomy, University College London, Gower Street, London, WC1E 6BT, UK}

\author{S.~Lee}
\affiliation{Jet Propulsion Laboratory, California Institute of Technology, 4800 Oak Grove Dr., Pasadena, CA 91109, USA}
\affiliation{Department of Physics and Astronomy, Ohio University, Clippinger Labs, Athens, OH 45701}

\author{H.~Lin}
\affiliation{Fermi National Accelerator Laboratory, P. O. Box 500, Batavia, IL 60510, USA}

\author{J.~L.~Marshall}
\affiliation{George P. and Cynthia Woods Mitchell Institute for Fundamental Physics and Astronomy, and Department of Physics and Astronomy, Texas A\&M University, College Station, TX 77843,  USA}

\author{J. Mena-Fern{\'a}ndez}
\affiliation{Universit\'e Grenoble Alpes, CNRS, LPSC-IN2P3, 38000 Grenoble, France}
\affiliation{Aix Marseille Univ, CNRS/IN2P3, CPPM, Marseille, France}

\author{F.~Menanteau}
\affiliation{Center for Astrophysical Surveys, National Center for Supercomputing Applications, 1205 West Clark St., Urbana, IL 61801, USA}
\affiliation{Department of Astronomy, University of Illinois at Urbana-Champaign, 1002 W. Green Street, Urbana, IL 61801, USA}

\author{R.~Miquel}
\affiliation{Institut de F\'{\i}sica d'Altes Energies (IFAE), The Barcelona Institute of Science and Technology, Campus UAB, 08193 Bellaterra (Barcelona) Spain}
\affiliation{Instituci\'o Catalana de Recerca i Estudis Avan\c{c}ats, E-08010 Barcelona, Spain}

\author{J.~Muir}
\affiliation{Perimeter Institute for Theoretical Physics, 31 Caroline St. North, Waterloo, ON N2L 2Y5, Canada}
\affiliation{Department of Physics, University of Cincinnati, Cincinnati, Ohio 45221, USA}

\author{R.~L.~C.~Ogando}
\affiliation{Observat\'orio Nacional, Rua Gal. Jos\'e Cristino 77, Rio de Janeiro, RJ - 20921-400, Brazil}

\author{D.~Petravick}
\affiliation{Center for Astrophysical Surveys, National Center for Supercomputing Applications, 1205 West Clark St., Urbana, IL 61801, USA}

\author{A.~A.~Plazas~Malag\'on}
\affiliation{Kavli Institute for Particle Astrophysics \& Cosmology, P. O. Box 2450, Stanford University, Stanford, CA 94305, USA}
\affiliation{SLAC National Accelerator Laboratory, Menlo Park, CA 94025, USA}

\author{J.~Prat}
\affiliation{Nordita, KTH Royal Institute of Technology and Stockholm University, Hannes Alfv\'ens v\"ag 12, SE-10691 Stockholm, Sweden}
\affiliation{University of Copenhagen, Dark Cosmology Centre, Juliane Maries Vej 30, 2100 Copenhagen O, Denmark}

\author{M.~Raveri}
\affiliation{Department of Physics, University of Genova and INFN, Via Dodecaneso 33, 16146, Genova, Italy}

\author{E.~S.~Rykoff}
\affiliation{SLAC National Accelerator Laboratory, Menlo Park, CA 94025, USA}
\affiliation{Kavli Institute for Particle Astrophysics \& Cosmology, P. O. Box 2450, Stanford University, Stanford, CA 94305, USA}

\author{S.~Samuroff}
\affiliation{Institut de F\'{\i}sica d'Altes Energies (IFAE), The Barcelona Institute of Science and Technology, Campus UAB, 08193 Bellaterra (Barcelona) Spain}
\affiliation{Department of Physics, Northeastern University, Boston, MA 02115, USA}

\author{E.~Sanchez}
\affiliation{Centro de Investigaciones Energ\'eticas, Medioambientales y Tecnol\'ogicas (CIEMAT), Madrid, Spain}

\author{D.~Sanchez Cid}
\affiliation{Physik-Institut, University of Zürich, Winterthurerstrasse 190, CH-8057 Zürich, Switzerland}
\affiliation{Centro de Investigaciones Energ\'eticas, Medioambientales y Tecnol\'ogicas (CIEMAT), Madrid, Spain}

\author{T.~Shin}
\affiliation{Department of Physics and Astronomy, Stony Brook University, Stony Brook, NY 11794, USA}

\author{M.~Smith}
\affiliation{Physics Department, Lancaster University, Lancaster, LA1 4YB, UK}

\author{E.~Suchyta}
\affiliation{Computer Science and Mathematics Division, Oak Ridge National Laboratory, Oak Ridge, TN 37831}

\author{M.~E.~C.~Swanson}
\affiliation{Center for Astrophysical Surveys, National Center for Supercomputing Applications, 1205 West Clark St., Urbana, IL 61801, USA}

\author{G.~Tarle}
\affiliation{Department of Physics, University of Michigan, Ann Arbor, MI 48109, USA}

\author{D.~Thomas}
\affiliation{Institute of Cosmology and Gravitation, University of Portsmouth, Portsmouth, PO1 3FX, UK}

\author{C.~To}
\affiliation{Department of Astronomy and Astrophysics, University of Chicago, Chicago, IL 60637, USA}

\author{D.~L.~Tucker}
\affiliation{Fermi National Accelerator Laboratory, P. O. Box 500, Batavia, IL 60510, USA}

\author{V.~Vikram}
\affiliation{Central University of Kerala, Periye, Kerala 671320, India}

\author{M.~Yamamoto}
\affiliation{Department of Physics, Duke University Durham, NC 27708, USA}
\affiliation{Department of Astrophysical Sciences, Princeton University, Peyton Hall, Princeton, NJ 08544, USA}

\collaboration{DES Collaboration}

\label{firstpage}


\begin{abstract}
In this work, we derive and calibrate the redshift distribution of the \maglimpp lens galaxy sample used in the Dark Energy Survey Year 6 (DES Y6) 3$\times$2pt cosmology analysis. The 3x2pt analysis combines galaxy clustering from the lens galaxy sample and weak gravitational lensing. The redshift distributions are inferred using the \textsc{SOMPZ} method -- a Self-Organizing Map framework that combines deep-field multi-band photometry, wide-field data, and a synthetic source injection (\texttt{Balrog}) catalog. Key improvements over the DES Year 3 (Y3) calibration include a noise-weighted SOM metric, an expanded \texttt{Balrog} catalogue, and an improved scheme for propagating systematic uncertainties, which allows us to generate O($10^8$) redshift realizations that collectively span the dominant sources of uncertainty. These realizations are then combined with independent clustering-redshift measurements via importance sampling. The resulting calibration achieves typical uncertainties on the mean redshift of 1--2\%, corresponding to a 20--30\% average reduction relative to DES Y3. We compress the $n(z)$ uncertainties into a small number of orthogonal modes for use in cosmological inference. Marginalizing over these modes leads to only a minor degradation in cosmological constraints. This analysis establishes the \maglimpp sample as a robust lens sample for precision cosmology with DES Y6 and provides a scalable framework for future surveys.
\end{abstract}

\preprint{DES-2025-0911}
\preprint{FERMILAB-PUB-25-0641-V}
\maketitle
\setcounter{footnote}{1}




\section{Introduction}
\label{sec:intro}








The success of modern cosmology builds upon a combination of probes that measure the geometry and structure growth of the Universe \citep{Weinberg2013}. For structure growth, the most constraining and robust structure measurements in modern galaxy surveys relies on the two-point auto-and cross-correlation between galaxy density and galaxy weak lensing. This combination, colloquially referred to as the ``3$\times$2pt`` probes, consistently combines cosmic shear (the auto-correlations of galaxy weak lensing), galaxy-galaxy lensing (the cross-correlation between galaxy density and galaxy weak lensing), and galaxy clustering (the auto-correlation of galaxy density). The three probes have different degeneracy directions in the model and share a number of nuisance parameters; this unique feature allows it to self-consistently calibrate the nuisance parameters and result in tighter and more robust constraints. Many modern galaxy surveys have demonstrated the effectiveness of the 3$\times$2pt probes \citep{Heymans2021, y3-3x2ptkp, Sugiyama2023}.

One of the key ingredients of the $3\times 2$pt probes is the distance, or redshift, to the galaxies used for the measurements. In a $3\times 2$pt analysis we define two galaxy samples according to their role: a \textit{lens} sample, used as a tracer of the galaxy density field, and a \textit{source} sample, used as a tracer of the weak lensing shear field. While these samples are often constructed to be disjoint, in principle any galaxy sample can serve as either, and some analyses use the same galaxies in both roles \citep{Salcedo_2025}. Due to the different nature of the samples, the treatment in how we derive the distances to these samples often differs. This paper focuses on deriving and characterizing the redshift distribution for the lens galaxy sample, the \maglimpp sample, which is defined in \citep*{y6-maglim} and will be used in the DES Year 6 (Y6) 3$\times$2pt cosmology analysis \citep{y6-3x2pt}.

The lens galaxy sample, used in both the galaxy clustering and the galaxy-galaxy lensing measurements, is designed to be a relatively bright galaxy sample that traces the dark matter density in a way that is easy to model. Since lens galaxy clustering probes the matter density field through a redshift kernel set by the width of the lens sample’s $n(z)$ (as opposed to weak lensing, which is sensitive to a much broader kernel), a desired property for the sample is to have narrow redshift distributions \citep{Tanoglidis2020} so as to maximize the information in the radial direction. In addition, as shown in \cite{annamaglim} and \cite{Tanoglidis2020}, for current datasets the precision with which we can calibrate the redshift distribution of the lens sample can have a greater impact on cosmological constraints than simply increasing its number density. In other words, improving $n(z)$ calibration often yields larger gains than adding more galaxies, since inaccurate redshift estimates can bias the results even in a high-density sample.

As the lens samples are typically bright galaxies, historically not as much attention has been placed on understanding their redshift distributions compared to the source sample. In the DES Y3 analysis, a large amount of work went into developing a new methodology for redshift estimation of the source sample, combining both photometric information \citep*[SOMPZ,][]{pitpz}, clustering information \citep*[WZ,][]{y3-sourcewz}, and geometric information \citep*[shear ratio,][]{y3-shearratio}. The SOMPZ framework is able to characterize the redshift distribution and uncertainty of the source sample in a fully Bayesian way, utilizing three sets of photometry from the DES wide field, the DES deep field and the spectroscopic sample (see Section~\ref{sec:data}); the WZ pipeline uses independent clustering information to further constrain the redshift distribution; finally the shear ratio technique provides an additional independent check on the high-redshift galaxies using weak lensing itself. 

Unlike the source sample, most previous work using a lens sample for either galaxy-galaxy lensing or galaxy-clustering opts to use spectroscopic samples if there is good overlap with lensing data \citep{lensspectra1, lensspectra2, lensspectra3}, since precise redshifts are especially beneficial. 
Although DES overlaps with several spectroscopic surveys, the coverage at this point is not sufficient to define a spectroscopic lens sample large enough for competitive cosmological analyses. We therefore construct photometric lens samples from bright galaxies with reliable photometric redshifts. In the fiducial DES Y3 analysis, we looked at a redshift-dependent magnitude-limited sample \maglim \citep{y3-2x2ptaltlensresults}. The redshift distribution of that sample was first determined using DNF \citep{DeVicente2016} and then calibrated using WZ \citep{y3-lenswz}. 

In a follow-up paper using a similar SOMPZ+WZ technology developed for the source sample \citep{Giannini2024}, we found that adopting the more principled redshift calibration method resulted in $\sim 0.4 \sigma$ shift of the cosmological constraints combining galaxy clustering and galaxy-galaxy lensing in the $S_8 - \Omega_{\rm m}$\footnote{$S_{8}\equiv \sigma_{8}\sqrt{\Omega_{\rm m}/0.3}$, where $\sigma_8$ is the normalization of the present-time, linear matter-power spectrum smoothed on $8h^{-1}$Mpc scales, and $\Omega_{\rm m}$ is the ratio of the present-time matter energy density to the critical energy density.} plane, highlighting the importance of Bayesian redshift calibration.     

The goal of this paper is to present the redshift distribution and its uncertainty for the DES Y6 \maglimpp lens sample: this will be the lens sample used in the legacy 3$\times$2pt analysis from DES. We describe the methodology as well as data updates to \citep{Giannini2024}, and assess their impact on cosmological constraints. The main updates include: 1) improvements in data products throughout the pipeline, including a cleaner lens sample (\maglimpp), a larger synthetic source injection catalog (\texttt{Balrog}, \citep*{y6-balrog}) that covers the entire footprint, and an enhanced redshift sample ; 2) the Y6 imaging is about twice as deep as Y3, which reduces magnitude errors, although for the relatively bright \maglimpp lenses ($i<22.2$) this effect is modest, it still provides more precise color measurements; 3) an updated SOMPZ algorithm that incorporates the photometric uncertainty when assigning galaxies into the SOM \citep{Sanchez2020, Campos2024}; 4) a more general prescription in marginalizing the uncertainties in the redshift sample; and 5) a more general parametrization for the uncertainties in the redshift distribution. 

             
            

\begin{table}
    \centering
    \caption{Reference table for the series of four papers that describe the full redshift treatment for the DES Y6 3$\times$2pt analysis.} 
    \label{tab:alias}
        \begin{tabular}{lp{1.6cm}p{4.9cm}}
            \hline 
            \hline
            Shorthand  & Reference & Description \\ 
        \hline
             \textsc{Y6ShearPZ} &   \cite{y6-sourcepz}   & Source photo-$z$ calibration  \\
             \textsc{Y6PosPZ} &   \textit{this work}  & Lens photo-$z$ calibration  \\
             \textsc{Y6WZ} &   \cite{y6-wz}   & Clustering redshift  \\
             \textsc{Y6Mode} &   \cite{y6-nzmodes}   & Redshift marginalisation method   \\
             \hline
             \hline
        \end{tabular}
\end{table}

\begin{figure*}
\includegraphics[width=\linewidth]{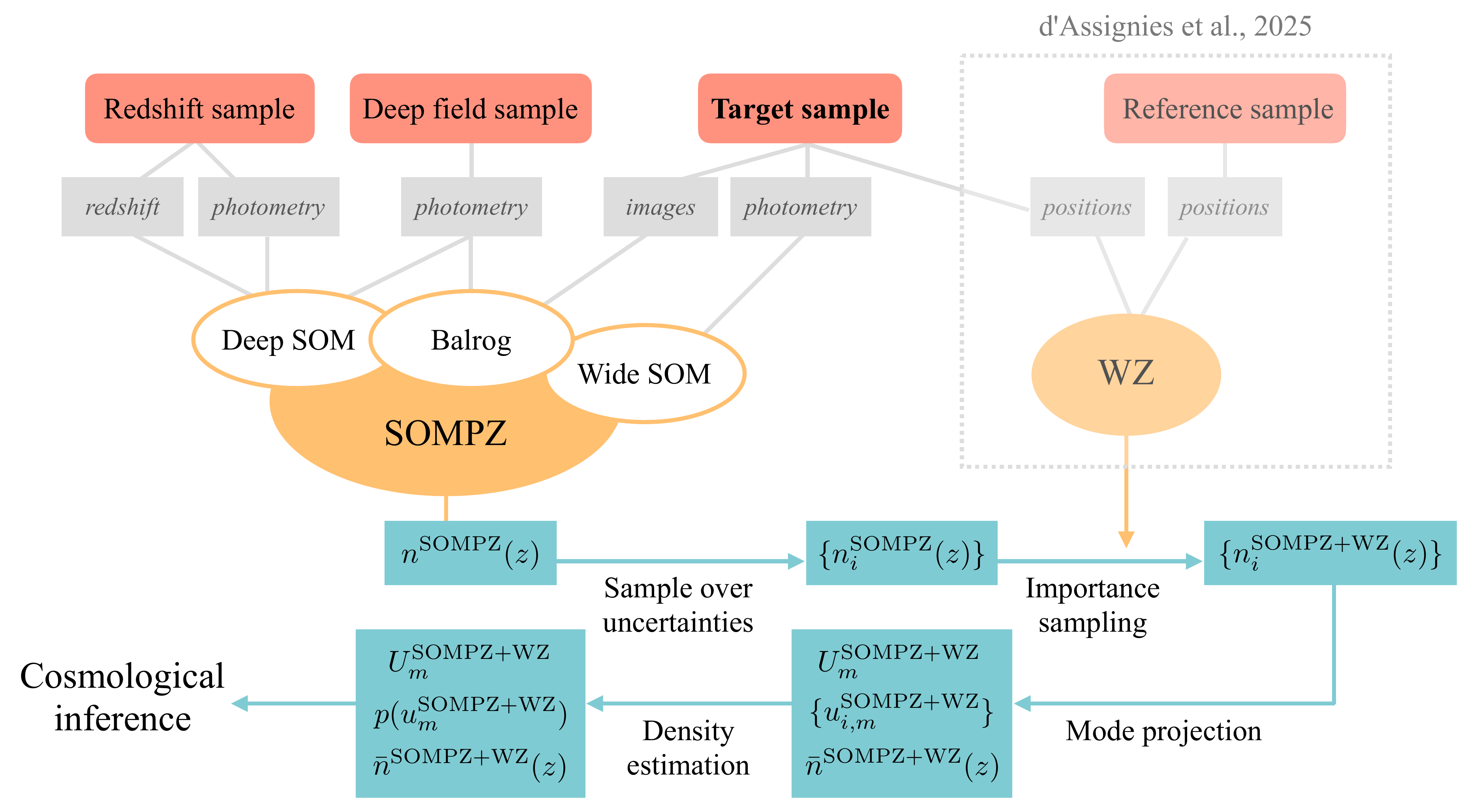}
\caption{Flowchart summarizing the DES Y6 3$\times$2pt redshift calibration pipeline. 
The target sample of this study is \maglimpp\ (Section~\ref{sec:data}), with redshift information provided by the redshift sample (Section~\ref{sec:redshiftcatalogs}) and deep fields combined with \texttt{Balrog} simulations (Section~\ref{sec:data}). 
Photometric data are then mapped using Self-Organizing Maps (SOMs) for deep and wide fields (Section~\ref{sec:methodology}), transferring the redshift information to the target galaxies. 
Sampling over the associated uncertainties (Section~\ref{sec:uncertainty}) produces realizations \(n_i^{\mathrm{SOMPZ}}(z)\). 
Clustering information (from \citepalias{y6-wz}) provides additional constraints via importance sampling (Section~\ref{sec:combine}), yielding \(n_i^{\mathrm{SOMPZ+WZ}}(z)\). 
These realizations are then projected onto modes (Section~\ref{sec:modes}; see also \citepalias{y6-nzmodes}), enabling accurate marginalization over the redshift distribution in cosmological inference. 
This paper details the SOMPZ derivation, its combination with WZ, the mode-projection framework, and robustness tests through cosmological inference, while the WZ measurement itself is presented in \citepalias{y6-wz}.}
\label{fig:flowchart}
\end{figure*}

This paper is one of the series of four papers that will describe the full redshift treatment for the DES Y6 3$\times$2pt analysis: 1) \citepalias{y6-sourcepz} describes the derivation of the SOMPZ redshift estimates for the source galaxy sample; 2) this paper describes the derivation of the SOMPZ redshift estimates for the lens galaxy sample; 3) \citepalias{y6-wz} describes how the clustering information (WZ) is incorporated into the redshift estimates of 1) and 2); and 4) \citepalias{y6-nzmodes} describes a novel approach used in the Y6 analysis to sample over the redshift uncertainty during the cosmological inference. Altogether, the four papers give the complete description of the Y6 redshift estimates. Figure \ref{fig:flowchart} illustrates the overall redshift workflow for the DES Y6 3$\times$2pt analysis. In this work, \maglimpp serves as the target sample in the flowchart. 

The paper is structured as follows. In Section~\ref{sec:data} we describe the data products used in this work, including both DES data as well as external redshift catalogs. In Section~\ref{sec:methodology} we describe the methodology we have adopted to estimate and calibrate the redshift distribution of the \maglimpp sample based on the SOMPZ method. Section~\ref{sec:uncertainty} describes how the different contributions to the uncertainties on the redshift distributions are estimated. 
In Section~\ref{sec:combine} we describe how the SOMPZ results are combined with the clustering redshift information presented in \citepalias{y6-wz}, and how we marginalize over the uncertainties following \citepalias{y6-nzmodes}. We discuss some of the major improvements of this work compared to DES Y3 and the impact on cosmological constraints in Section~\ref{sec:results}. We conclude in Section~\ref{sec:conclusions}.

\section{Data}
\label{sec:data}

This section outlines the datasets used in our analysis, including the photometric galaxy sample, redshift calibration inputs, and external datasets employed for validation and cross-correlation.

\subsection{DES data products}

The Dark Energy Survey (DES) is a photometric galaxy survey that took data with the Dark Energy Camera \citep[DECam,][]{Flaugher2015} on the 4-m Blanco Telescope at the Cerro Tololo Inter-American Observatory (CTIO) in Chile during the period of August 2013 to January 2019. This paper uses the full six years of data. We describe the main ingrediants used in this paper below.

\begin{itemize}
\item \textbf{DES Y6 \texttt{Gold} catalog:} Described in \citep*{y6-gold}, this catalog is the foundation for most of the Y6 data products. It includes the final coadded $grizY$ photometry, object classifications, masks, and quality flags. Of the different flux estimates provided in this catalog, we use the \texttt{Fitvd} fits to each object. The full catalog contains 448 million galaxies and 120 million stars. 
\item \textbf{DES Y3$^\star$ deep field catalog:} 
We use an updated version (hence the $^\star$) of the catalog originally presented in \citep{Hartley2022}. In particular, we start from the same set of detected objects as \citep{Hartley2022} and then re-measure the eight-band photometry for each object using an updated fitting algorithm. We note that the update is only in the fitting algorithm and \textit{not} in the images. The motivation for developing this updated pipeline is to enable its application to both the deep- and wide-field \texttt{Gold} fluxes. We apply a number of quality cuts to this deep-field sample, following \citep{Hartley2022}, to remove objects in regions of poor image quality or regions without eight-band coverage. Finally, following previous redshift calibration work in DES, we discard any deep-field galaxies that are never detected in our wide-field; this is determined using the synthetic source catalog described below. Figure~\ref{fig:redshift_samples_deep_fields} depicts the four deep fields and their overlap with redshift samples (see next section \ref{sec:redshiftcatalogs}).

\item \textbf{DES Y6 synthetic source catalog (\texttt{Balrog}):} In this work, we use the \texttt{Balrog} catalog derived in \citep{y6-balrog} to connect the deep-field photometry to the wide-field photometry which we refer to as the``transfer function'' in this work. Galaxies from the deep-field catalog are injected into the real CCD images, with each galaxy being injected multiple times, and the modified images are processed through the entire DES pipeline. As a result, for each deep-field object we can compute the distribution of its measured, wide-field properties, i.e., the transfer function. Compared to the dataset in DES Y3, the Y6 dataset spans the entire survey footprint \citep*[described in][]{RodriguezMonroy2025} and is further optimized for improving the precision on estimates of the transfer function. The underlying injection and processing pipeline remains largely unchanged from Y3, aside from minor updates such as the use of PIFF models \citep[PSFs In the Full FOV]{y6-piff} and an improved weighting scheme, which are described in Section 3.4 of \citep{y6-balrog}.

\item \textbf{The \maglimpp lens catalog:} The \maglimpp sample in DES Y6 is described in \citep*{y6-maglim}, and consists of the same basic magnitude limited selection from DES Y3 \citep{y3-2x2ptaltlensresults}, with additional refinements to increase sample purity and mitigate systematic contamination. 
We only summarize the main characteristics of the sample here, and refer the reader to \citep*{y6-maglim} for more details.

As in DES Y3, the primary selection is performed on galaxies in the \texttt{Gold} catalog using the following $i$-band limiting magnitude threshold that depends on the predicted mean photometric redshift of each galaxy, estimated from the Directional Neighborhood Fitting algorithm \citep[DNF,][]{DeVicente2016}:
\begin{equation} \label{eq:maglim_equation}
    i < 4 \times z_{\rm MEAN} + 18, \quad  
 i > 17.5.
\end{equation}
This cut is designed to create a magnitude-limited sample with a smooth redshift-dependent selection function, balancing between density and photometric redshift accuracy. The selection has been optimized in \cite{annamaglim}, with the goal of improving the cosmological constraints, as quantified by the figure of merit on $\Omega_{\rm m}$, $\sigma_8$, and the dark energy equation of state, $w$. We then split the sample in 6
tomographic bins from $z_{\rm MEAN} = 0.2$ to $z_{\rm MEAN} = 1.05$, with bin edges $[0.20, 0.40, 0.55, 0.70, 0.85, 0.95, 1.05]$.
For DES Y6, two additional selection criteria were introduced to 
further refine the sample. First, the redshift-bin optimized star-galaxy classifier of \citep{y6-maglim} is applied to remove residual stellar contamination. This leverages near-infrared (NIR) data from \texttt{unWISE} \citep{unwise2019} to identify the optimal color cuts in $(r-z,\ z-W1)$ to separate stars and galaxies in each redshift bin, complementing the morphological \texttt{EXTMASH} star-galaxy classifier in the \texttt{Gold} sample, and removing $\sim0.6 - 3.3\%$ of objects from different bins.



Second, \citep*{y6-maglim} apply a novel non-parametric selection based on Self-Organizing Map (SOM). Briefly, the SOM serves as an unsupervised clustering and dimensionality reduction algorithm that maps the four-dimensional color space onto a 2D grid, see Section \ref{sec:methodology} for details. It is trained on the four-band \textit{griz} photometry of the sample to remove objects that occupy compact regions of color space with large photo-$z$ dispersion. Such dispersion comes from low precision in photo-$z$ estimation in this region as a result of degeneracies in the color-redshift relation or limitations in the training data. In either case, redshifts of galaxies in such regions are more likely to be mismodeled and thus function as contaminants to the sample. 

Three complementary measures are used to identify compact regions in the SOM with high levels of photo-$z$ dispersion at either the galaxy or cell level, and objects in these cells are removed from the selection (amounting to $2.4\%$ of the total). Cross-matching these removed objects with the VISTA survey reveals that they are primarily QSOs \citep[Quasi-Stellar Objects,][]{McCracken2012} .




These two improvements in \maglimpp result in a cleaner lens sample in three ways: 1) reducing stellar contamination using a combination of morphological and redshift-bin optimized NIR color-based criteria, 2) removing contamination from redshift-interloper regions through precise color cuts of regions with unreliable photometric redshifts, and 3) sharpening the clustering signal by removing objects more likely to populate the tails of the $n(z)$.

These refinements lead to a more homogeneous sample with reduced systematic uncertainties, enhancing its suitability for cosmological analyses.

\end{itemize}

\subsection{The redshift catalogs}\label{sec:redshiftcatalogs}
\begin{figure}
\includegraphics[width=\linewidth]{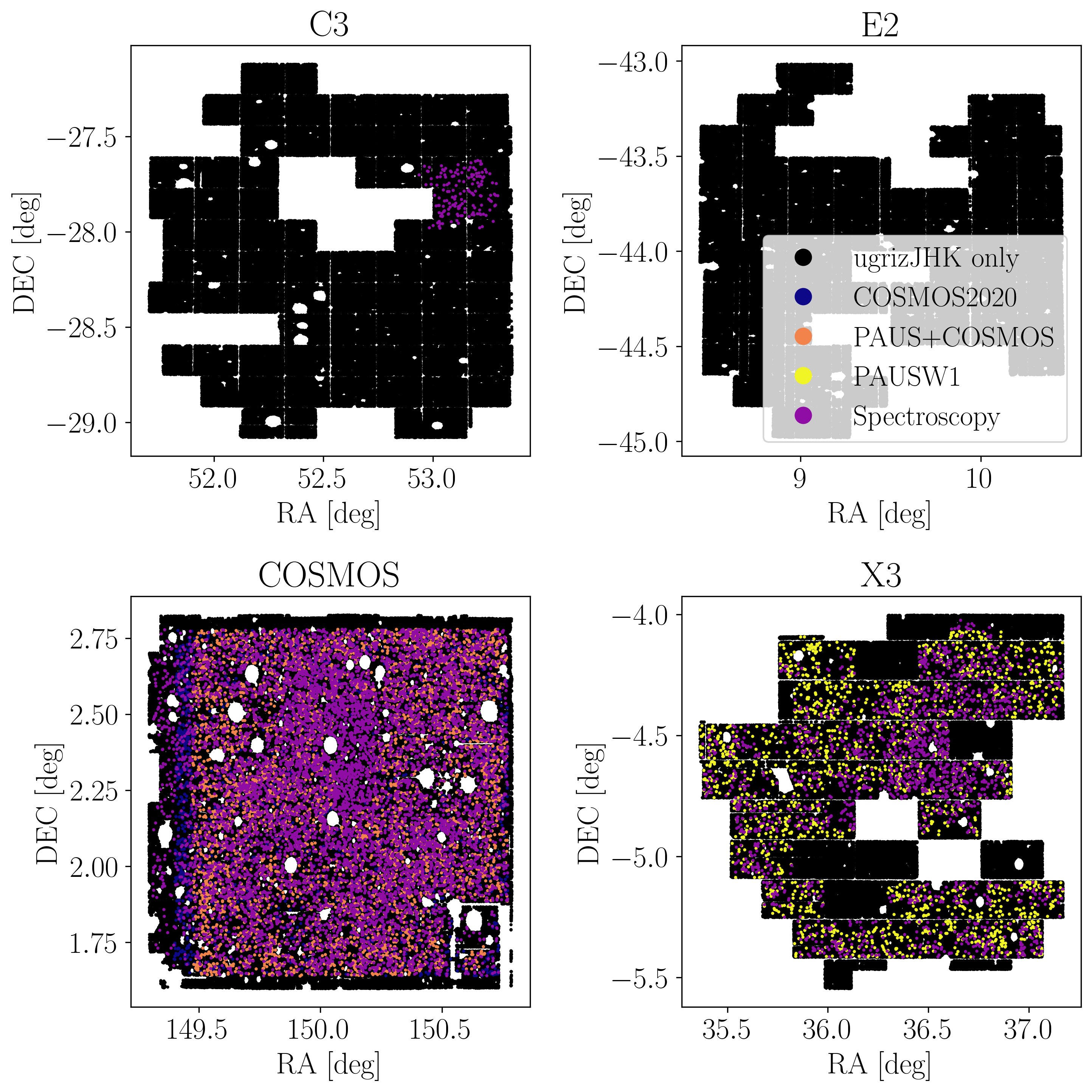}
\caption{The four DES deep fields used for our redshift analysis, which includes overlapping deep DES ugriz bands and VIDEO or UltraVISTA JHK bands, as compiled from \citep{Hartley2022}. Black points indicate DES deep-field galaxies with no redshift information, while points with colors show galaxies with spectroscopy or from COSMOS2020, PAUS+COSMOS or PAUSW1, as indicated in the legend.}
\label{fig:redshift_samples_deep_fields}
\end{figure}

\begin{figure} 
\includegraphics[width=\linewidth]{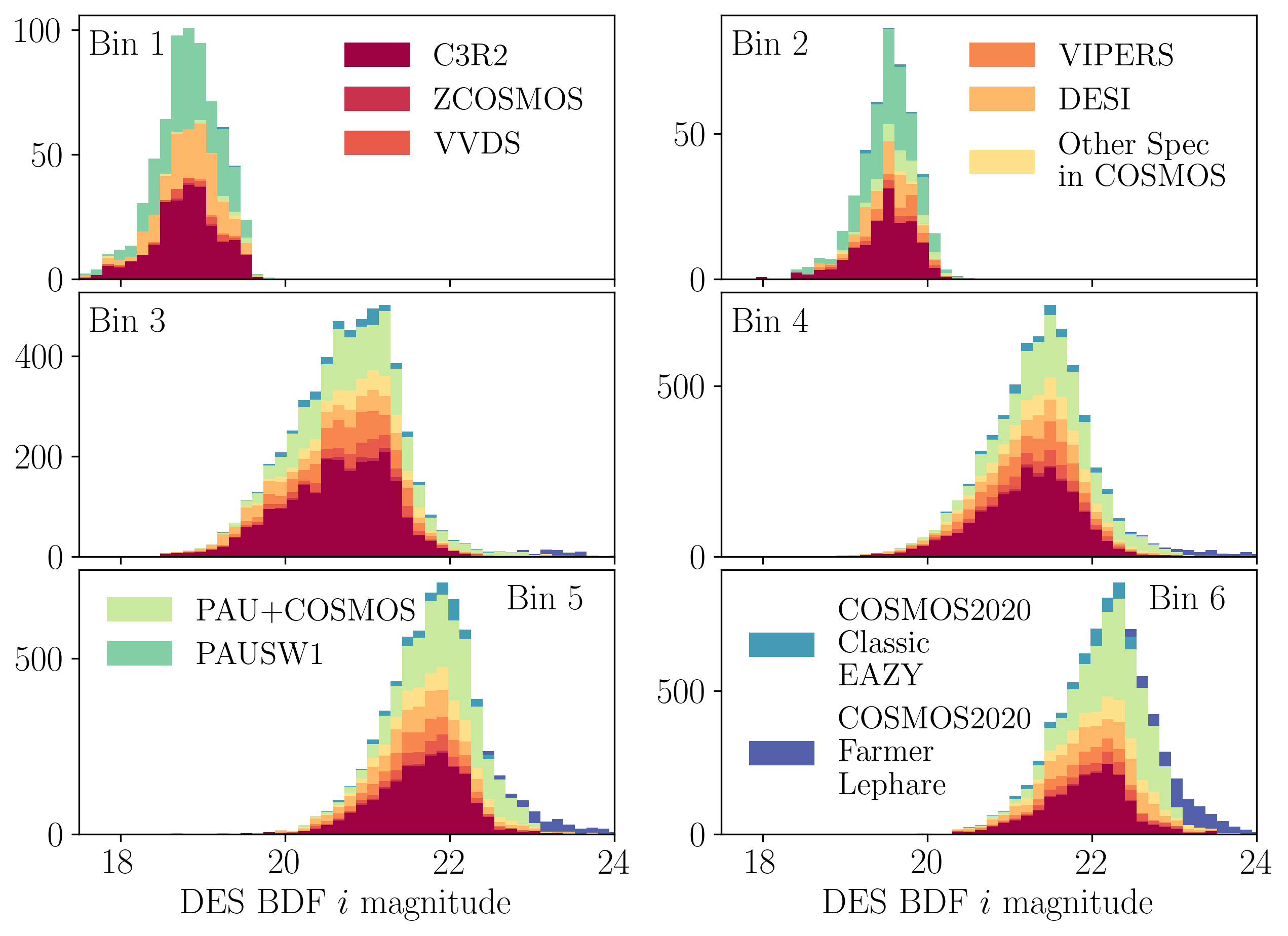}
\caption{Distribution of redshift samples as a function of the deep field DES $i$-band magnitude. Each galaxy in these stacked histograms is weighted by the \texttt{Balrog} probability of detection and selection into each Maglim sample (different rows). For details on the definition of the ‘Bulge Plus Disk, Fixed Ratio’ (BDF) galaxy profile see \citep{Hartley2022}.}
\label{fig:redshift_samples}
\end{figure}

Our analysis relies on the use of galaxy samples with known redshift and deep-field photometry. To this end, we use catalogues of both high-resolution spectroscopic and multi-band photometric redshifts, and combine them into a redshift calibration sample, prioritizing information coming from spectroscopic surveys. In addition, we develop a model to propagate the uncertainty in our redshift calibration arising from residual biases and uncertainty in the original redshift catalogues (quantified in Section~\ref{sec:redshiftsampleunc}). 

\begin{figure}
\includegraphics[width=\linewidth]{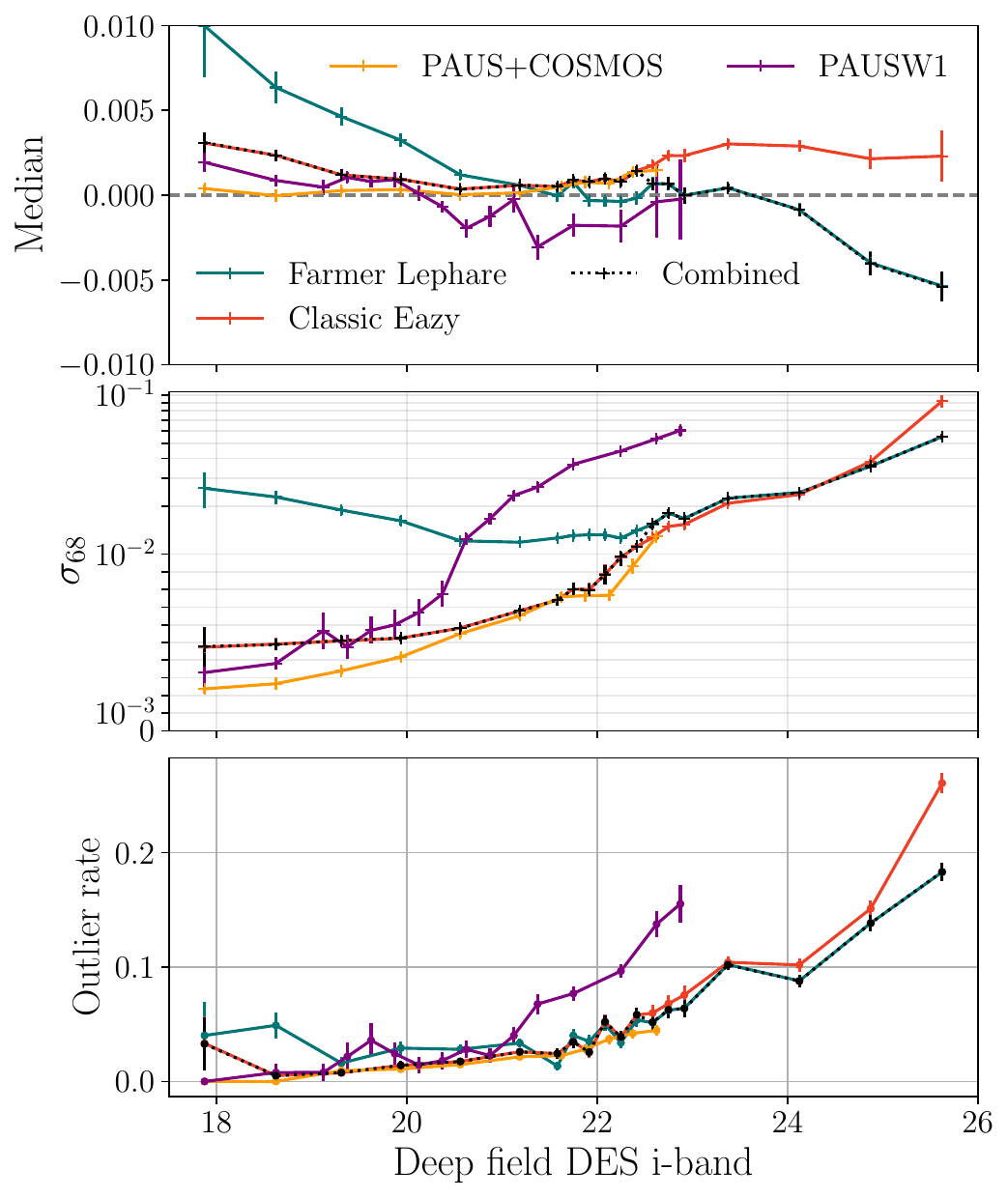}
\caption{Redshift quality of different redshift catalogues used for the redshift calibration of \maglimpp samples, relative to spec-z measurements for the same objects, using $\Delta_z/(1+z_{\rm spec})$. We define outliers as having $|\Delta_z|/(1+z_{\rm spec})>0.1$. The redshift catalogues shown are PAUS+COSMOS; PAUS+W1; and then Classic-EAZY and Farmer-Lephare from the COSMOS2020 release (see text). Relative to each other, Classic-EAZY performs better at $i<22.5$, while Farmer-Lephare performs better at $i>22.5$, in terms of median bias and photo-z error (top and middle panels). The PAUS-W1 is used for $i<20.5$, where it shows excellent photo-$z$ precision and low bias.}
\label{fig:redshift_samples_quality}
\end{figure}

Figure~\ref{fig:redshift_samples_deep_fields} shows all the DES deep fields and the distribution of different photometric and spectroscopic datasets that provide redshift information for the \maglimpp sample. Primarily, we use redshift information coming from the COSMOS and X3 DES deep fields. The spectroscopic catalog we use contains both public and private spectra from multiple surveys compiled by \citep{Gschwend2018}, and references therein. A majority of these spectra come from: zCOSMOS \citep{Lilly09zcosmos}, C3R2 \citep{Masters2017,C3R2_DR2, C3R2_DR3}, VVDS \citep{vvds}, VIPERS \citep{scodeggio2018}, and DESI EDR \citep{desiedr}. In addition to spectroscopic surveys, we use three multi-band photo-$z$ catalogues: 1) the COSMOS2020 photometric redshift catalogue \citep{cosmos2020}, which includes 30 broad, intermediate, and narrow bands covering the UV, optical, and IR regions of the electromagnetic spectrum in the COSMOS field \citep{Scoville2007_COSMOS}; 2) the PAUS+COSMOS 66-band photometric redshift catalogue \citep{Alarcon2020}, also in the COSMOS field, from the combination of PAU Survey data  in 40 narrow-band filters \citep{paucam,eriksen2019} and 26 COSMOS2015 bands excluding the mid-infrared \citep{Laigle2016}; and finally 3) the PAUS-W1 \citep{PAUSW1} catalogue, which is part of the $\sim 50\,{\rm deg}^2$ complete PAUS survey that overlaps with the CFHTLS-Wide W1 field and the DES X3 deep field.

Figure~\ref{fig:redshift_samples} shows the $i$-band magnitude distribution of the galaxies in the redshift sample, split by tomographic bin (panels) and color-coded by the dataset providing the redshift. Each galaxy is weighted by its \texttt{Balrog}-based detection and selection probability into the corresponding \maglimpp tomographic bin. This weighting ensures that the redshift sample reflects the same selection as the lensing sample, thus enabling unbiased calibration. The figure also highlights the diversity of redshift sources used, including both spectroscopic surveys (e.g., C3R2, VIPERS, ZCOSMOS, DESI) and high-quality photometric catalogs (e.g., PAUS, COSMOS2020). Notably, the transition in redshift source composition with increasing bin number and magnitude indicates the varying depth and availability of redshift information across the sample.

An important consideration when building a redshift calibration sample is its completeness and redshift quality. An incomplete redshift sample will result in a biased $n(z)$ inference due to selection effects \citep[e.g.][]{Hartley2020, gruen2016, photozreview} that can be particularly challenging to forward model. Only galaxies with reliable redshifts -- whether spectroscopic or from COSMOS2020 photo-$z$ catalogs -- are included. We emphasise that we restrict our calibration to the regions where these samples are complete across the full magnitude range of interest, minimizing the risk of selection biases that would otherwise impact the inferred $n(z)$. For this reason, to construct the redshift sample we only use areas in the deep fields with complete redshift coverage for all galaxies that could be observed as \maglimpp galaxies. This results in excluding fields such as C3 and E2, which contain very little redshift information, as visible in Figure~\ref{fig:redshift_samples_deep_fields}. In contrast, both the COSMOS and X3 fields have a large spatial coverage. In particular, we use COSMOS-field galaxies to inform all tomographic bins, while we only use X3-field galaxies to inform the first two tomographic bins. This is motivated by the quality of the redshift information available in each field; while COSMOS has high quality photometric redshifts for all galaxies down to $i\sim25$, in the X3-field this is only true down to $i\sim20.5$. For reference, the maximum (observed) magnitude for each tomographic bin is $i_{\rm max} = [19.6, 20.2, 20.8, 21.4, 21.8, 22.2]$  (Eq.~\ref{eq:maglim_equation}), which motivates only using the X3-field for the first two tomographic bins.

Figure~\ref{fig:redshift_samples_quality} shows the redshift performance of the different photo-$z$ catalogues used in the COSMOS and X3 fields as a function of deep DES BDF $i$-band magnitude.  To evaluate this performance, we compute $(z_{\rm phot} - z_{\rm spec}) / (1 + z_{\rm spec})$ for each object and derive the median, the scatter (denoted as $\sigma_{68}$), and the outlier rate of the resulting distributions. The top panel shows the median redshift bias, the middle panel displays the scatter $\sigma_{68}$, and the bottom panel reports the outlier rate — defined as the fraction of galaxies for which $|z_{\rm phot} - z_{\rm spec}| > 0.1(1 + z_{\rm spec})$.
When we have redshift information from more than one source, we prioritize the most reliable one, as detailed in Figure~\ref{fig:redshift_samples_quality}. In the COSMOS field, we first use spectroscopic redshifts, then PAUS+COSMOS, and finally COSMOS2020. Within the COSMOS2020 release there are two types of photometric catalogs (Classic and Farmer) each providing two redshift estimates, from \texttt{LePhare} and \texttt{EAZY}. 

Based on our assessment of the redshift quality in Figure~\ref{fig:redshift_samples_quality}, we use COSMOS2020 Classic-EAZY for galaxies with $i < 22.5$, and Farmer-LePhare for galaxies with $i > 22.5$. The Classic-EAZY combination shows the lowest median bias and smallest scatter at bright magnitudes, while Farmer-LePhare performs better at the faint end, at least up to $i < 24$, which is the relevant deep magnitude range for \maglimpp. In the X3 field, we prioritize spectroscopic redshifts, followed by PAUS-W1. PAUS-W1 galaxies are only used for $i < 20.5$, where their photometric redshifts show low bias and excellent precision. As a result, the X3 field contributes to the first two \maglimpp tomographic bins alongside COSMOS. For the remaining four \maglimpp bins, we rely solely on the COSMOS field, as X3 lacks reliable redshift coverage at fainter magnitudes (see Figure~\ref{fig:redshift_samples}).

\section{SOM-based Redshift Inference Methodology}
\label{sec:methodology}

Our calibration process to determine the redshift distributions of the \maglimpp galaxies in DES Y6 incorporates both SOM-based photometric redshift estimates and clustering redshift constraints. This section provides a comprehensive account of the Self-Organizing Map Photometric Redshift (SOMPZ) method, refining the approach explored by \cite{carrascosom}, and  previously employed in Y3 for the source galaxies \citep{y3-sompzbuzzard}\citep*{y3-sompz}, and later applied to the lens galaxies \citep{Giannini2024}. The SOMPZ effort for the \maglimpp sample is fully documented in this work. 

\subsection{Self-Organizing Maps for Photometric Redshift Calibration}\label{sec:soms}

The Self-Organizing Map (SOM) is an unsupervised machine learning algorithm that projects high-dimensional data onto a two-dimensional grid while maintaining the relative proximity of similar data points. Its key strength is in capturing the structure of complex datasets by organizing inputs with similar properties into neighboring cells.

In this work, the input to the SOM consists of galaxy flux measurements across multiple photometric bands. Each cell on the map represents a set of galaxies with similar observed fluxes. Owing to redshift degeneracies, a given cell does not correspond to a single redshift value but instead maps to a distribution of redshifts. Beyond clustering galaxies with similar fluxes, the SOM provides a topological representation of the data, preserving relationships between neighboring cells. This structure allows us to visualize transitions between different galaxy populations and identify distinct features such as Lyman breaks or contamination from stars. These properties help refine data selection and assess potential systematics.

\begin{figure*}
\includegraphics[width=0.49\linewidth]{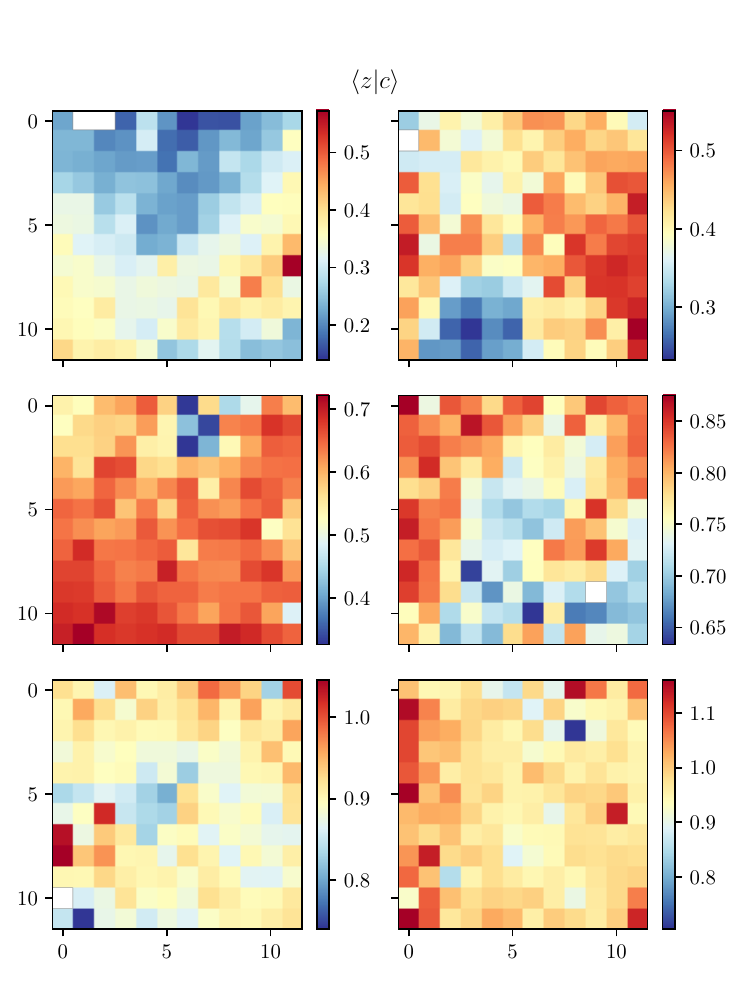}
\includegraphics[width=0.49\linewidth]{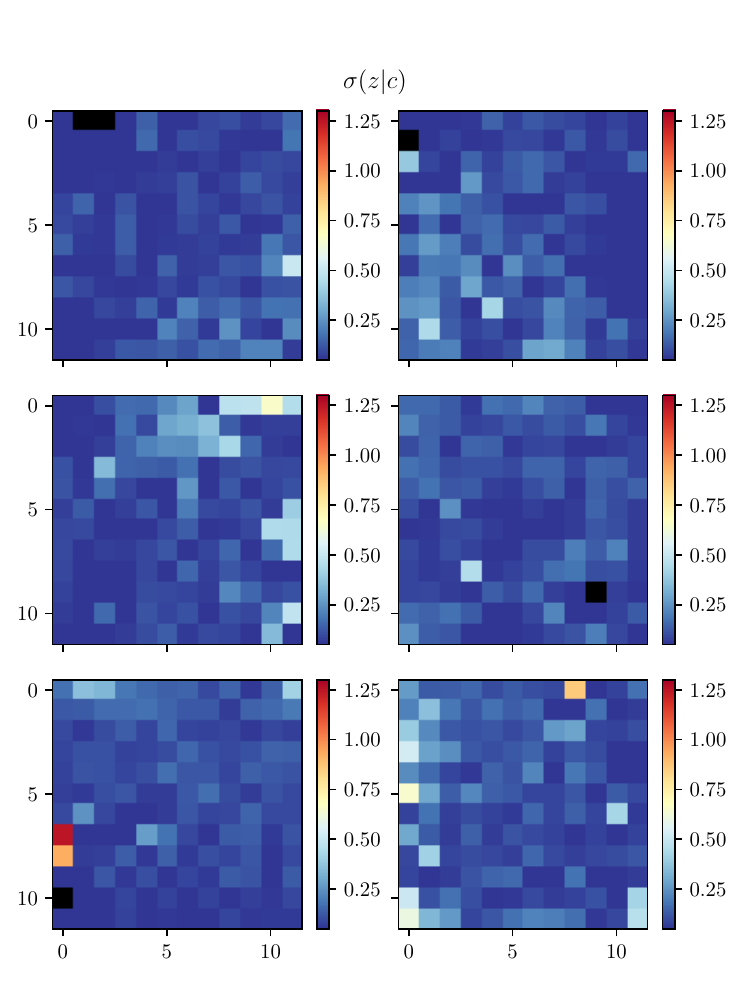}
\caption{Visualisation of the 6 deep SOMs, one for each tomographic bin (ordered from left to right and top to bottom, i.e., 1–2 on the first row, 3–4 on the second, 5–6 on the third),  each composed of 144 cells (12$\times$12). Left: average redshift for each deep SOM cell $c$. Right: standard deviation on the redshift distribution for each deep SOM cell $c$. The black cells in the deep SOM are due to the lack of spectroscopic information in those regions of the color space, i.e., there are no galaxies in the redshift samples that were assigned to those cells.}
\label{fig:deepsom}
\end{figure*}

\begin{figure*}
\includegraphics[width=0.49\linewidth]{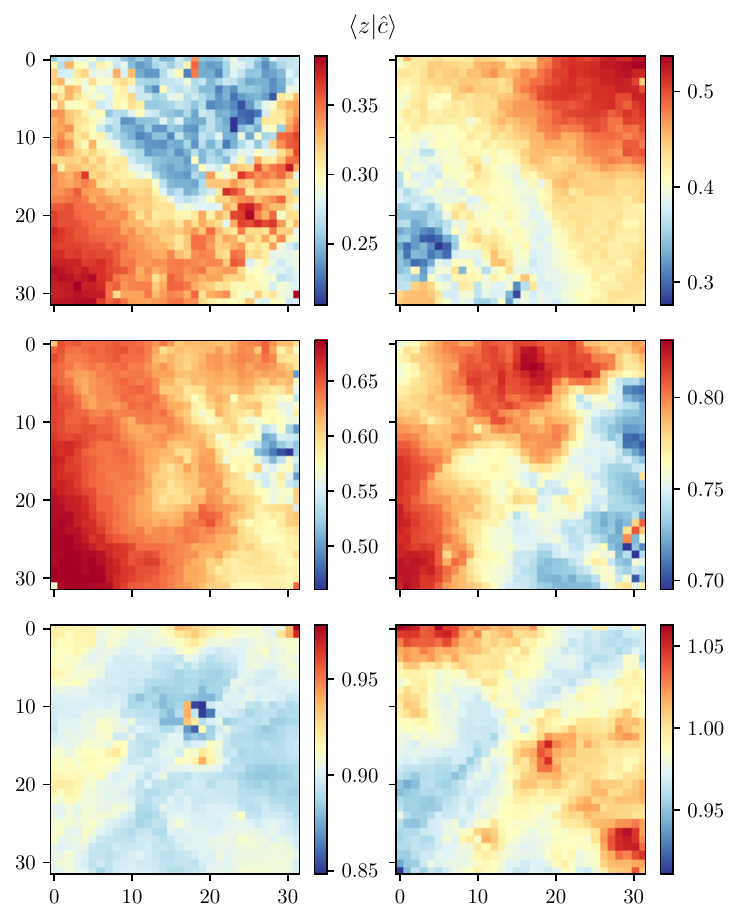}
\includegraphics[width=0.49\linewidth]{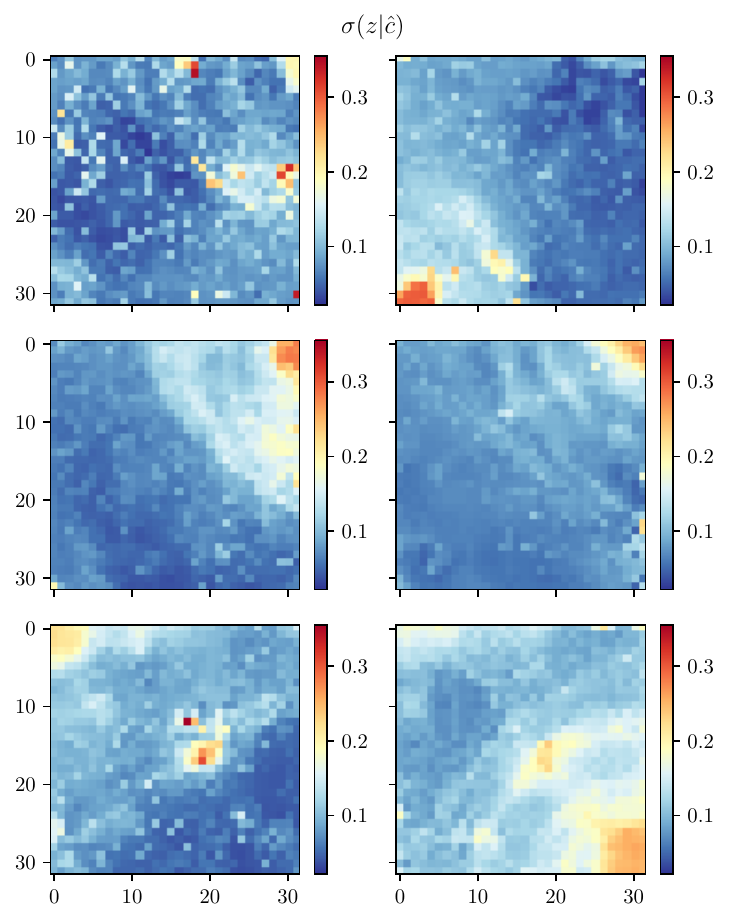}
\caption{Visualisation of the 6 wide SOMs, one for each tomographic bin, each composed of 1024 cells (32$\times$32). Left: average redshift for each wide SOM cell $\hat{c}$. Right: standard deviation on the redshift distribution for each wide SOM cell $\hat{c}$.}
\label{fig:widesom}
\end{figure*}

The SOM process consists of two stages: 

\begin{enumerate}
    \item \textbf{Training:} The SOM is initialized with a grid of cells, each with a representative set of flux values. During training, galaxies are sequentially mapped to the closest-matching cell based on a predefined metric. The winning cell and its neighboring cells are then updated to better match the input flux values, with updates decreasing over time to stabilize the mapping. This results in a structured representation where neighboring cells contain similar galaxy populations.
    
    \item \textbf{Assignment:} Once trained, the SOM can be used to classify new galaxies by assigning them to the closest-matching cell, effectively categorizing galaxies into different phenotypes based on their photometric properties.
\end{enumerate}

To determine which SOM cell most closely matches a galaxy, DES Y6 adopts the revised distance metric of \cite{Sanchez2020}, applied to data in \citep{Campos2024}.  
Unlike the Y3 formulation, which compared raw color–magnitude vectors at equal weight, the Y6 metric works directly in flux space and explicitly accounts for photometric uncertainties: bands with low signal-to-noise ratio (S/N) are down-weighted, so noisy measurements cannot dominate the match.  
This improvement is crucial for faint sources and still beneficial, though subtler, for the comparatively bright \maglimpp\ sample. This construction boosts the influence of high-quality bands, suppresses noisy measurements, and remains insensitive to absolute flux differences.

Besides the new metric, the Y6 SOM employs non-periodic map boundaries, better reflecting real flux behaviour, and adds the \(g\) band to the \(riz\) set used in Y3, further sharpening the color space mapping \citep{Campos2024}. Consequently, the DES Y6 SOM yields a smoother and more physically meaningful organisation of galaxies in color space, improving the stability of photometric-redshift calibration.

\subsubsection{Redshift Calibration via Deep and Wide SOMs}

The SOMPZ 
methodology was first developed by DES in \cite{y3-sompzbuzzard}. This technique utilizes SOMs to classify galaxies into distinct phenotypes based on their observed properties (e.g., flux, colors). The redshift calibration of each phenotype is achieved by incorporating galaxies with high-quality redshift measurements into the same SOM. This approach employs two separate SOMs: one representing the color space of the sample needing calibration (\textit{wide SOM}) and another constructed from the DES deep fields (\textit{deep SOM}). A direct calibration of \maglimpp using a single SOM would be inadequate, as broad-band DECam photometry lacks the precision needed to resolve certain redshift-type degeneracies.The overlap of the DES deep fields with spectroscopic and narrow-band photometric surveys provides additional wavelength coverage, including near-ultraviolet and near-infrared bands (\textit{u}, $J$, $H$, $K_s$), which helps mitigate these degeneracies. More details are provided in Section \ref{sec:data}.

To bridge the deep SOM and the wide SOM, an additional intermediary is needed to transfer redshift information across different observational conditions. This transfer function is constructed using \texttt{Balrog} simulations described in section \ref{sec:data} and implemented through a Bayesian framework, ensuring consistency between the deep and wide observations. 

The two-tiered SOM redshift calibration method was applied to observational data for the first time using the Y3 dataset \citep*{y3-sompz, Giannini2024}. The calibration pipeline follows these steps:

\begin{enumerate}
    \item Train a deep SOM using deep-field photometry ($ugrizJHK_s$ bands) and a wide SOM with wide-field photometry ($griz$ bands);
    \item Assign galaxies from the deep sample to the deep SOM and those from the wide sample to the wide SOM, yielding probability distributions $p(c)$ and $p(\hat{c})$ of a galaxy being assigned to a specific deep cell $c$ or wide cell $\hat{c}$, respectively;
    \item Cross-match galaxies with known redshifts from spectroscopic and high-quality photometric samples (e.g., COSMOS, PAUS) to deep SOM cells, deriving the redshift distribution of each deep cell $p(z|c)$;
    \item Assign \texttt{Balrog} galaxies to the wide SOM, connecting the deep and wide photometric spaces through $p(c|\hat{c})$.
\end{enumerate}

Combining this information, the final redshift distribution for each tomographic bin is computed as:
\begin{equation}
    p(z) = \sum_{\hat{c}} \sum_c  p(z | c) p(c) p(c | \hat{c}) p(\hat{c}),
\end{equation}
where 
We can rewrite this equation in terms of the data products used: 
\begin{equation}
\label{eqn:pz_uncertainty}
p(z) = \sum_{\hat{c}} \sum_c 
\underbrace{p(z|c)}_{\text{Redshift}} 
\underbrace{p(c)}_{\text{Deep}} 
\underbrace{\frac{p(c, \hat{c})}{p(c) p(\hat{c})}}_{\text{Balrog}} 
\underbrace{p(\hat{c})}_{\text{Wide}}.
\end{equation}
This formulation effectively reweights the deep-field redshift sample to match the selection function of the wide-field survey, ensuring a representative redshift distribution for weak lensing analyses.

The choice of SOM resolution follows the approach adopted in DES Y3: since the \texttt{MagLim++} sample did not change significantly in color or depth, we apply the same configuration as in Y3—using a $12 \times 12$ grid for each deep SOM and a $32 \times 32$ grid for each wide SOM. Because the SOM is trained and populated using the full deep photometric sample, some cells—especially those corresponding to rare or extreme color–magnitude combinations—may contain only galaxies without redshift measurements, even though the redshift sample itself is complete within its footprint. These empty cells are therefore a consequence of incomplete color–space overlap between the photometric and redshift samples, not of missing redshift measurements for galaxies in the latter.

This setup balances the need for resolution with the requirement that the fraction of galaxies mapping to deep cells lacking redshift information remains below 1\%, a condition we verified also holds in Y6.

Figure~\ref{fig:deepsom} and Figure~\ref{fig:widesom} showcase the six deep SOMs and six wide SOMs, respectively. 
The left panel of Figure~\ref{fig:deepsom} displays the mean redshift $(\langle z|c \rangle$) associated with each deep SOM cell for each tomographic bin. As expected, the redshift structure across the SOM reflects the underlying color–magnitude distribution of galaxies: contiguous regions in SOM space typically correspond to galaxies with similar photometric properties and, consequently, similar redshifts. Each tomographic bin highlights distinct regions of the SOM, illustrating how the binning aligns with the SOM’s learned structure. The progression of redshift patterns across bins confirms the consistency of the binning strategy and demonstrates the SOM’s effectiveness in organizing photometric observables in redshift space. The right panel of Figure~\ref{fig:deepsom} instead presents the corresponding redshift scatter, $\sigma(z|c)$, for each deep SOM cell. Cells with low $\sigma(z|c)$ indicate regions of high redshift precision, where the photometric observables tightly constrain the redshift distribution. Conversely, cells with high redshift scatter often trace color--magnitude degeneracies or lower signal-to-noise regimes, where photometric redshifts become less reliable. These maps are useful for identifying regions of high redshift uncertainty that may be down-weighted or excluded in the final analysis.

To better understand the origin and impact of these high-scatter cells, we examine their occupancy by both the deep \texttt{Balrog} photometric sample and the spectroscopic redshift sample in Figure~\ref{fig:deep_som_occupancy}. This analysis shows that high-$\sigma(z|c)$ cells are not generally associated with low occupancy or poor spectroscopic sampling. Instead, they reflect intrinsic limitations in the photometric data: in these regions, galaxies with similar photometric properties have broad or multimodal redshift distributions that cannot be cleanly separated, even with the deep multi-band photometry. We further quantify the effect of these cells by showing, in Figure~\ref{fig:som_occupancy_vs_sigmaz}, the number of galaxies per $\sigma(z|c)$ bin for both the deep and wide samples. These histograms confirm that high-scatter cells  are extremely rare and contribute negligibly to the overall redshift calibration. For more details, see Appendix~\ref{app:soms}.

Figure~\ref{fig:widesom} shows the corresponding information for the wide SOMs. The left panel displays the average redshift per each wide SOM cell, \(\langle z|\hat{c} \rangle\), while the right panel shows the redshift scatter \(\sigma(z|\hat{c})\). Although the wide data have increased photometric noise, clear redshift gradients are still visible across SOM space, indicating that the wide photometry retains significant discriminatory power. Compared to the deep SOMs, the wide SOMs show smoother and more coherent features, with well-defined regions of increasing redshift. The redshift scatter \(\sigma(z|\hat{c})\) in the right panel shows that the highest uncertainties tend to lie at the edges and corners of the SOMs—regions that typically correspond to low-density or degenerate areas of the color--magnitude space.

We further assess the potential impact of these high-\(\sigma(z|\hat{c})\) regions in Figure~\ref{fig:som_occupancy_vs_sigmaz}, which shows the number of wide-field galaxies per value of \(\sigma(z|c)\). This distribution confirms that only a small fraction of the wide sample lies in SOM cells with large redshift scatter, reinforcing the conclusion that the overall redshift uncertainty remains well controlled even in the wide data.

In this work, the training phase was performed using a subset of the deep sample, which differed from the fiducial one by 1\%. Catastrophic outliers were removed by eliminating objects with non-physical colors (i.e. colors required to be $>-1$). This step was necessary to prevent these extreme cases from disproportionately influencing the training process. The entire deep sample was then assigned to the deep SOM, ensuring that redshift characterization remained stable. Although this process may introduce double peaks in some SOM cells, the impact is expected to be smaller than the risk of completely misclassified cells.

\section{Characterization of Sources of Uncertainty}
\label{sec:uncertainty}

\begin{figure*}
    \centering
    \includegraphics[width=0.9\linewidth]{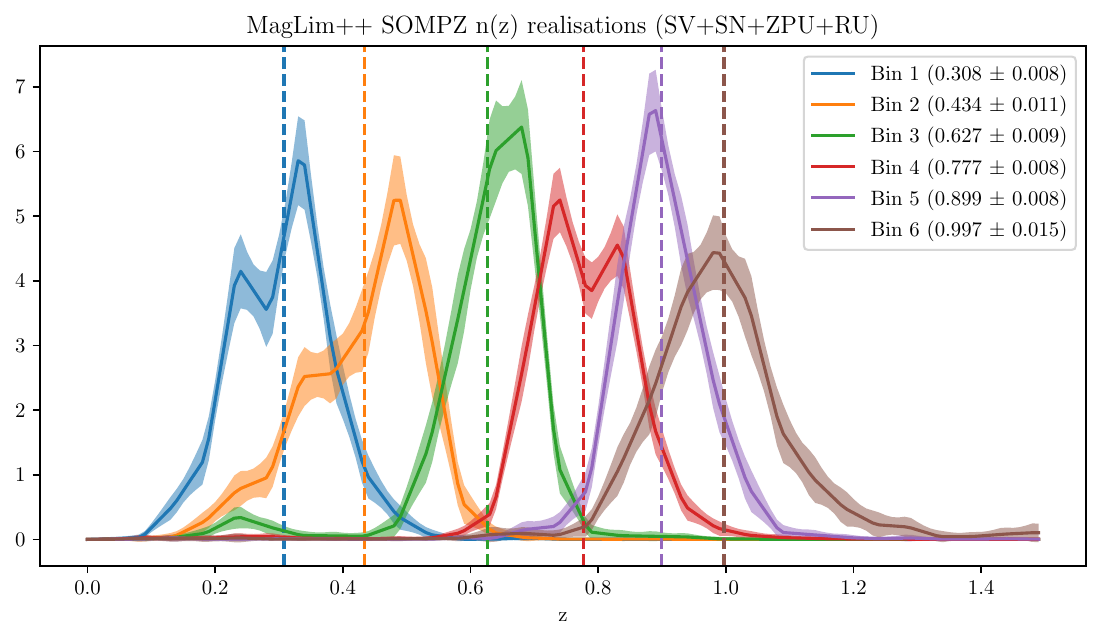}
    \caption{Ensemble of 100M $n(z)$ distributions for the \maglimpp lens sample, for each of the six tomographic bins. The spread across realizations reflects the four sources of uncertainty included in cosmological analyses: sample variance, shot noise, zero point, and redshift uncertainty. The legend reports also the mean redshift and its uncertainty.}
    \label{fig:nzplot}
\end{figure*}

In this section we describe the process we employ to characterize the uncertainties in the SOMPZ redshift distribution. We consider the following four sources of uncertainties: 
\begin{itemize}
    \item \textit{Sample variance (SV)} of the deep and redshift samples, due to the limited area of the deep fields ($\sim 9\,\mathrm{deg}^2$). This is modeled via the \textit{three-step Dirichlet} (3sDir) analytical model described in Section~\ref{sec:3sdir}.
    \item \textit{Shot noise (SN)}, arising from the finite number of galaxies in the deep and redshift samples. This is also captured within the 3sDir framework.
    \item \textit{Photometric calibration uncertainty (ZPU)}, introduced by errors in the deep-field photometric zero-points. This is modeled by perturbing the deep photometry and rerunning the SOMPZ pipeline, as described in Section~\ref{sec:zero pointunc}.
    \item \textit{Redshift sample uncertainty (RU)}, which accounts for residual biases in the redshift sample itself. This is modeled via coherent shifts in magnitude–redshift bins based on spectroscopic comparisons, as described in Section~\ref{sec:redshiftsampleunc}.
\end{itemize}

Compared to the DES Y3 framework in \citep{Giannini2024}, two sources of uncertainty are no longer included in this work. The uncertainty introduced by discretizing the continuous color space was gauged to be negligible in Y3. Similarly, the uncertainty in \texttt{Balrog}, which was also estimated as negligible in Y3, is even less significant given that the Y6 \texttt{Balrog} sample is approximately five times larger than in Y3 (see Section~\ref{sec:data}). Potential systematic effects—e.g., from mismatches between injected and real sources—are expected to be subdominant, as the injection and measurement pipeline is largely unchanged from Y3 aside from minor improvements in PSF modeling and weighting.

\begin{figure*}
    \includegraphics[width=0.49\linewidth]{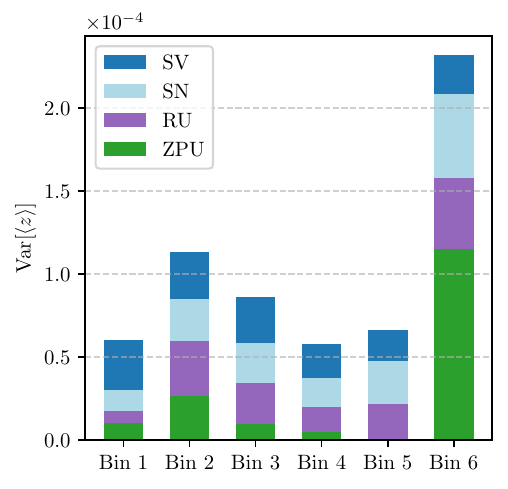}
    \includegraphics[width=0.49\linewidth]{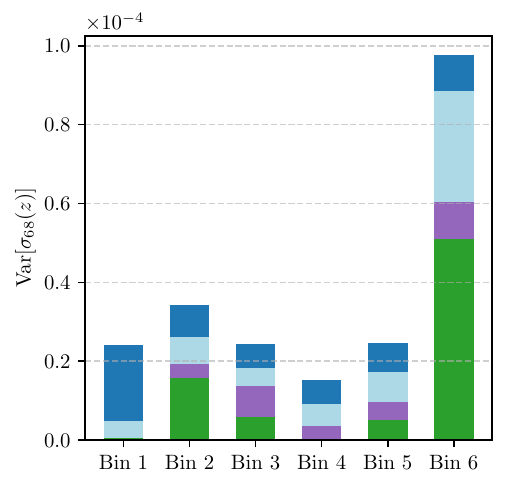}
    \caption{The relative contribution of each source of uncertainty for the mean (left) and the width (right) for each of the six tomographic bins. We show the individual contributions from sample variance (SV, blue), shot noise (SN, light blue), redshift sample uncertainty (RU, purple), and photometric zeropoint uncertainty (ZPU, green). No single source dominates all bins or metrics, underscoring the need to propagate all uncertainty components when modeling \(n(z)\). In bin 6, where the redshift distribution is broader and more sensitive to magnitude-dependent shifts, ZPU drives a particularly large uncertainty on the width. The magnitude of the effect varies by source and bin, and in some cases only the mean or only the width shows a significant increase (e.g., \(>10\%\)). This reflects the non-linear nature of uncertainty propagation at the SOM \(n(z)\) level, which impacts the full distribution shape, with the mean and width being affected only indirectly as derived quantities. To illustrate the relative magnitude of each source of uncertainty in each bin, and its variation with redshift, we rescale the total variance in this figure to match the combined uncertainty (see Table~\ref{tab:unc}).
}
    \label{fig:unc_bar}
\end{figure*}

Following its application to Y3 data, the 3sDir method models the sample variance and shot noise uncertainties on the $n(z)$, by explicitly sampling fluctuations in both the redshift calibration sample and the deep field, based on finite galaxy counts in each. This results in a set of $n(z)$ realizations that reflect expected statistical fluctuations, without additional systematic shifts in redshift or color space. ZPU and RU are then incorporated on top of 3sDir, but differently than the latter, their handling has been changed in Y6: ZPU is modeled by perturbing deep-field photometry and recomputing SOM assignments, affecting how galaxies are mapped to redshift. RU is modeled by applying empirical redshift shifts of the narrow-band photometric redshift sample derived from spectroscopic comparisons, coherently across magnitude–redshift bins.

These perturbations can alter the effective shape of \(n(z)\) in a way that either broadens or narrows it, depending on the direction and magnitude of the induced shifts. In some bins, they may introduce coherent changes that reduce the high-redshift tails or shift galaxies toward narrower redshift distributions.  
Adding an extra category of uncertainties (e.g., redshift, zero-point) generally increases the uncertainty on the mean and width of the redshift distribution. The magnitude of this effect can vary by uncertainty type and may differ between the two metrics: in some cases, only the mean or only the width shows a significant increase (e.g., \(>10\%\)). This is because uncertainty propagation is a non-linear process, encoded at the SOM \(n(z)\) level rather than directly at the level of the mean and width. As a result, it impacts the full shape of the redshift distribution, with the mean and width being affected only indirectly as derived quantities. This highlights the limitation of using individual summary statistics—such as the mean or width—to represent the total uncertainty in a redshift distribution, as these metrics can fail to capture coherent variations in shape introduced by different uncertainty sources. In contrast, ensembles of \(n(z)\) realizations fully represent these variations and their impact on cosmological inference, as illustrated in Figure~\ref{fig:modes} and Table~\ref{tab:cosmology}, where RU and ZPU shift the dominant modes and their priors.

The final ensemble of $n(z)$ realizations from SOMPZ is shown in Figure~\ref{fig:nzplot}, which collectively spans the region where the true redshift distribution is likely to land. We generate these by generating 100 realizations for each of ZPU and RU sampled values which parametrise the corresponding uncertainty, and then generating 1,000,000 additional realizations for each of the 100 SOMPZ $n(z)$ estimates using the 3sDir model. This yields a total of 100,000,000 redshift distributions realizations per tomographic bin, capturing the full propagated uncertainties for the \maglimpp lens sample.

Figure~\ref{fig:unc_bar} presents the variance in the mean redshift $\langle z \rangle$ (left panel) and width parametrized as $\sigma_{68}$\footnote{\(\sigma_{68}\) is defined as half the width of the interval that encloses the central 68.25\% of the redshift probability distribution. It is computed as \((z_{84.075} - z_{15.825}) / 2\), where \(z_p\) is the redshift at the \(p\)th percentile of the cumulative distribution. This provides a robust measure of the redshift scatter, analogous to the standard deviation for non-Gaussian or asymmetric distributions.} (right panel) across six tomographic bins, decomposed by the source of the uncertainty. This comparison allows us to assess the relative contribution of sample variance (SV), shot noise (SN), redshift sample uncertainty (RU), and zero point uncertainty (ZPU) to the total redshift uncertainty. The contribution of each uncertainty source to the variance of the mean and width is computed by differencing the variance of ensembles of \(n(z)\) realizations with and without that source included (e.g., \(\sigma^2_{\mathrm{SV}} = \mathrm{Var}[\mathrm{SV+SN}] - \mathrm{Var}[\mathrm{SN}]\)), with any residual between the total variance and the sum of individual terms assigned to a cross-term accounting for non-linear interactions. In the following subsections, we describe the methodology used to model each component and interpret its behavior across bins in the context of the plots and numerical values.

In what follows, we describe in detail how each of the four uncertainty components—SV, SN, ZPU, and RU—is modeled and propagated.

\subsection{Sample Variance and Shot Noise}\label{sec:3sdir}

Most redshift distributions in photometric surveys are inherently uncertain due to two dominant sources of error: sample variance and shot noise. Sample variance arises from the fact that redshift calibration samples are typically drawn from small patches of the sky, meaning they do not perfectly represent the global distribution of galaxies due to fluctuations from large-scale structure. Shot noise, on the other hand, is a consequence of the finite number of galaxies in a given calibration sample, leading to Poisson fluctuations in redshift distributions. To properly propagate these uncertainties into cosmological analyses, we employ the \textit{three-step Dirichlet resampling (3sDir)} method developed in \cite{Sanchez2020}, which generates realizations of the redshift distribution that include both sources of uncertainty. This method was applied to the DES Y3 dataset for the calibration of the source galaxy redshifts \citep*{y3-sompz} and of the lens galaxy redshifts \citep{Giannini2024}. 

The 3sDir approach treats redshift calibration as a hierarchical process, in which the overall redshift distribution of a galaxy sample is built by combining the redshift distributions of smaller, distinct sub-populations. These sub-populations are typically defined based on observational or photometric properties (such as color or magnitude), and each is associated with its own underlying redshift distribution. The final, observed redshift distribution is then obtained by appropriately weighting and summing these individual components, reflecting the relative abundance of each sub-population in the full sample. Instead of assuming a fixed redshift distribution for a given calibration sample, we perturb it by drawing three sequential Dirichlet samples, each capturing a different contribution to the total uncertainty. The Dirichlet distribution is chosen because it is a natural way to model probability distributions while ensuring that all values remain positive and normalized to one. Each sampled realization of the redshift distribution is thus self-consistent, properly incorporating both shot noise and sample variance in a statistically robust manner.

The three steps of 3sDir correspond to different levels of uncertainty propagation, where the output of each step serves as the input for the next. These also require to define a \textit{phenotype} as a single SOM cell and a \textit{superphenotype} as a broader region of the SOM constructed by merging adjacent cells with similar color-redshift properties. The superphenotypes are defined by grouping deep SOM cells that are close in redshift, such that the superphenotypes become nearly disjoint in redshift space \citep[details in][]{Sanchez2020}.
The three steps are:
\begin{itemize}
    \item Superphenotype sampling $(p(T))$: First, we sample the fraction of galaxies in each superphenotype $T$. This step introduces large-scale sample variance by allowing fluctuations in the relative abundance of different galaxy populations across the sky. A Dirichlet sample is drawn for $p(T)$, giving a realization of how many galaxies belong to each superphenotype.
    
    \item Redshift given superphenotype $(p(z|T))$: Next, for each superphenotype, we sample the redshift distribution within that category. This introduces sample variance within each galaxy type, capturing the fact that different realizations of the universe will have fluctuations in redshift distributions due to large-scale structure. The effect of large-scale structure is implicitly encoded in the redshift distributions derived from a specific calibration field (e.g., COSMOS), which reflect the density fluctuations present in that region. Drawing a second Dirichlet sample for $p(z|T)$ introduces variation consistent with the expected cosmic variance from limited-area observations, while ensuring the sampled distributions remain properly normalized.
    
    \item Phenotype given redshift $(p(t|z,T))$: Finally, we sample the phenotype $t$ distribution within each redshift bin. Here, the term "redshift bin" refers to fine-grained bins ($\Delta z = 0.05$) used internally in the construction of the redshift distribution. The redshift separation of $\Delta z = 0.05$ is chosen to minimize the correlation between bins, which allows us to sample each redshift bin independently in 3sDir \citep[see e.g. Fig. 14 in][] {Sanchez2020}.  This step captures shot noise, as it represents the Poisson fluctuations in the number of galaxies assigned to each fine-scale classification. A third Dirichlet sample is drawn for $p(t|z,T)$, distributing galaxies within each redshift bin among the phenotypes while preserving normalization. 
\end{itemize}

Putting all components together, the 3sDir method generates a full redshift-phenotype joint distribution modeled as:
\begin{equation}
f(z, t) = \sum_T p(T)\, p(z\,|\,T)\, p(t\,|\,z, T),
\end{equation}
where each component is sampled via a Dirichlet distribution as described above. This hierarchical factorization enables a structured propagation of uncertainty from large-scale sample variance (through $p(T)$ and $p(z|T)$) and shot noise (through $p(t|z,T)$). 

Since we are interested in the marginalized redshift distribution, we sum over all phenotypes:
\begin{equation}
f(z) = \sum_t f(z, t) = \sum_T p(T)\, p(z\,|\,T).
\end{equation}
This marginalization ensures that the final $f(z)$ properly reflects both sample variance and shot noise while remaining agnostic to the internal phenotypic structure.


Since each realization of the redshift distribution is generated sequentially in this manner, every sampled realization includes all sources of uncertainty simultaneously; it is not the case that some samples contain only sample variance and others only shot noise. Instead, across an ensemble of realizations, we recover the full range of fluctuations in the redshift distribution.
The expectation value of these realizations converges to the redshift distribution measured in the redshift sample, meaning that 3sDir introduces a statistically consistent uncertainty around the input distribution without biasing it. This ensures that the propagated uncertainties reflect both sample variance and shot noise, while preserving the original mean $z$. 

By applying 3sDir to redshift calibration samples, we obtain a statistically robust set of resampled redshift distributions that can be directly incorporated into cosmological analyses. The hierarchical nature of 3sDir makes it particularly well-suited for methods that rely on multi-tiered calibration steps, such as our SOMPZ method. 

The 3sDir method, as applied here, assumes that the redshift sample used for calibration originates entirely from the COSMOS field. In practice, high-quality redshift information is also available in other deep fields, particularly in X3, which contributes spectroscopic and narrow-band redshifts at the bright end. However, these external redshifts often come from heterogeneous surveys with complex, color- and magnitude-dependent selection functions that are difficult to model accurately. To avoid introducing potential biases in the calibration due to these poorly characterized selection effects, we conservatively choose to exclude non-COSMOS redshifts from the uncertainty model. While this increases the sample variance—especially for the brighter \maglimpp lens galaxies, for which the COSMOS coverage is less dominant (62\%, 65\%, 65\%, 75\%, 85\%, 87\%, from lowest to highest bin)—it ensures that the estimated uncertainty in $p(z|c)$ remains robust. The choice therefore leads to a more conservative (i.e., broader) uncertainty estimate by avoiding the risk of underestimating redshift uncertainties due to unmodeled systematics in non-COSMOS fields.

We expect the impact of this approximation on the uncertainty estimate to remain small, since the dominant source of variance captured by 3sDir arises from the limited area of COSMOS itself. Nonetheless, future extensions of 3sDir could incorporate multiple deep fields with redshift coverage, allowing a more accurate characterization of the sampling variance and potential cross-field systematics.

As shown in Figure~\ref{fig:unc_bar}, sample variance (SV, dark blue) dominates the uncertainty on the mean redshift and width of $n(z)$ at low redshift, especially in Bin 1. In contrast, shot noise (SN, light blue) becomes increasingly important with redshift and overtakes SV in the higher bins—most notably in Bin 6. This is likely due to a combination of broader redshift PDFs and lower effective number densities in high-redshift regions of the SOM, especially after selection and matching cuts. The relative balance of SV and SN also differs slightly between the mean and width of $n(z)$, with SN contributing more prominently to the width in the high-redshift bins. These trends support the use of 3sDir as a practical and flexible method for propagating statistical uncertainties from the calibration sample, since it naturally captures the interplay between sample variance and shot noise across tomographic bins.

\subsection{Deep Fields Photometric Calibration}
\label{sec:zero pointunc}

Accurate photometric redshift estimation relies on precise flux measurements across multiple bands, making it sensitive to small systematic errors in photometric calibration—commonly referred to as photometric zero point uncertainties (ZPU). In DES Y6, these uncertainties arise from imperfect calibration across the deep fields (COSMOS, X3, C3, E2), leading to systematic offsets in the measured fluxes. Such offsets can distort the flux-based mapping between deep and wide field galaxies in the SOMPZ pipeline, introducing biases in the inferred redshift distributions and, consequently, in cosmological constraints. To account for this, we simulate zero point shifts in each deep field and propagate their impact through the entire photo-$z$ estimation pipeline. The resulting variations in the lens galaxy redshift distributions are then marginalized over in the cosmological analysis. 

In practice, we fix the zero point of the COSMOS field as a reference and apply independent shifts to the zero points of the X3, C3, and E2 fields. For each realization, we draw independent zero point shifts in all bands ($ugrizYJHK_s$), perturb the corresponding galaxy fluxes and flux errors, and reassign the galaxies to the deep SOM based on their perturbed fluxes. The SOM-to-redshift mapping is then recomputed to produce a new $n(z)$. This reassignment captures how calibration errors can distort the mapping from flux space to SOM cell and, in turn, to redshift.

A simplifying assumption in this procedure is that the dominant impact of zero point uncertainties arises from the $u$-band. This assumption is motivated by the fact that the $u$-band has the largest calibration uncertainties and is not included in either the wide field observations or the \texttt{Balrog} simulations used to derive the deep-to-wide transfer function. The zero-point uncertainties (in magnitude space) are: 0.055 for the $u$ band; 0.005 for the $g$, $r$, $i$, and $z$ bands; and 0.008 for the $J$, $H$, and $K_s$ bands (see \cite{Hartley2022} for details).
As a result, we do not recompute \texttt{Balrog} for each perturbed realization, assuming that ZPU-induced changes to the transfer function are negligible.

To efficiently sample the multidimensional space of possible zero point shifts, we use Latin Hypercube Sampling (LHS) in quantile space. LHS is a stratified, space-filling sampling technique that divides the range of each parameter into equal-probability intervals and draws one sample per interval, ensuring uniform coverage along each dimension. Unlike simple random sampling, this method avoids clustering and enables efficient exploration of high-dimensional parameter spaces with relatively few realizations. It balances efficiency and diversity in the combinations of shifts applied across different bands and fields.
We generate 100 such realizations from the LHS sampling, each corresponding to a unique combination of zero point shifts across the deep fields and bands, perturbing the redshift distributions $p(z|c)$ and transfer functions $p(c, \hat{c})$. For each of these 100 realisations, we generate 1,000,000 3sDir realizations, obtaining the 100 million.

The contribution of photometric zero point uncertainties (ZPU, green) shows a striking redshift dependence, as seen in Figure~\ref{fig:unc_bar}. Bin 6 exhibits the largest ZPU-driven variance in both $\langle z \rangle$ and $\sigma_{68}$—by a wide margin—while Bin 2 shows a moderate contribution, particularly to the width. This pattern suggests that at high redshift, small shifts in photometric calibration more readily cause galaxies to cross SOM cell boundaries with steep redshift gradients, magnifying the impact on the inferred redshift distribution. The increased sensitivity in Bin 6 may also reflect the complex flux–redshift degeneracies and broader intrinsic scatter in the SOM mapping at faint magnitudes. By contrast, the larger uncertainties seen in Bin~2 are primarily driven by photo-$z$ degeneracies around $z\!\sim\!0.4$, coinciding with the $g$–$r$ filter transition, where small color changes can shift galaxies between competing redshift solutions.


Interestingly, this trend is not monotonic. In Bin 5—despite its proximity in redshift to Bin 6—the ZPU contribution is anomalously low. In fact, we observe that the standard deviation of the redshift realizations in the 3sDir+ZPU case is slightly smaller than in the 3sDir-only baseline. This counterintuitive result likely reflects non-linear interactions in the SOM assignment, where flux perturbations lead to a narrower redshift distribution by chance. Since uncertainty components do not combine linearly under this mapping, and the apparent reduction is not physically meaningful, we conservatively set the ZPU contribution to zero in Bin 5 when reporting decomposed uncertainties.

\subsection{Photometric Redshift Biases}
\label{sec:redshiftsampleunc}

\begin{figure*}
    \centering
    \includegraphics[width=\linewidth]{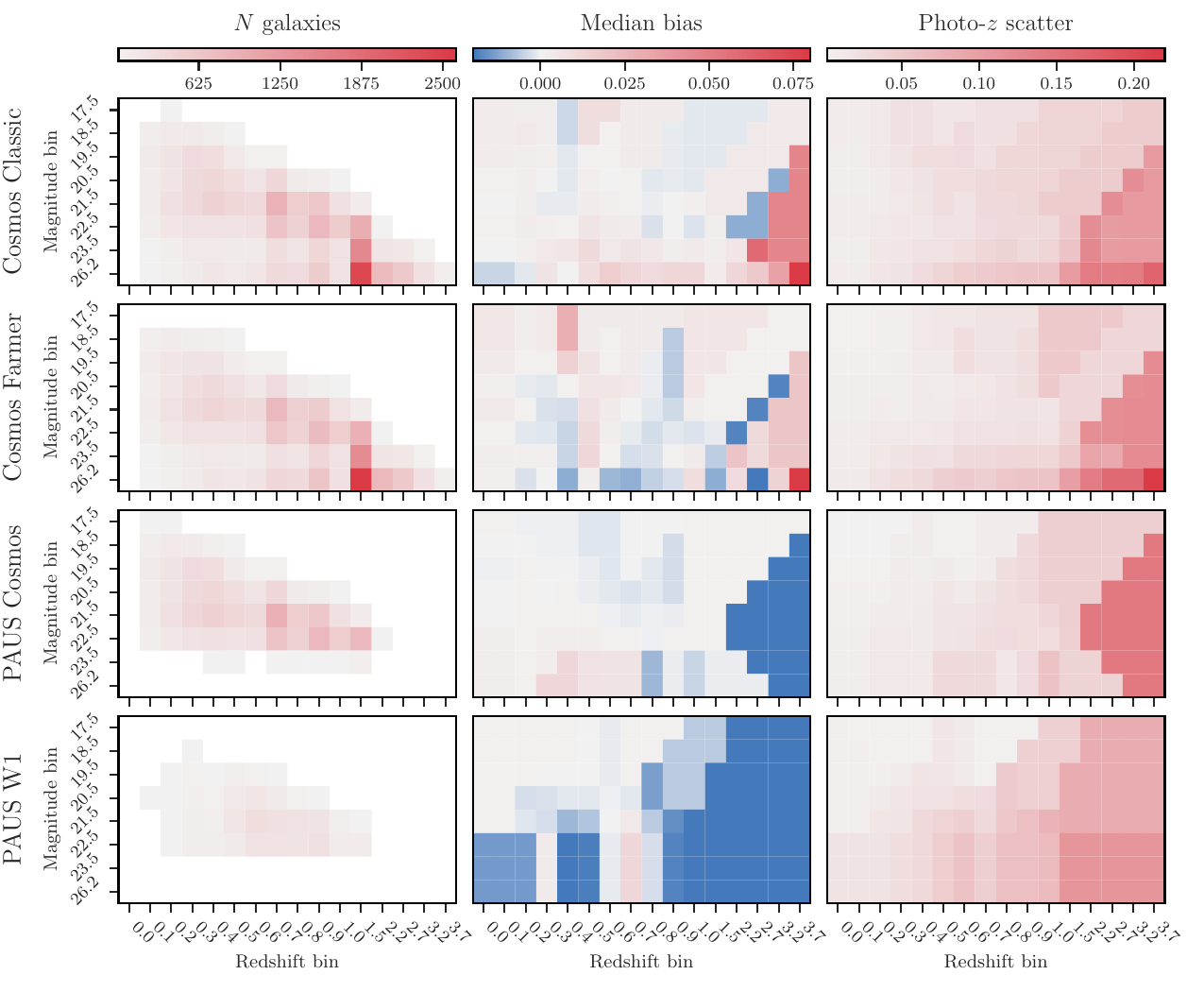}
    \caption{Distribution and photometric redshift performance across magnitude and redshift bins, for the four photometric redshift samples Cosmos Classic, Cosmos Farmer, PAUS Cosmos, and PAUS W1. Left: Number of galaxies (with spectra) in each magnitude–redshift bin. Center: Median photometric redshift bias $(z_{\mathrm{photoz}} - z_{\mathrm{specz}})$. Right: Scatter of photometric redshift estimates.
    On the left column, we indicate with a white color cells that have less than 10 galaxies. For those cells, we extrapolate the values of median bias and photo-z scatter from the value of the closest (magnitude, redshift) cell with more than 10 galaxies, which explains the patterns shown in the and right-most column. For example, the relatively large and constant median biases seen at high redshift in most panels, come from the 1-2 mag-z cells. Upon deeper inspection, these large median biases come from those cells being dominated by large outlier fractions, coming from confusing the Balmer and Lyman break.
    }
    \label{fig:redshift_unc}
\end{figure*}

Achieving a representative and complete redshift sample for calibration is essential for accurate estimation of the redshift distribution $n(z)$. A sample that does not fully populate the color-magnitude space of the target population can lead to biased estimates due to selection effects \citep{Hartley2020}. In DES Y6, we mitigate this by leveraging high-quality photometric redshifts from narrow-band imaging surveys, as described in Section~\ref{sec:redshiftcatalogs}. While these narrow-band photo-$z$ estimates are typically less precise than spectroscopic redshifts on a per-galaxy basis, their uncertainties remain small compared to the much broader redshift resolution of the wide-field \textit{griz} photometry. Crucially, these samples offer high completeness and wide coverage in the relevant color space.

Nevertheless, photometric redshifts are inherently subject to systematic uncertainties, arising from measurement errors, mismatches in template libraries, and limitations in the photo-$z$ algorithms themselves \citep{hildebrandt2010, beck17, photozreview}. If not accounted for, such biases can distort the inferred $n(z)$ and propagate into the cosmological inference. We therefore implement a redshift uncertainty calibration procedure designed to model systematic redshift shifts in a way that reflects these uncertainties while preserving the statistical properties of the underlying redshift distributions.

In DES Y3, the lens redshift distributions were derived using a combination of three heterogeneous calibration catalogs (spectroscopic compilations, PAU+COSMOS, and COSMOS30), with three realizations constructed by ranking the catalogs differently and assigning the redshift from the highest-ranked source available for each galaxy \citep{y3-sompz, Giannini2024}. The final ensemble of realizations was obtained by averaging over these three samples with equal prior weight, which provided coverage across multiple calibration sets but did not explicitly propagate systematic uncertainties associated with incomplete or heterogeneous information.

To address this limitation, we adopt a different strategy for propagating redshift uncertainties. Rather than perturbing the redshifts of individual galaxies independently---which would mainly introduce random fluctuations that tend to average out across the large number of objects and therefore underestimate the impact of systematic effects---we instead apply coherent shifts defined at the level of magnitude--redshift bins.  
This method captures correlated redshift errors among galaxies with similar properties and more faithfully reflects the kind of biases expected from photo-$z$ systematics. The procedure begins by matching galaxies from each narrow-band catalog (Cosmos Classic, Cosmos Farmer, PAUS COsmos, PAUS W1, see section \ref{sec:redshiftcatalogs} for more details) to available spectroscopic redshifts. The matched galaxies are then binned in both magnitude and redshift. For each bin $i$, we compute the median redshift bias and the redshift scatter using the distribution of $dz \equiv z_{\mathrm{photoz}} - z_{\mathrm{specz}}$ values in that bin:
\begin{align}
    b(i) &= \mathrm{median}(dz), \\
    s(i) &= \left\langle \sigma(dz^{(k)}) \right\rangle,
\end{align}
where $\sigma(dz^{(k)})$ is the sample standard deviation of the $k$-th bootstrap resampling of the $dz$ values in bin $i$, and the average $\langle \cdot \rangle$ is taken over 1000 bootstrap realizations. This approach provides a robust estimate of the scatter of the distribution, mitigating the effects of sampling noise and accounting for uncertainty due to limited statistics in each bin.

The bias and scatter values are used to define the range of systematic shifts applied to galaxies in each bin. Specifically, we draw shifts $\Delta z$ from a Gaussian prior characterized by:
\begin{equation*}
    \Delta z \sim \mathcal{N}(\mu = b(i),\ \sigma = s(i)),
\end{equation*}
where $\mu$ is the median redshift bias and $\sigma$ is the scatter estimated as described above. All galaxies in a given bin receive the same shift.

We apply the same Latin Hypercube Sampling strategy described in Section~\ref{sec:zero pointunc} to sample coherent shifts across all magnitude–redshift bins. This approach ensures broad and efficient coverage of the uncertainty space while avoiding redundant or biased combinations of perturbations. Each realization of the redshift uncertainty is constructed as follows:

\begin{enumerate}
    \item A systematic shift $\Delta z$ is drawn for each magnitude-redshift bin using LHS.
    \item All galaxies in a given magnitude–redshift bin are perturbed by the corresponding $\Delta z$.
    \item The resulting redshift distribution $n(z)$ is computed from the modified sample.
\end{enumerate}

This process is repeated $N$ times (in our case, $N=100$), generating an ensemble of perturbed redshift distributions. The final $n(z)$ uncertainty is characterized by the spread across these realizations.

Figure~\ref{fig:redshift_unc} summarizes the properties of the photometric redshift samples used in this calibration. It shows the number of galaxies per magnitude–redshift bin (left column), the median photometric redshift bias (center column), and the scatter in photometric redshift estimates (right column). Each row corresponds to one of the photometric redshift samples. The median bias and scatter values shown in the figure are used to define the means and widths of the Gaussian prior from which the systematic shifts $\Delta z$ are drawn via LHS.

The choice of $\sigma = s(i)$, i.e., using the photo-$z$ scatter in each bin to set the width of the Gaussian prior, is intentionally conservative. This is especially important for faint galaxies, where spectroscopic redshift coverage is sparse and the true systematic uncertainty may be underconstrained. For brighter galaxies with more robust spectroscopic calibration, alternative or tighter priors could be used, but in this work we apply a uniform treatment across all magnitude bins for consistency.

By applying systematic shifts at the joint magnitude–redshift bin level, this method avoids artificially underestimating uncertainties that can result from random, uncorrelated perturbations canceling out. This approach differs with the methodology adopted in DES Y3, where redshift sample uncertainty was modeled by constructing three separate redshift realizations based on different catalogs (spectroscopic, PAUS+COSMOS, and COSMOS30), and propagating them through the full pipeline with equal weights. In that framework, the systematic uncertainty was captured by comparing results across different samples, each assumed to be independently valid. While this method accounted for catalog-level systematics, it treated galaxy redshifts as fixed once assigned from a particular catalog and did not explicitly model correlated redshift errors as a function of galaxy properties. In contrast, the DES Y6 approach implements a continuous and flexible model of redshift uncertainty, applying coherent perturbations across magnitude–redshift bins and sampling the space of possible biases via Latin Hypercube Sampling. This forward-modeling-compatible framework allows for a more granular treatment of systematic shifts and can be easily extended or refined. In future work, this could include assigning more informative priors to specific bins based on spectroscopic coverage or combining coherent and stochastic perturbations to capture both structured and residual uncertainties.

Redshift sample uncertainty (RU, purple in Figure~\ref{fig:unc_bar}) shows a non-trivial pattern across tomographic bins, with distinct differences in its impact on the mean and width of $n(z)$. For the mean redshift, RU contributes most significantly in Bins 2 and 6, while in Bin 1 the estimated change in variance relative to the 3sDir baseline is slightly negative. The prominent RU contribution in Bin~2 is again consistent with the difficulties of resolving the $z\!\sim\!0.4$ degeneracies associated with the $g$–$r$ filter transition, which amplify the sensitivity of this bin to limited redshift-sample statistics.

For the width $\sigma_{68}$, the strongest contributions appear in Bins 3 and 6, where magnitude–redshift degeneracies and limited spectroscopic coverage make the inferred distributions more sensitive to redshift calibration shifts. In contrast, RU has minimal impact on the width in the other bins. The small negative values observed in Bin~1 do not correspond to a true variance---since variance is by definition non-negative---but arise because we compute differential contributions relative to the 3sDir baseline. Owing to the non-linear nature of the SOM mapping, coherent perturbations can sometimes redistribute galaxies in a way that reduces the apparent spread of the redshift distribution. These negative values have no impact on the propagated uncertainties though, which are always derived from the full set of realizations. We only set the Bin~1 contribution to zero when displaying the relative breakdown in Figure~\ref{fig:unc_bar}, to avoid misinterpretation.

\section{Final Redshift Distributions and Uncertainties for \maglimpp}
\label{sec:combine}

The output of the SOMPZ framework described in the previous section is 100 million realizations of the redshift distributions for each of the 6 tomographic bins, consistent with our estimated uncertainties from the photometric information. These are shown in Figure \ref{fig:nzplot}. To derive the final redshift distributions and uncertainties used for the cosmological inference, we first incorporate additional information from spatial clustering (Section~\ref{sec:wz}), and then derive a representation of the final realizations that can be efficiently sampled in an MCMC framework (Section~\ref{sec:modes}). The resulting calibration achieves typical uncertainties on the mean redshift at the 1--2$\%$ level, corresponding to an average improvement of $\sim20$--$30\%$ relative to DES Y3. The uncertainties on the width are likewise reduced in most bins, with improvements ranging from $\sim20\%$ to $\sim45\%$. This improvement does not arise from a single component, but rather from the combination of several methodological and data-related advances in the Y6 analysis.

At the SOMPZ level, improvements are primarily driven by reductions in methodological uncertainty terms relative to Y3. The substantially larger Balrog sample reduces shot noise in the estimation of the deep-to-wide transfer function, particularly in sparsely populated regions of the deep SOM. In addition, the spectroscopic portion of the redshift calibration sample was revised to better match the requirements of this analysis, restricting the calibration to a subset of spectroscopic surveys and redshift-quality selections deemed most robust for this application. The Y6 analysis also adopts a more flexible treatment of redshift-sample uncertainty than Y3, modeling coherent redshift systematics as a function of galaxy properties through perturbations applied in magnitude--redshift bins. Additionally, the updated construction of the \maglimpp sample yields a cleaner and more homogeneous galaxy selection, improving the mapping between the deep and wide samples used in SOMPZ. Comparing the SOMPZ-only calibration between Y3 and Y6, we find an average reduction of $\sim22\%$ in the uncertainty on the mean redshift, while the uncertainty on the distribution width is reduced by $\sim8\%$ on average. These improvements are primarily driven by reductions in the methodological uncertainty contributions, while the sample variance component remains largely unchanged.

The reduction is not fully uniform across tomographic bins. The highest-redshift bin already had the largest uncertainties in Y3 and remains the least constrained bin in Y6. This likely reflects the intrinsic difficulty of calibrating this redshift range, together with the more complete uncertainty propagation adopted in Y6. In particular, the Y3 calibration for this bin relied on a combination of heterogeneous catalogs with equal prior weight (see \ref{sec:redshiftsampleunc}), which may have led to a less conservative estimate of the uncertainty. We therefore interpret the slightly larger Bin 6 error bars in Y6 as a more faithful accounting of the limitations in this regime, rather than a loss of constraining power due to data quality.

Additional information is then incorporated from clustering-redshift (WZ) measurements, which further improves the uncertainties of both the mean redshift and the overall shape of the distributions. The impact of WZ on the mean-redshift uncertainties is comparable to that observed in DES Y3, providing an additional reduction of $\sim20\%$ on average across tomographic bins. In Y6, however, the WZ constraint on the distribution widths is stronger, with typical improvements of $\sim40\%$. This sensitivity to the width is likely aided by the extended redshift coverage of the reference sample, including QSOs, which help constrain the high-redshift tail of the distributions. The WZ information is incorporated consistently within the importance-sampling framework mentioned in the following sections.

Table~\ref{tab:components} provides an overview of the different elements entering the \textsc{MagLim++} redshift calibration pipeline, indicating for each step whether it is introduced here or in an accompanying DES Y6 analysis paper.

\begin{table}
    \centering
    \caption{Summary of the main components of this work and their relation to companion DES papers within the Y6 $3\times2$pt analysis.}\label{tab:components}
    \begin{tabular}{p{5.25cm}p{3.03cm}}
        \hline 
        \hline
        Component & Defined/performed in \\
        \hline
        \textsc{MagLim++} lens sample definition & \cite{y6-maglim} \\
        Redshift sample definition & This work \\
        Deep samples & \cite{y3-deepfields} \\
        Deep and wide SOMs & This work \\
        SOMPZ transfer function & This work, from the Balrog catalog \cite{y6-balrog} \\
        \textsc{MagLim++} SOMPZ calibration & This work \\
        SOMPZ uncertainty estimation & This work \\
        WZ measurement  & \citepalias{y6-wz} \\
        Combination of WZ and SOMPZ $n(z)$ & \citepalias{y6-wz} \\
        Mode projection methodology & \citepalias{y6-nzmodes} \\
        Mode projection of the combined $n(z)$ & This work \\
        Tests on simulated data vectors & This work \\
        \hline
        \hline
    \end{tabular}
\end{table}


\subsection{Incorporating Clustering Information}
\label{sec:wz}

As outlined in the introduction, the final redshift calibration methodology consists of the statistical combination of photometric and clustering information. In fact, we leverage cross-correlations between lens galaxies and spectroscopic reference samples to refine the redshift distributions from SOMPZ further. The clustering redshift (WZ) technique relies on the spatial association of photometric sources with spectroscopic samples. For example, we measure the small-scale angular cross-correlations between our target sample and data from BOSS \citep[Baryonic Oscillation Spectroscopic Survey]{Smee2013,boss} and eBOSS \citep[extended-Baryon Oscillation Spectroscopic Survey]{eboss2016,dr16,Alam2020}.

The full implementation of this method for both lens and source samples—including the modeling framework, systematic treatment, and application to DES Y6—was developed and carried out in \citepalias{y6-wz}, on which this work builds. We summarize the key components here for completeness.

We model the observed angular correlation function $w_{ur}$ between photometric unknown  samples (indexed by $\rm u$) and spectroscopic reference samples (indexed by $\rm r$) as:
\begin{equation}
    \hat{w}_{\rm ur} = f(b_{\rm u}, \, b_{\rm r},\,  n_{\rm r}(z), \, n_{\rm u}(z), \, s_{\rm u}),
\end{equation}
where $b_x$ represents galaxy bias, $n_x(z)$ denotes redshift distributions, and $s_{\rm u}$ encapsulates systematic effects such as magnification. The likelihood function for clustering redshifts is then given by:
\begin{equation}
    \ln\mathcal{L}(w_{\rm ur} | b_{\rm u}, b_{\rm r}, n_{\rm r}, n_{\rm u}, s_{\rm u}) \propto (w_{\rm ur} - \hat{w}_{\rm ur})^T C_w^{-1} (w_{\rm ur} - \hat{w}_{\rm ur}),
\end{equation}
where $C_w$ is the covariance of the cross-correlation measurements. We obtain posterior constraints on $n_{\rm u}(z)$ by reweighting the SOMPZ realizations with the clustering-redshift likelihood, which already accounts for nuisance parameters such as galaxy biases and magnification.

To statistically combine the photometric modeling from SOMPZ with the external constraints from WZ, we adopt an importance sampling scheme that reweights the SOMPZ realizations based on their consistency with the clustering-redshift measurements. This approach treats the ensemble of SOMPZ redshift realizations as a prior over plausible $n(z)$ shapes, and uses the WZ-derived likelihood to assign weights to each realization.

While clustering-based redshift estimates have been used in the past to construct $n(z)$ directly, this typically requires assumptions about galaxy bias evolution and can suffer from limited redshift resolution. In contrast, importance sampling allows us to incorporate WZ information without reconstructing the $n(z)$ from scratch or introducing additional parametric modeling. Thus, importance sampling offers a principled and computationally efficient way to combine these two sources of information. The resulting reweighted ensemble preserves the complex, non-Gaussian uncertainty structure of SOMPZ while integrating empirical constraints from large-scale structure. This yields a statistically consistent posterior distribution that reflects both photometric and clustering information. The final $n(z)$ distributions used in cosmological inference are drawn from this reweighted set.

\begin{table*}
    \centering
    \begin{tabular}{l c c c c c c}
        & & & Mean $\langle z \rangle$ \\
        \hline
        \hline
        Uncertainty & Bin 1 & Bin 2 & Bin 3 & Bin 4 & Bin 5 & Bin 6\\
         & z $\in$ [0.2, 0.4] &  z $\in$ [0.4, 0.55] &  z $\in$ [0.55, 0.7] &  z $\in$ [0.7, 0.85] &  z $\in$ [0.85, 0.95] &  z $\in$ [0.95, 1.05] \\
      
        \hline
        SOMPZ & 0.3079 $ \pm\ $ 0.0078 & 0.4343 $ \pm\ $ 0.0106 & 0.6270 $ \pm\ $ 0.0093 & 0.7771 $ \pm\ $ 0.0076 & 0.8992 $ \pm\ $ 0.0081 & 0.9970 $ \pm\ $ 0.0152\\
        SOMPZ+WZ & 0.3059 $ \pm\ $ 0.0063 & 0.4352 $ \pm\ $ 0.0077 & 0.6235 $ \pm\ $ 0.0073 & 0.7777 $ \pm\ $ 0.0065 & 0.9027 $ \pm\ $ 0.0062 & 1.0112 $ \pm\ $ 0.0131 \\
        \hline

        \\
        & & & Width $\sigma_{\rm 68}$ \\

        \hline
        \hline
        SOMPZ & 0.0769 $\pm\ $ 0.0049 & 0.0947 $\pm\ $ 0.0059 & 0.0637 $\pm\ $ 0.0049 & 0.0756 $\pm\ $ 0.0039 & 0.0653 $\pm\ $ 0.0050 & 0.0970 $\pm\ $ 0.0099 \\
        SOMPZ+WZ &  0.0793 $\pm\ $ 0.0027 & 0.0916 $\pm\ $ 0.0040 & 0.0605 $\pm\ $ 0.0021 & 0.0692 $\pm\ $ 0.0021 & 0.0604 $\pm\ $ 0.0028 & 0.0889 $\pm\ $ 0.0088\\
        \hline

    
    \end{tabular}
    \caption{Summary of values for systematic uncertainties and center values for mean ($\langle z \rangle$, top panel) and width ($\sigma_{\rm 68}$, bottom panel) for the \nz\ distributions. The various components are computed as described in Section \ref{sec:uncertainty} and as they are not completely independent they do not sum up to the total value. The values related to SOMPZ and SOMPZ+WZ refer to Figure \ref{fig:nzplot}. The final \nz\ which has been used in the cosmological analysis is the bottom line.  }
    \label{tab:unc}
\end{table*}

\begin{figure}
    \centering
    \includegraphics[width=\linewidth]{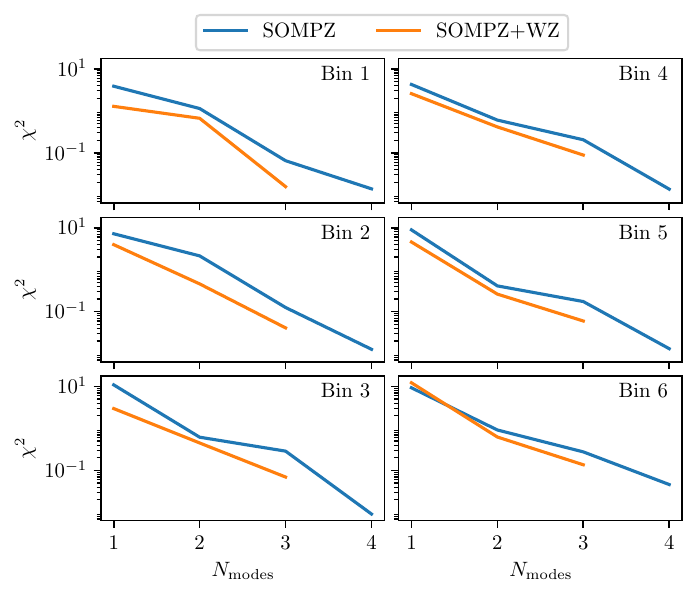}
    \caption{Cumulative discarded \(\chi^2\) as a function of the number of retained modes. The sharp decline indicates that most of the cosmologically relevant variance is captured in just a few modes, justifying our truncation choice.}
    \label{fig:scree}
\end{figure}

\begin{figure*}
    \centering
    \includegraphics[width=\linewidth]{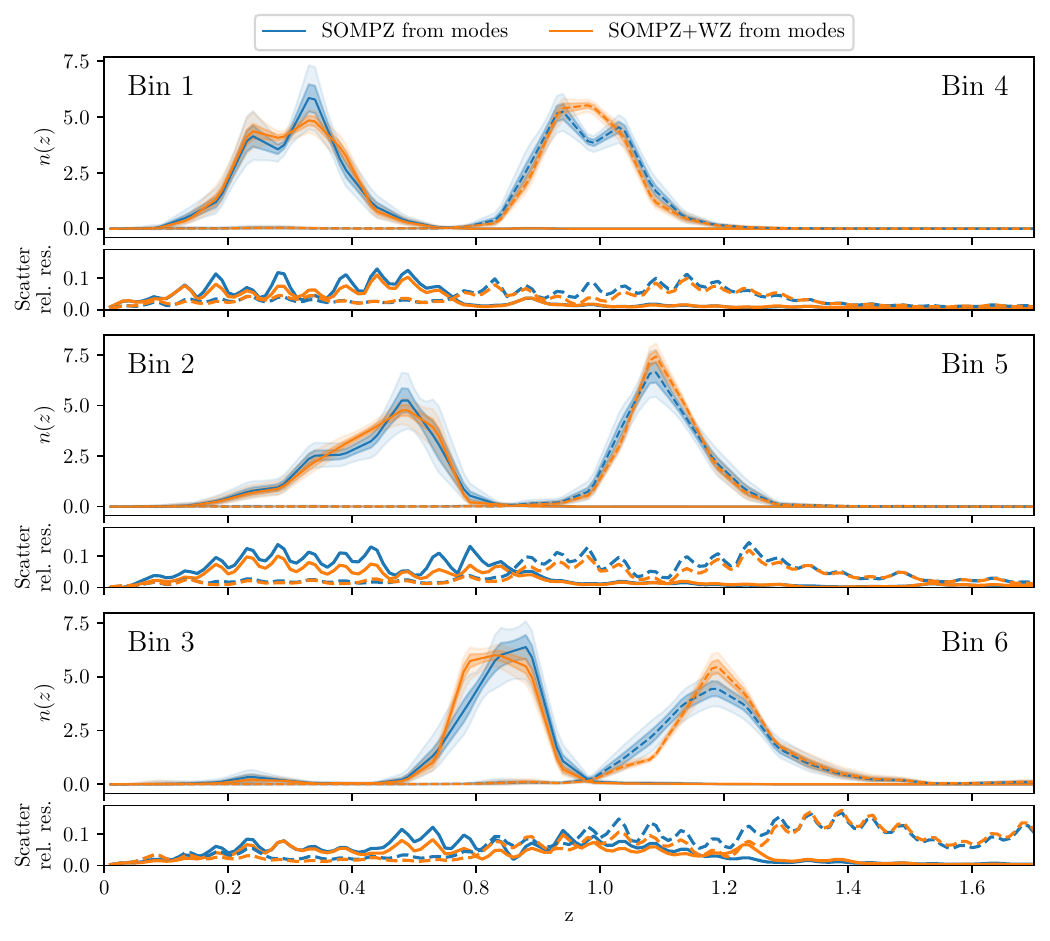}
    \caption{Comparison of the redshift distributions for the six tomographic bins of \maglimpp, reconstructed from the modes. The blue lines represent the average SOMPZ $n(z)$ reconstructed from the modes, with the $\pm 1\sigma$ and $\pm 2\sigma$ bands showcasing the realizations. The orange lines and bands represent the SOMPZ+WZ $n(z)$ reconstructed from modes. Solid and dashed styles are used solely to differentiate tomographic bins in regions where the curves overlap. These are the distributions marginalised over in the cosmological inference. The lower panels show the scatter of the relative residuals across realizations, defined as $(n_{\mathrm{recon}} - n_{\mathrm{orig}})/(1 + \langle n_{\mathrm{orig}} \rangle)$, which suppresses artificial amplification in low-density regions. The reconstructed distributions successfully preserve the key statistical features of the original samples while significantly reducing dimensionality. The small wiggles visible in the residuals are not due to the mode compression but are introduced during the rebinning of $n(z)$ using the triangle kernel representation.}
    \label{fig:nzmodes}
\end{figure*}

\subsection{Redshift Distribution Preprocessing}\label{sec:pileup}

To avoid unnecessarily increasing the runtime of the cosmological inference, we ``pile up'' the high-redshift tail by assigning all probability at $z > 3$ to a single bin at $z = 3$. This removes a negligible and noisy component of the distribution that would otherwise slow down the Monte Carlo chains without contributing useful information. This truncation removes the numerically unstable, low-weight tail of $n(z)$ and ensures all realizations are defined on a common redshift grid, simplifying the subsequent cosmological inference without significantly affecting the clustering signal, since the clipped high-redshift tail contributes negligibly to the galaxy–galaxy correlations.

Secondly, the redshift distributions are then rebinned onto a basis of triangular kernels, which we called \textit{sawtooth}. This transformation replaces the original binned $n(z)$ with a piecewise-linear approximation formed by overlapping triangular functions. The resulting representation is continuous and differentiable, which facilitates accurate computation of the derivatives of the cosmological data vector with respect to redshift shifts. While this rebinning introduces small artifacts in the form of low-amplitude wiggles, particularly near sharp features in the $n(z)$, these effects are negligible and do not impact the resulting cosmological inference. The triangular basis also ensures normalization and consistency across all realizations used in the compression.

\subsection{Efficient Sampling for the Final Redshift Distributions}\label{sec:modes}

The estimation of photometric redshift distributions for galaxy samples introduces a high-dimensional set of nuisance parameters, as each redshift realization corresponds to a different possible shape of the redshift distribution \( n(z) \). Directly sampling over these high-dimensional realizations in a cosmological analysis is computationally infeasible, both because of the sheer dimensionality and the difficulty of defining a smooth, physically meaningful prior over such a space. 

To address this, we employ a mode projection technique that reduces the effective dimensionality of the problem by identifying a basis of variations in \( n(z) \) that significantly affect cosmological observables—such as the two-point correlation functions—while projecting out directions that leave the likelihood nearly unchanged. This approach ensures that only those variations in the redshift distribution that matter for cosmology are retained, greatly improving sampling efficiency without sacrificing accuracy.

In DES Y3, a similar dimensionality reduction was attempted using the \texttt{Hyperrank} method \citep{y3-hyperrank}, which reparameterizes the space of redshift realizations using a Gaussian Process latent space and ranks directions by their cosmological importance. While effective, \texttt{Hyperrank} requires learning a mapping from redshift realizations to a latent space and back, which can introduce approximations and technical complexity. In contrast, the mode projection method used here directly operates in the observable space and selects modes based on their impact on the likelihood, allowing for a more transparent and targeted marginalization over redshift uncertainty. 

This approach follows the method outlined and validated in \citepalias{y6-nzmodes}, where the high-dimensional set of nuisance parameters is compressed into a much lower-dimensional set of modes. Specifically, the redshift distribution is expressed as a weighted sum of basis functions:
\begin{equation}
    n(z) = \sum_{k} n_k\, b_k(z),
\end{equation}
where the coefficients \( n_k \) are treated as nuisance parameters. Given a set of sampled realizations of \( n(z) \), we determine a transformation that maps these into a low-dimensional space spanned by a set of modes \( u \), such that only the variations that significantly affect cosmological observables are retained. The compression follows a linear transformation:
\begin{equation}
    u = E\, (n - \bar{n}),
\end{equation}
where \( \bar{n} \) is the mean redshift distribution across realizations, and \( E \) is a projection matrix that isolates the directions in redshift space to which the observables (our two point correlation functions) are most sensitive. The transformation matrix is determined from the Fisher information matrix of the observables (i.e., the data vector), such that the retained modes are those that are best constrained by the data given its covariance. These modes correspond to directions in observable space with the highest signal-to-noise ratio, which may or may not coincide with the directions that maximize constraints on cosmological parameters. This distinction arises because the procedure optimizes information content in the data space rather than directly in parameter space, although in practice many of the best-constrained data modes also carry significant cosmological information.
The original redshift distribution can be reconstructed from the compressed representation as:
\begin{equation}
    n(z) \approx \bar{n}(z) + \sum_{i=1}^{M} u_i\, U_i(z),
\end{equation}
where \( U_i(z) \) are the dominant modes of variation. The number of retained modes, \( M \), is chosen to ensure that the discarded directions contribute negligibly to the likelihood function.

The role of the mode projection technique is illustrated in Figure~\ref{fig:scree} and Figure~\ref{fig:nzmodes}, which show the structure and impact of the dominant modes used in the redshift uncertainty model. Additional examples, including the visual representation of the modes themselves, are provided in Appendix~\ref{app:modes}.

Figure~\ref{fig:scree} illustrates the number of modes retained in our analysis by showing the cumulative discarded \(\chi^2\) as a function of the number of modes included. This metric quantifies the statistical significance of any biases in the model for the observables that can be caused by the fluctuations in $n(z)$ that are in the discarded modes. We retain only the number of modes necessary to keep the cumulative discarded \(\chi^2\) across tomographic bins below a defined threshold of $\chi^2<0.15$. A threshold $\ll 1$ is chosen to ensure that any biases or variability in the inferred parameters are well below the $1\sigma$ variation attributable to measurement errors. The rapid drop in \(\chi^2\) with increasing mode count demonstrates that only a small number of modes are needed to preserve nearly all sensitivity to redshift uncertainties, justifying the truncation strategy and confirming the efficiency of the compression. For the SOMPZ-only uncertainty, this criterion leads to 4 retained modes, while for the SOMPZ+WZ case, only 3 modes are needed.

Figure~\ref{fig:nzmodes} compares the original sampled \(n(z)\) distributions to those reconstructed using only the retained modes for both SOMPZ-only and SOMPZ+WZ uncertainties. The close agreement demonstrates that the dominant statistical properties of the redshift distributions are preserved despite a strong dimensionality reduction: for example, keeping \(M=3\) modes per tomographic bin reduces the number of free parameters per realization from \(6 \times 300 = 1800\) (one per histogram bin) to \(6 \times 3 = 18\) mode amplitudes, a factor of \(\sim 100\) reduction. The mean and modes basis (each of length 300) are fixed and stored once; only the mode coefficients vary between realizations. These reconstructions visually compare to the uncompressed \(n(z)\) shown in Figure~\ref{fig:nzplot}. The bottom panels show the residuals \((n_{\mathrm{recon}} - n_{\mathrm{orig}}) / [1 + \langle n_{\mathrm{orig}} \rangle]\). A value of 0.1 in this metric corresponds to differences well below the statistical uncertainty in the cosmological analysis, and is therefore tolerable. A small wiggle is visible in the residuals: as previously mentioned, this feature is introduced during the conversion to the sawtooth binning and does not originate from the mode compression itself. Its amplitude is negligible relative to the overall uncertainty and has no discernible impact on cosmological constraints.




This method provides a way to account for redshift uncertainty while significantly reducing computational complexity. Instead of defining a prior directly over the high-dimensional space of redshift distributions \( n(z) \), we parametrise the uncertainty in the cosmological inference using the $u$ coefficients. A prior over \( u \) can be efficiently learned using density estimation techniques (e.g., Gaussian mixtures, kernel density estimation) applied to a suite of \( n(z) \) realizations. This low-dimensional representation preserves the dominant modes of variation in \( n(z) \), allowing the inference pipeline to marginalize over redshift uncertainty with significantly reduced computational cost, while retaining all relevant cosmological information.

For source redshifts, the formalism naturally accommodates correlations between tomographic bins. For lens redshifts, where the \texttt{MagLim++} binning was explicitly designed to produce independent redshift distributions, we confirm that the mode amplitudes remain uncorrelated in the cosmological posteriors, justifying the per-bin treatment of uncertainties. This ensures an accurate and efficient propagation of redshift calibration errors without introducing unnecessary nuisance correlations. Additional validation of this uncorrelated treatment is provided in Appendix~\ref{app:modes}.



%

\section{Impact on Cosmology}\label{sec:results}

In this section, we quantify how different assumptions about the redshift distributions and their associated uncertainties affect the inference of cosmological parameters. To this end, we use noiseless synthetic data vectors constructed of the three two-point functions—galaxy clustering $w(\theta)$, galaxy–galaxy lensing $\gamma_{t}(\theta)$, and cosmic shear $\xi_{\pm}(\theta)$—generated using real data redshift distributions. By repeating the inference with different $n(z)$ estimates or different marginalisation schemes for redshift uncertainties, we can isolate their impact on the cosmological constraints.

In particular, we evaluate the impact of the redshift uncertainty parametrisation, contrasting the fiducial Y6 mode-based approach against the simpler shift-and-stretch model adopted in the Y3 analysis (Section~\ref{sec:cosmology_modes}). We also compare the use of lens redshift distributions estimated from \texttt{SOMPZ} alone to those obtained from the fiducial \texttt{SOMPZ+WZ} combination (Section~\ref{sec:cosmology_calib_method}).

Each synthetic data vector is computed at fixed cosmological parameters and includes all relevant two-point functions evaluated using the fiducial angular scale cuts defined in \citep{SanchezCid2025}. Cosmological inference is then performed using the \texttt{CosmoSIS} pipeline \citep{cosmosis}, assuming a flat $\Lambda$CDM cosmology with five free parameters:
$\left\{ \Omega_{\rm m}, \sigma_8, h, n_s, \Omega_{\rm b} \right\}$, and fixing the sum of neutrino masses $\sum m_\nu$ to the minimum allowed value of 0.06 eV. We adopt a linear galaxy bias model for the lens sample and include the full set of nuisance parameters used in the DES Y6 analysis:
\begin{itemize}
  \item one linear galaxy bias parameter per lens bin;
  \item intrinsic alignment parameters from the Tidal Alignment and Tidal Torquing \citep[TATT,][]{tatt} model: $\{a_1, a_2, \eta_1, \eta_2, b_{\rm TA}\}$;
  \item one shear calibration parameter $m_i$ per source bin;
  \item one magnification slope parameter $\alpha_i$ per lens bin;
  \item source sample redshift uncertainty parameters: seven mode coefficients (for all bins);
  \item lens redshift uncertainty parameters: either three orthogonal redshift mode coefficients per tomographic bin (fiducial case), or one shift and one stretch parameters per lens tomographic bin, depending on the test.
\end{itemize}

\begin{table}
	\centering	
	\caption{The parameters and their priors used in the
		fiducial  \maglim $\Lambda$CDM and $w$CDM analyses. The parameter $w$ is fixed to $-1$ in $\Lambda$CDM.  Square brackets denote a flat prior, while parentheses denote a Gaussian prior of the form $\mathcal{N}(\mu,\sigma)$. The priors on the calibration parameters of the source galaxy samples—namely the multiplicative shear bias $m$ and the source redshift distribution modes—are correlated. }

	\begin{tabular}{ccc}
		\hline
		Parameter & Fiducial & Prior\\\hline
		\multicolumn{3}{c}{\textbf{Cosmology}} \\
		$\Omega_{\rm m}$ &  0.31 &[0.1, 0.6] \\ 
		$A_{\rm s} 10^{9}$ & 1.831 & [$0.5$, $5.0$]  \\ 
		$n_{\rm s}$ & 0.965 & [0.93, 1.00]  \\
		$w$ &  -1.0  &  fixed \\
		$\Omega_{\rm b}$ & 0.051 &[0.03, 0.07]  \\
		$h_0$  & 0.69  &[0.58, 0.80]   \\
		$m_\nu\ [eV]$ & 0.77 & [0.06, 0.6]
		\\\hline

		\multicolumn{3}{c}{\textbf{Linear galaxy bias  } }	 \\
		$b_{i}$  & 1.54 & [0.8,3.0]\\
		$b_{i}$  & 1.81 & [0.8,3.0]\\
		$b_{i}$  & 1.85 & [0.8,3.0]\\
        $b_{i}$  & 1.76 & [0.8,3.0]\\
        $b_{i}$  & 1.93 & [0.8,3.0]\\
        $b_{i}$  & 1.90 & [0.8,3.0]\\\hline		
		
		\multicolumn{3}{c}{\textbf{Lens
				magnification } }  \\
		$C_{1} $ & 0.43 &   $\mathcal{N}$(1.58, 0.51)\\ 
		$C_{2} $ & 0.30 & $\mathcal{N}$(0.30, 0.48)\\
		$C_{3} $ & 1.75 & $\mathcal{N}$(1.75, 0.39) \\
		$C_{4} $ &  1.94 & $\mathcal{N}$(1.94,  0.35)\\
	    $C_{5} $ &  1.56  & $\mathcal{N}$(1.56, 0.71) \\
		$C_{6} $ &  2.96 & $\mathcal{N}$(2.96,  0.95) \\
		\hline

		\multicolumn{3}{c}{\textbf{Lens photo-z}}	 \\
        $u^{l}_{i, m} \; (i \in [1, 6], \; m \in [1, 3])$  & 0 & $\mathcal{N}(0, 1) \in [-3,3]$ \\

        \hline  
		\multicolumn{3}{c}{\textbf{Source photo-z}}	 \\
        $u^{l}_{m} \; ( m \in [1, 7])$  & 0 & $\mathcal{N}(0, 1) \in [-3,3]$ \\
        \hline
		\multicolumn{3}{c}{{\bf
				Intrinsic alignment}} \\
        $z_{0}$ & 0.3 & fixed \\
        $A_1$ & 0.13 & [-1, 3]  \\
        $A_2$ & -0.2 & [-3, 3] \\
        $\eta_1$ & 2.0 & $\mathcal{N}(0.0,3.0) \in [-5,5]$ \\
        $\eta_2$ & 2.0 &  $\mathcal{N}(0.0,3.0) \in [-5,5]$ \\
        $b_{\rm TA}$ & 1.0 & fixed \\
		\hline
		\multicolumn{3}{c}{{\bf Shear calibration}} \\
		$m^1$ & 0.0  & [-0.1, 0.1]\\
		$m^2$ & 0.0  & [-0.1, 0.1]\\
		$m^3$ & 0.0  & [-0.1, 0.1]\\
		$m^4$ & 0.0  & [-0.1, 0.1]\\
		\hline
	\end{tabular}
    \label{tab:parameters}
\end{table}

We adopt flat priors for all cosmological parameters and informative priors for nuisance parameters, listed in Table \ref{tab:parameters}. In particular, the priors on magnification and intrinsic alignments follow those derived in the DES Y6 modeling framework \citep{SanchezCid2025}.  The priors on the calibration parameters of the source galaxy samples, the multiplicative shear bias $m$ and the source redshift distribution modes, are treated as correlated. We assume the redshift distributions and corresponding priors as estimated from data: for the \maglimpp\ lens sample, we use the results from this work; for the \texttt{metadetect} source sample \citep*{y6-metadetect}, we use the calibration from \citepalias{y6-sourcepz}. In both cases, the redshift distributions are combined with WZ constraints following the procedure described in \citepalias{y6-wz}. The covariance matrix is computed using \texttt{CosmoCov} \citep{cosmocov} at the same cosmology as the data vector, and includes contributions from shape noise and galaxy shot noise. Parameter inference is performed using the nested sampler \texttt{nautilus} \citep{nautilus}, and convergence is assessed using internal stopping criteria based on the stability of the likelihood bounds and the effective sample size, as implemented by default in \texttt{CosmoSIS}.

\begin{table*}
    \centering
    \begin{tabular}{l l l c c c}
        Analysis & Method & Marginalisation & $\sigma_8$ & $S_8$ & $\Omega_{\rm m}$\\
        \hline
        3x2pt & SOMPZ+WZ & fixed &   0.732 $\pm$ 0.024 & 0.763 $\pm$ 0.009 &  0.326 $\pm$ 0.017 \\
        3x2pt & SOMPZ+WZ & shift\&stretch &   0.726 $\pm$ 0.026 & 0.762 $\pm$ 0.010 &  0.331 $\pm$ 0.019 \\ 
        3x2pt & SOMPZ & modes &   0.726 $\pm$ 0.031 & 0.762 $\pm$ 0.011 &  0.332 $\pm$ 0.023 \\
        3x2pt & SOMPZ+WZ & modes &   0.726 $\pm$ 0.030 &  0.762 $\pm$ 0.011 &  0.331 $\pm$ 0.022 \\

        \\
        2x2pt & SOMPZ+WZ & fixed &   0.719 $\pm$ 0.031 &  0.758 $\pm$ 0.015 &  0.334 $\pm$ 0.020 \\
        2x2pt & SOMPZ+WZ & shift\&stretch &  0.711 $\pm$ 0.032 &  0.753 $\pm$ 0.017 &  0.338 $\pm$ 0.022 \\   
        2x2pt & SOMPZ & modes &   0.714 $\pm$ 0.038 &  0.756 $\pm$ 0.018 &  0.338 $\pm$ 0.026 \\
        2x2pt & SOMPZ+WZ & modes &   0.713 $\pm$ 0.033 & 0.756 $\pm$ 0.017 &  0.339 $\pm$ 0.022 \\

    \hline
    
    \end{tabular}
    \caption{Marginalised constraints on the cosmological parameters  $\sigma_8$, $S_8$, and $\Omega_{\rm m}$ for the simulated analysis described in sections \ref{sec:cosmology_calib_method} and \ref{sec:cosmology_modes}. For each parameter we report the mean of the posterior and the 68 per cent confidence interval. }
    \label{tab:cosmology}
\end{table*}

\subsection{Characterization of the Redshift Uncertainties} 
\label{sec:cosmology_modes}

In the DES Y3 analysis, the uncertainty in the \maglim redshift distributions was modeled using a two-parameter approach. 
Specifically, for each tomographic bin, we used a fiducial redshift distribution $\bar{n}(z)$, derived by averaging over the ensemble of redshift distribution realizations.
The uncertainty was parameterized in terms of a shift in the mean redshift, $\Delta z$, and a stretch in the width of the distribution, $\Delta w$. The priors on these parameters were derived by fitting the fiducial $n(z)$ to the $n(z)$ obtained from a clustering redshift measurement. These parameters were sampled from Gaussian distributions:
\begin{equation}
    \Delta_z^i \sim \mathcal{N}(0, \Delta_z), \quad \sigma_z^i \sim \mathcal{N}(0, \sigma_z),
\end{equation}
where $\Delta_z$ and $\sigma_z$ were estimated from the scatter among the realizations. 
The fiducial $\bar{n}(z)$ was then modified at each step of the inference via:
\begin{gather}
    n^i_{\rm shifted}(z) = \bar{n} (z - \Delta_z^i), \\
    n^i_{\rm shifted\&stretched}(z) = \frac{1}{\sigma_z^i}n^i_{\rm shifted}(z) \left(\frac{z - \left<z\right>}{\sigma_z^i} + \left<z\right> \right).
\end{gather}

While this method captured leading-order uncertainties in the redshift distributions, it imposed restrictive assumptions on their form, allowing only global shifts and rescalings. This can overlook more structured or localized variations in $n(z)$ that arise in realistic scenarios.

In Y6, we move beyond the shift/stretch model and adopt the more flexible mode-based marginalization framework described in Section~\ref{sec:modes} and detailed in \citepalias{y6-nzmodes}. This approach decomposes redshift uncertainties into a set of orthogonal modes $U_i$ derived from an ensemble of redshift realizations, which represent principal directions of variation in the redshift distributions. For each tomographic bin of the SOMPZ+WZ $n(z)$, we marginalise over 3 corresponding mode coefficients $u_{i,j}$ (4 modes for SOMPZ only) that parametrize the amplitude of variation along each mode. These mode coefficients serve as redshift nuisance parameters in the cosmological inference. This enables us to capture a much broader class of redshift uncertainties without imposing strong priors on their functional form. The priors and posteriors of the leading mode coefficients $u_{i,0}$ are shown in Appendix~\ref{app:modes}. As a minor numerical artifact of the linear mode combination, $n(z)$ can become slightly negative in the very low-density tails. We verified that clipping these excursions to zero has a negligible impact on cosmological constraints; see Appendix~\ref{app:clipped}.

In Figure~\ref{fig:cosmology_marg}, we compare the cosmological constraints obtained using these two approaches. Specifically, we marginalize over the source redshift uncertainties using the SOMPZ+WZ modes, and compare the resulting $S_8$–$\Omega_{\rm m}$ posteriors when applying either the shift/stretch method or the mode-based marginalization for the lens sample. For the 2$\times$2pt analysis, the joint analysis of galaxy clustering and galaxy-galaxy lensing, the contours are nearly identical, whereas for 3$\times$2pt the shift/stretch method yields slightly tighter constraints—likely due to a mild underestimation of uncertainties by the parametric model. The marginalized posterior values are listed in Table~\ref{tab:cosmology}. The close agreement between the two approaches offers a useful consistency check, but the mode-based framework is ultimately preferred: it captures a broader range of redshift variations, ensures consistency across samples and probes, and provides a more robust basis for future analyses.

\begin{figure*}
    \centering
    \includegraphics[width=0.49\linewidth]{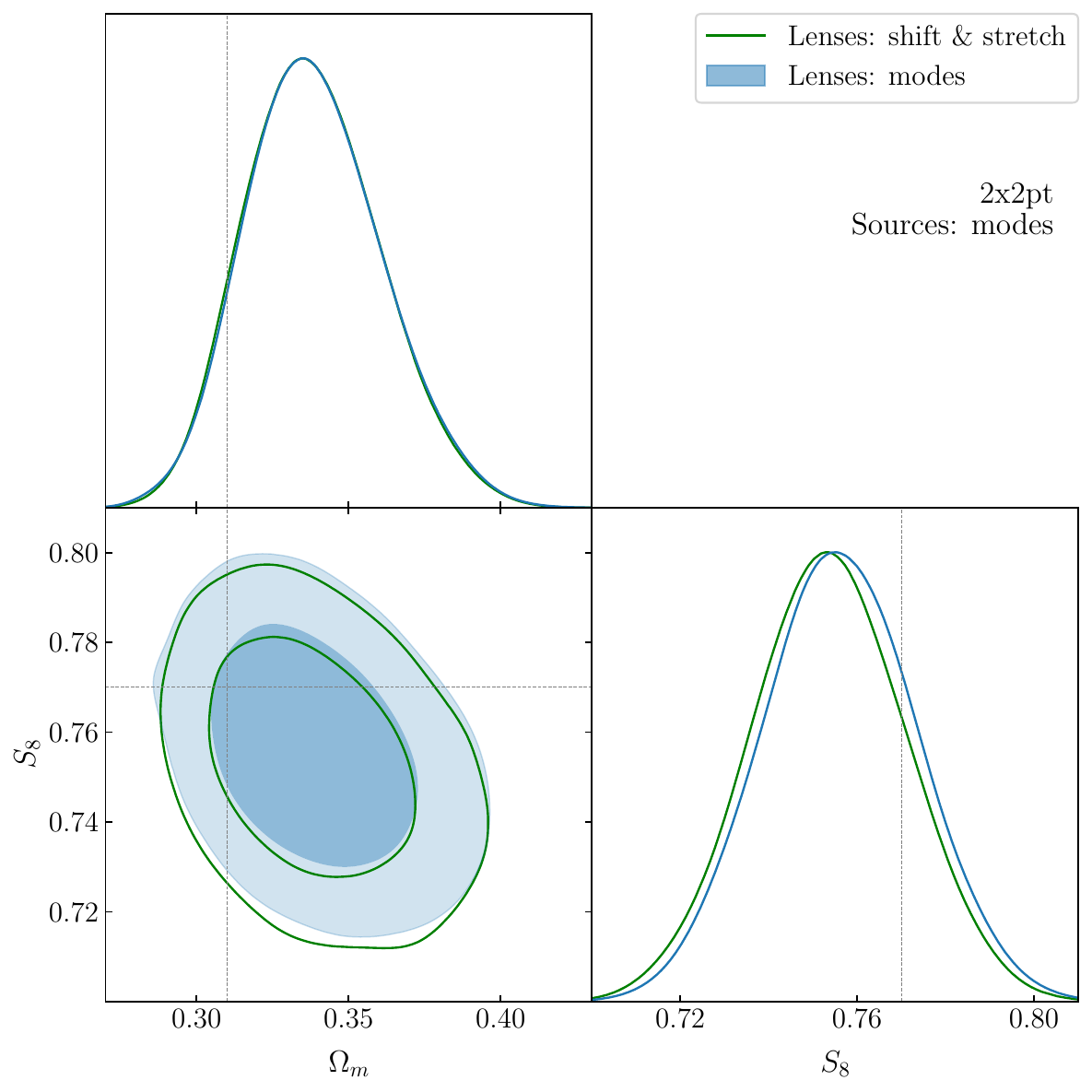}
    \includegraphics[width=0.49\linewidth]{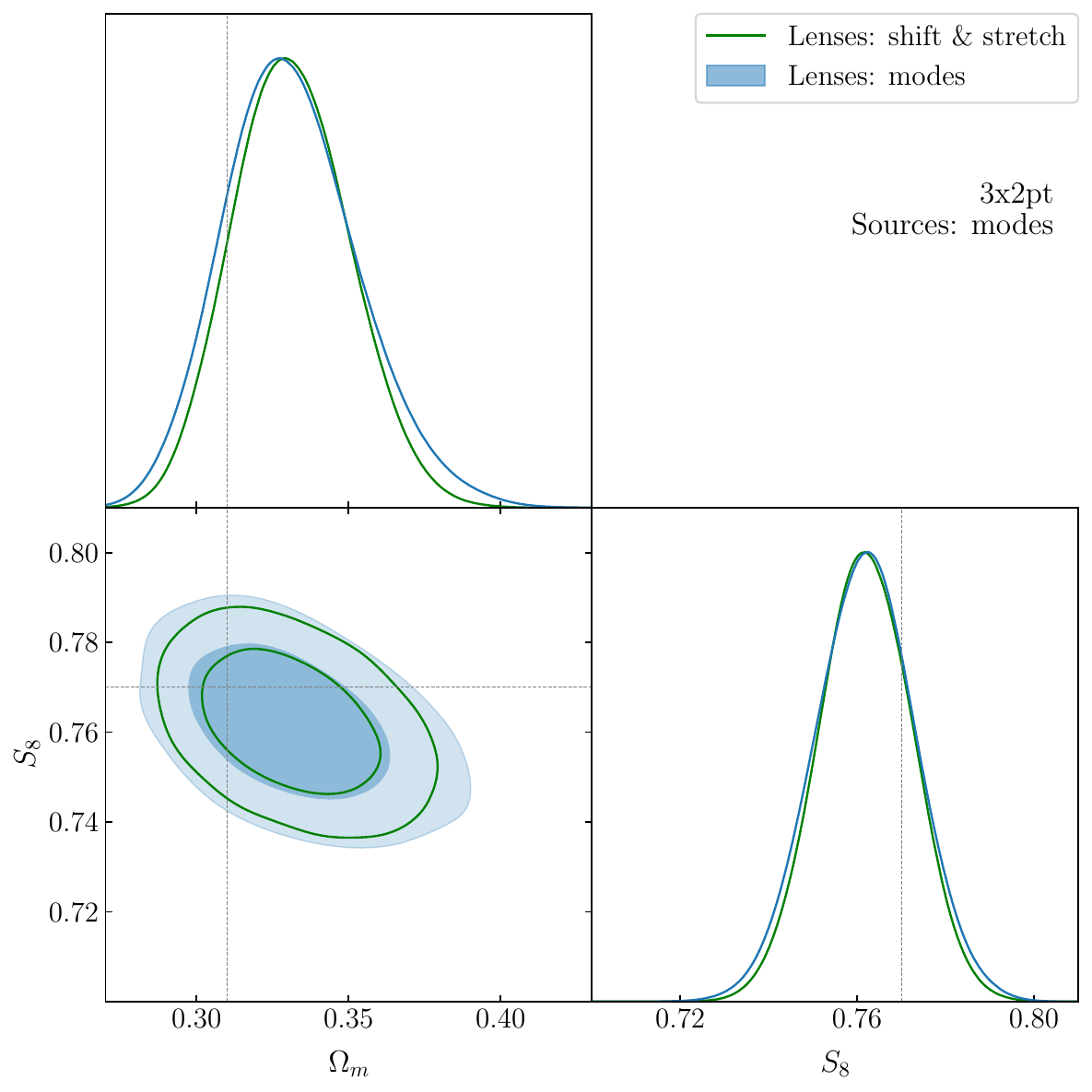}
    \caption{Comparison of marginalization methods for photometric redshift uncertainties for 2$\times$2pt (left) and 3x2pt (right). The figure contrasts the shift and stretch approach (green) with the mode-based marginalization method used in the fiducial analysis (blue). While the resulting $S_8$–$\Omega_{\rm m}$ constraints for 2$\times$2pt are very similar, for 3x2pt the shift and stretch method yields slightly tighter contours, suggesting that it may underestimate the true redshift uncertainties, as by construction the mode-based framework captures a broader range of redshift distribution variations in a data-driven and flexible manner, providing a more accurate and robust marginalization approach for current and future analyses.
}
    \label{fig:cosmology_marg}
\end{figure*}

\subsection{Impact on Cosmology from the \maglimpp Redshift Uncertainties}
\label{sec:cosmology_calib_method}

The accuracy of photometric redshift estimates directly impacts cosmological parameter constraints in the 2$\times$2pt analysis. In this section, we quantify how uncertainties in the \texttt{MagLim++} redshift distribution propagate into cosmological inference. As described in the previous section, we perform this test using a simulated data vector, enabling us to isolate the impact of redshift uncertainties in a controlled setting.

\begin{figure*}
    \centering
    \includegraphics[width=0.49\linewidth]{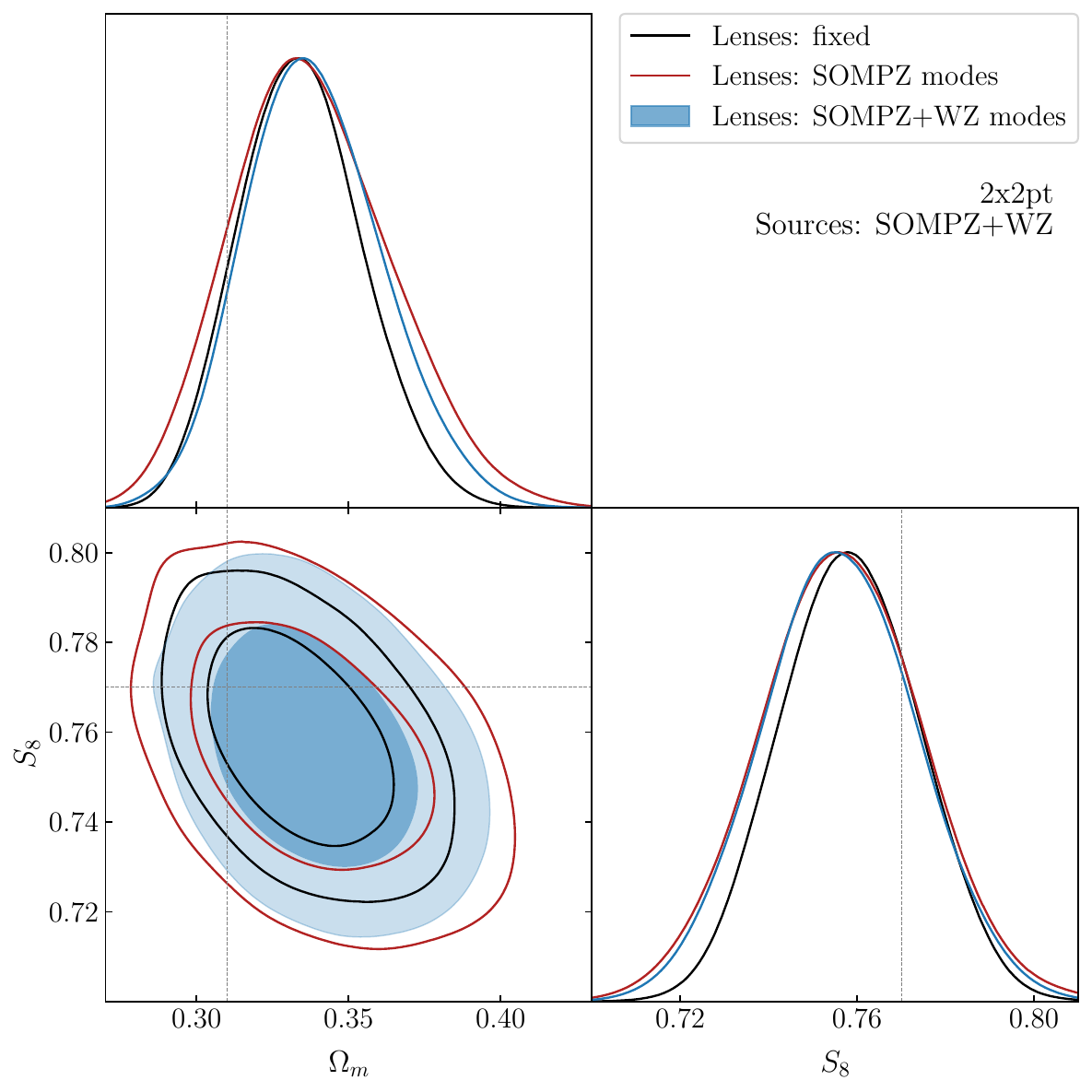}
    \includegraphics[width=0.49\linewidth]{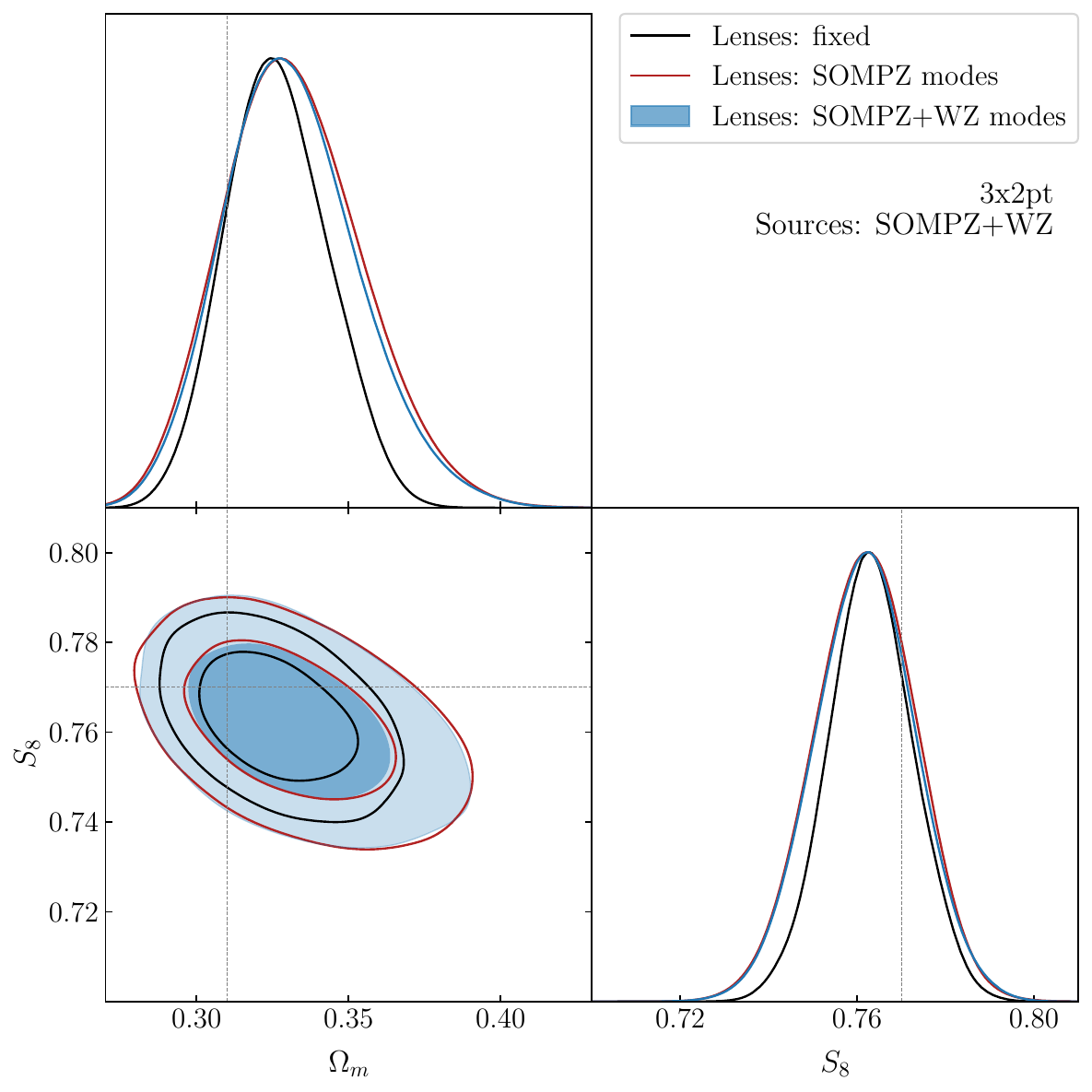}
    \caption{Impact of photo-z uncertainty modeling on cosmological parameter constraints ($\Omega_{\rm m}$, $\sigma_8$, and $S_8$) in 2$\times$2pt analysis. The different contours show the effect of fixing the n(z) for both lenses and sources, fixing n(z) only for lenses while marginalizing over SOMPZ+WZ modes for sources, using SOMPZ-only modes for lenses, and marginalizing over SOMPZ+WZ modes for both lenses and sources. Increasing marginalization freedom shifts and broadens the posteriors, highlighting the impact of redshift uncertainties on parameter constraints.}
    \label{fig:cosmology_unc}
\end{figure*}

Figure~\ref{fig:cosmology_unc} shows constraints in the $(\Omega_{\rm m}, S_8)$ plane under three different treatments of the lens redshift distribution: (i) assuming a fixed $n(z)$; (ii) marginalizing over SOMPZ-derived modes; and (iii) marginalizing over SOMPZ+WZ modes. In all cases, the source redshift distribution is treated using SOMPZ+WZ modes.

The three chains are centered around a common region in parameter space, but the posterior widths depend on the treatment of redshift uncertainties. As expected, the fixed-$n(z)$ case yields the tightest constraints, since it assumes perfect knowledge of the lens redshift distribution. Marginalizing over SOMPZ modes introduces additional uncertainty and broadens the contours. Including WZ information alongside SOMPZ typically helps recover some constraining power. This pattern is clearly observed in the 2$\times$2pt analysis. For the 3$\times$2pt case (right panel), the constraints from SOMPZ and SOMPZ+WZ are nearly identical. We note that, given the already good calibration of the lens $n(z)$, the cosmological constraints are not very responsive to variations in the lens $n(z)$, rendering the additional WZ information redundant relative to the SOMPZ-only case. The corresponding marginalized constraint values for each case are listed in Table~\ref{tab:cosmology}.


We also note that the posteriors are not centered exactly on the true cosmology of the simulation. This is expected, and likely reflects projection effects in the inference process. In a high-dimensional parameter space, even when the true model lies within the allowed region, marginalized posteriors in low-dimensional projections (such as the $(\Omega_{\rm m}, S_8)$ plane) may appear offset from the truth. 

These results illustrate both the importance of properly treating redshift uncertainties and the utility of combining complementary redshift calibration methods. Including WZ information not only improves the fidelity of the redshift model, but also helps recover constraining power that would otherwise be lost through marginalization.

\begin{figure}
    \centering
    \includegraphics[width=\linewidth]{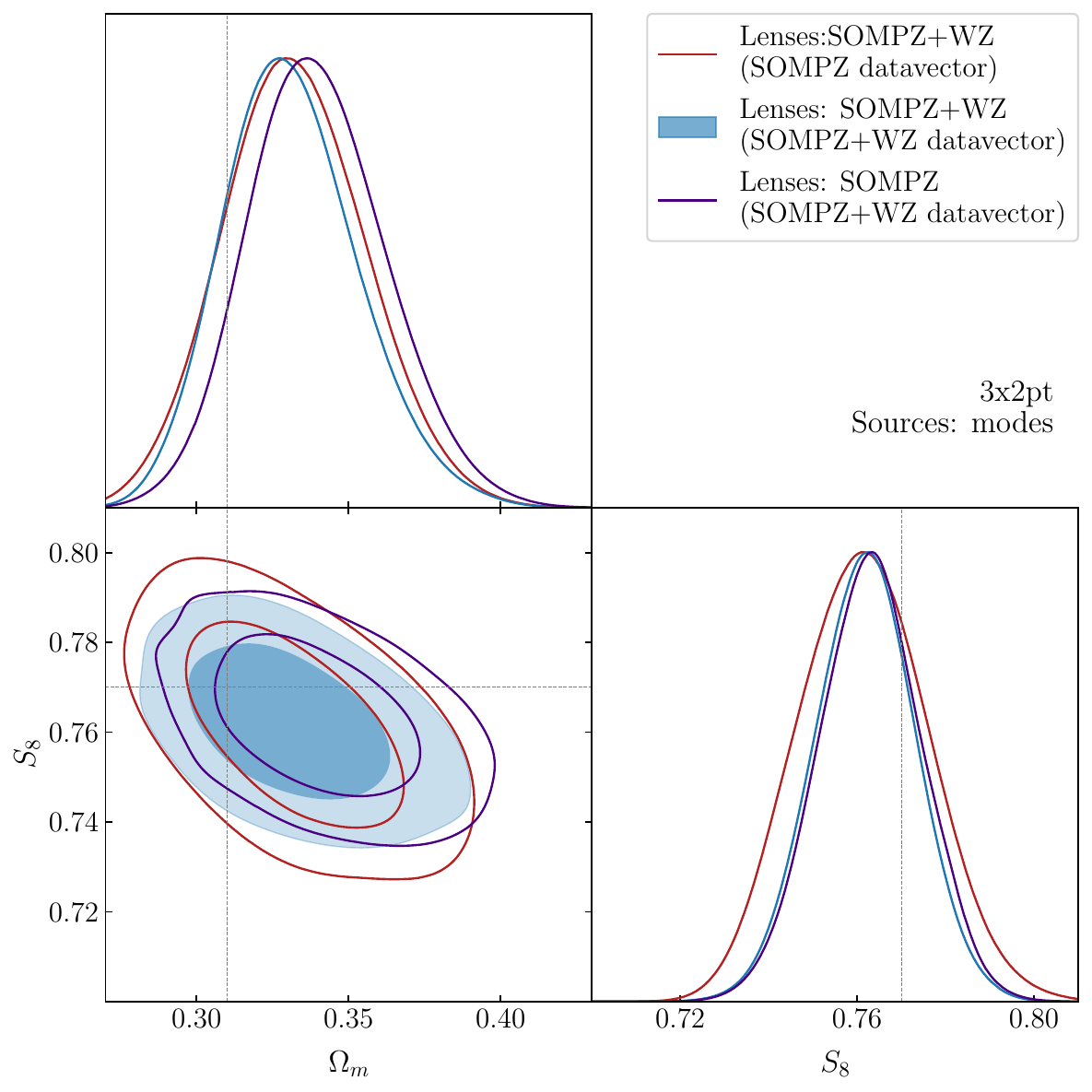}
    \caption{Impact of mismatches between the redshift distributions used in the theory data vector and the redshift modes used for data compression on cosmological constraints. 
    We show the $\Omega_{\mathrm{m}}$–$S_8$ posterior contours for three runs: 
        using \texttt{SOMPZ} redshift distributions with \texttt{SOMPZ+WZ} compression modes (red), 
        using \texttt{SOMPZ+WZ} redshift distributions with \texttt{SOMPZ} modes (green), 
        and using \texttt{SOMPZ+WZ} redshift distributions with matched \texttt{SOMPZ+WZ} modes (blue), 
        which represents our fiducial analysis. 
        We find that the cosmological constraints are insensitive to moderate mismatches between the redshift model used to generate the theory vector and the one used to construct the compression basis. 
        In particular, the combination of \texttt{SOMPZ} redshifts with \texttt{SOMPZ+WZ} modes yields results consistent with the fiducial run, confirming the robustness of the method.
    }
    \label{fig:dv_sensitivity}
\end{figure}

As an additional test, we verify the robustness of the compressed likelihood pipeline to inconsistencies between the redshift distributions used to generate the theory data vector and the redshift modes used for compression. We perform a set of internal validation tests.

We consider three configurations:
\begin{itemize}
    \item The baseline configuration, where both the theory data vector and the compression modes are computed using the \texttt{SOMPZ+WZ} redshift distributions.
    \item A test configuration where the theory data vector is computed using \texttt{SOMPZ} redshift distributions, while the compression modes are derived from \texttt{SOMPZ+WZ}.
    \item A second test configuration where the theory data vector uses \texttt{SOMPZ+WZ}, while the compression modes are derived from \texttt{SOMPZ}.
\end{itemize}

Figure~\ref{fig:dv_sensitivity} compares the $\Omega_{\mathrm{m}}$–$S_8$ constraints obtained with three different choices for the lens redshift distribution used in the data vector and in the theory predictions.  
The fiducial case, in which both the compressed data vector and the model employ the \texttt{SOMPZ+WZ} redshifts, delivers the tightest contours.  
If the data vector is built with \texttt{SOMPZ} redshifts while the theory still uses \texttt{SOMPZ+WZ} $n(z)$, the posteriors broaden modestly, with a distance of $0.14 \sigma$ between the two contours, indicating that a less informative compression basis degrades the cosmological precision.  
A similar, though slightly smaller, widening appears when the roles are reversed—i.e.\ the data vector retains \texttt{SOMPZ+WZ} while the theory adopts \texttt{SOMPZ} $n(z)$—with the corresponding distance reduced to $0.008\sigma$.   

We have also verified that the nuisance parameters most directly coupled to the lens $n(z)$—the galaxy bias and magnification slopes—remain stable across these tests. 
The corresponding posteriors are shown in Appendix~\ref{app:bias_mag}.

\section{Discussion}
\label{sec:conclusions}


Photometric redshift (photo-\(z\)) estimation for lens galaxies in the Dark Energy Survey (DES) 3x2pt analysis has undergone substantial refinement over the course of the survey. In the Science Verification (SV) and Year 1 (Y1) analyses, the lens galaxy sample was based on \texttt{redMaGiC} \citep{Rozo2016}, a selection of red-sequence galaxies with tightly controlled photo-\(z\) estimates. The \texttt{redMaGiC} algorithm provides high-precision photometric redshifts, calibrated using spectroscopic samples and characterized by a typical scatter of \(\sigma_z / (1+z) \sim 0.015\), making it well-suited for early cosmological analyses.

In DES Year 3 (Y3), a new sample of lens galaxies known as \texttt{MagLim} was introduced \citep{annamaglim}, defined by magnitude and optimized for cosmological signal-to-noise in 2$\times$2pt and 3$\times$2pt analyses. For this sample, the fiducial photo-\(z\) estimation was performed using the Directional Neighbourhood Fitting (DNF) algorithm \citep{DeVicente2016}, with redshift distributions constructed for each tomographic bin. To account for biases in the DNF estimates, the redshift distributions were empirically corrected using clustering redshift (WZ) measurements, by applying a shift and stretch in each bin to align the DNF \(n(z)\) with the WZ estimates \citep{y3-lenswz}. These corrected distributions, along with Gaussian priors on the shift and stretch parameters, were propagated through the cosmological inference pipeline.
A similar approach was taken for the DES Y6 BAO analysis \citep{bao}, where the DNF $n(z)$ served as the baseline shape, and was shifted and stretched to match WZ (at $z_{\rm ph}<1$) or VIPERS (at higher $z$) \citep{juanmena}.

In Y3, sources and lenses photo-$z$ calibrations approaches were different. In fact, the SOMPZ+WZ methodology had been developed and validated specifically for the source galaxies only \citep*{y3-sompzbuzzard, y3-sompz}. Motivated by its success in the source sample, this approach was applied for the first time to a lens galaxy sample in a dedicated reanalysis of the \texttt{MagLim} redshift distributions \citep{Giannini2024}. We found that using SOMPZ+WZ instead of the fiducial DNF+\texttt{WZ}-corrected redshifts resulted in a shift of approximately \(0.4\sigma\) in the \(\Omega_{\rm m}\)-\(S_8\) parameter plane, indicating the importance of more robust redshift calibration. Given that the SOMPZ+WZ method provides a more comprehensive and flexible estimation of \(n(z)\) with a clearer uncertainty model, it was adopted as the fiducial redshift calibration approach for the DES Year 6 (Y6) cosmological analysis.

In this paper, we calibrate the redshift distributions and associated uncertainties for the Y6 lens galaxy sample, which enter the modeling of galaxy clustering and galaxy-galaxy lensing observables and represent a key input to the DES Y6 3$\times$2pt cosmological analysis \citep[in prep.]{y6-3x2pt}; we also evaluate their impact on parameter constraints using simulated likelihood analyses. In Y6, there have been even further refinements to both the upstream data products, and the methodology, which together contribute to better redshift calibration. In terms of improved data samples:
\begin{itemize}
\item the Y6 imaging is about twice as deep as in Y3, resulting in lower magnitude errors. For the bright MagLim lenses ($i<22.2$) this has only a minor effect on the photo-$z$ calibration, but it provides more precise color measurements and contributes to the overall robustness of the redshift inference;
\item there have been many advances in the curation of the \maglimpp sample described in \citep*{y6-maglim}, including less contamination from stars and QSOs, better control of photo-$z$ outliers, and better masking \citep*{RodriguezMonroy2025};
\item the number of objects in the \texttt{Balrog} simulations generated from synthetic source injection is almost 5 times larger than Year 3 and now covers the entire footprint. This translates into a more precise and less biased transfer function between the deep and the wide fields;
\item significant improvements to the redshift calibration sample were achieved by prioritizing the most accurate redshift per object and restricting to deep-field regions with complete coverage, minimizing selection biases.
\end{itemize}

In addition, this work incorporates significant advances in the methodology over the previous version from Y3:
\begin{itemize}
\item an improved SOM algorithm is used for redshift calibration \citep{Campos2024}, with better handling of photometric uncertainties, non-periodic boundaries, and $g$-band inclusion—leading to a more realistic and robust color space representation;
\item we introduced a more comprehensive framework to estimate redshift sample uncertainties, capturing residual biases and catalog incompleteness;
\item a new approach based on \citepalias{y6-nzmodes} is taken to parametrize the uncertainty in the redshift distributions that allows for more flexible and general variations in the distributions, rather than relying on simple shift-and-stretch parametrizations.

\end{itemize}

Figure~\ref{fig:nzmodes} displays the mode-reconstructed redshift distributions, $n(z)$, that serve as the effective inputs to the cosmological parameter inference.
Compared with the redshift distributions of the \maglim sample used in \citep{Giannini2024}, the relative contribution of each source of uncertainty differs significantly. In Y3, sample variance and shot noise were typically the dominant sources of uncertainty on the mean redshift across all bins, with subdominant contributions from redshift sample incompleteness and photometric zeropoint variations. In Y6, the breakdown is more nuanced: no single source dominates across all bins or metrics, with redshift sample and zero point uncertainties becoming more relevant in Y6, in particular the latter is the dominant uncertainty in bin~6.  This more complex structure reflects the improved modeling of systematic effects in Y6, as well as the use of a deeper and more representative redshift sample, resulting in a more complete and accurate propagation of redshift uncertainties across the tomographic bins.

We investigate the impact of the specific parametrization used to model redshift uncertainties by comparing the traditional shift-and-stretch approach to our new framework, which projects uncertainties onto modes that most affect cosmological inference. We find that the resulting cosmological constraints are similar in both the 2$\times$2pt and 3$\times$2pt analyses, though the shift-and-stretch method yields slightly tighter constraints in 3$\times$2pt. This suggests that the shift-and-stretch parametrization may underestimate the true redshift uncertainties. In contrast, the mode-based approach offers a more flexible and principled treatment that is better suited for capturing complex, data-driven variations in the redshift distributions and for extending to joint analyses.

We include these redshift distributions in a DES Y6-like simulated likelihood analysis. We analyze either the 2$\times$2pt (galaxy clustering + galaxy-galaxy lensing) or 3$\times$2pt (2$\times$2pt + cosmic shear) data vectors under $\Lambda$CDM cosmology. Fixing everything else to the fiducial analysis choice and only varying the treatment of the lens redshift, we find that overall, the lens redshift uncertainties enlarges the overall constraints on $S_8$ by $20\%$ for both 2$\times$2pt and 3$\times$2pt when including SOMPZ information; including WZ yields tighter constraints on $S_8$ of $5\%$ for 2$\times$2pt, while it remains stable for 3$\times$2pt. The marginalised posteriors values are summarised in Table \ref{tab:cosmology}.


This work is part of a coordinated effort to calibrate redshifts across all galaxy samples used in the DES Y6 $3\times2pt$ cosmological analysis. In addition to the lens redshift calibration presented here, companion papers detail the calibration of source galaxy redshifts using the SOMPZ+WZ methodology \citepalias{y6-sourcepz}, the derivation of clustering redshift measurements for both source and lens samples \citepalias{y6-wz}, and the development of the mode-based uncertainty parametrization \citepalias{y6-nzmodes}. Together, these studies provide a unified and self-consistent framework for treating redshift uncertainties across the full 3$\times$2pt data vector.

In summary, the DES Y6 redshift calibration for the \maglimpp{} lens sample marks a significant advance in the precision and robustness of photometric redshift treatment for cosmological analyses. Through improved data products, a carefully curated redshift sample, and a principled framework that propagates uncertainties directly into cosmological inference, we ensure that redshift errors are accurately modeled and consistently integrated into the analysis. This approach strengthens the reliability of the DES Y6 3$\times$2pt results and establishes a scalable foundation for redshift calibration in next-generation multi-probe surveys such as the Vera C. Rubin Observatory's Legacy Survey of Space and Time and the Euclid mission.

\section*{Acknowledgements}

\textbf{Author Contributions:} All authors contributed to this paper and/or
carried out infrastructure work that made this analysis possible. GG
performed almost all analysis and manuscript preparation. AAl contributed to analysis and manuscript preparation, and prepared the redshift sample. CC contributed with manuscript preparation. GB and Troxel developed the modes pipeline. WA ran the clustering-z measurement. GB also performed the importance sampling. BY and AAm contributed to analysis interpretation as members of the DES Y6 redshift WG.
NW, JM, AP contributed to
manuscript preparation as collaboration internal reviewers.
The remaining authors have made contributions to this paper that include,
but are not limited to, the construction of DECam and other aspects
of collecting the data; data processing and calibration; catalog creation; developing
broadly used methods, codes, and simulations; running the pipelines
and validation tests; and promoting the science analysis

Funding for the DES Projects has been provided by the U.S. Department of Energy, the U.S. National Science Foundation, the Ministry of Science and Education of Spain, 
the Science and Technology Facilities Council of the United Kingdom, the Higher Education Funding Council for England, the National Center for Supercomputing 
Applications at the University of Illinois at Urbana-Champaign, the Kavli Institute of Cosmological Physics at the University of Chicago, 
the Center for Cosmology and Astro-Particle Physics at the Ohio State University,
the Mitchell Institute for Fundamental Physics and Astronomy at Texas A\&M University, Financiadora de Estudos e Projetos, 
Funda{\c c}{\~a}o Carlos Chagas Filho de Amparo {\`a} Pesquisa do Estado do Rio de Janeiro, Conselho Nacional de Desenvolvimento Cient{\'i}fico e Tecnol{\'o}gico and 
the Minist{\'e}rio da Ci{\^e}ncia, Tecnologia e Inova{\c c}{\~a}o, the Deutsche Forschungsgemeinschaft and the Collaborating Institutions in the Dark Energy Survey. 

The Collaborating Institutions are Argonne National Laboratory, the University of California at Santa Cruz, the University of Cambridge, Centro de Investigaciones Energ{\'e}ticas, 
Medioambientales y Tecnol{\'o}gicas-Madrid, the University of Chicago, University College London, the DES-Brazil Consortium, the University of Edinburgh, 
the Eidgen{\"o}ssische Technische Hochschule (ETH) Z{\"u}rich, 
Fermi National Accelerator Laboratory, the University of Illinois at Urbana-Champaign, the Institut de Ci{\`e}ncies de l'Espai (IEEC/CSIC), 
the Institut de F{\'i}sica d'Altes Energies, Lawrence Berkeley National Laboratory, the Ludwig-Maximilians Universit{\"a}t M{\"u}nchen and the associated Excellence Cluster Universe, 
the University of Michigan, NSF NOIRLab, the University of Nottingham, The Ohio State University, the University of Pennsylvania, the University of Portsmouth, 
SLAC National Accelerator Laboratory, Stanford University, the University of Sussex, Texas A\&M University, and the OzDES Membership Consortium.

Based in part on observations at NSF Cerro Tololo Inter-American Observatory at NSF NOIRLab (NOIRLab Prop. ID 2012B-0001; PI: J. Frieman), which is managed by the Association of Universities for Research in Astronomy (AURA) under a cooperative agreement with the National Science Foundation.

The DES data management system is supported by the National Science Foundation under Grant Numbers AST-1138766 and AST-1536171.
The DES participants from Spanish institutions are partially supported by MICINN under grants PID2021-123012, PID2021-128989 PID2022-141079, PID2023-153229NA-I00, SEV-2016-0588, CEX2020-001058-M and CEX2020-001007-S, some of which include ERDF funds from the European Union. IFAE is partially funded by the CERCA program of the Generalitat de Catalunya.

We  acknowledge support from the Brazilian Instituto Nacional de Ci\^encia
e Tecnologia (INCT) do e-Universo (CNPq grant 465376/2014-2).

Part of this research was carried out at the Jet Propulsion Laboratory, California Institute of Technology, under a contract with the National Aeronautics and Space Administration (80NM0018D0004).

This document was prepared by the DES Collaboration using the resources of the Fermi National Accelerator Laboratory (Fermilab), a U.S. Department of Energy, Office of Science, Office of High Energy Physics HEP User Facility. Fermilab is managed by Fermi Forward Discovery Group, LLC, acting under Contract No. 89243024CSC000002.

AA acknowledges support from the Spanish Ministerio
de Ciencia, Innovacion y Universidades (MICIU), project
PID2023-153229NA-I00. The project that gave rise to
these results received the support of a fellowship from
“la Caixa” Foundation (ID 100010434). The fellowship
code is LCF/BQ/PI23/11970028.

Funding for the Sloan Digital Sky Survey IV has been provided by the Alfred P. Sloan Foundation, the U.S. Department of Energy Office of Science, and the Participating Institutions. SDSS acknowledges support and resources from the Center for High-Performance Computing at the University of Utah. The SDSS web site is www.sdss.org.

SDSS is managed by the Astrophysical Research Consortium for the Participating Institutions of the SDSS Collaboration including the Brazilian Participation Group, the Carnegie Institution for Science, Carnegie Mellon University, Center for Astrophysics | Harvard \& Smithsonian (CfA), the Chilean Participation Group, the French Participation Group, Instituto de Astrofisica de Canarias, The Johns Hopkins University, Kavli Institute for the Physics and Mathematics of the Universe (IPMU) / University of Tokyo, the Korean Participation Group, Lawrence Berkeley National Laboratory, Leibniz Institut f{\"u}r Astrophysik Potsdam (AIP), Max-Planck-Institut f{\"u}r Astronomie (MPIA Heidelberg), Max-Planck-Institut f{\"u}r Astrophysik (MPA Garching), Max-Planck-Institut f{\"u}r Extraterrestrische Physik (MPE), National Astronomical Observatories of China, New Mexico State University, New York University, University of Notre Dame, Observatorio Nacional / MCTI, The Ohio State University, Pennsylvania State University, Shanghai Astronomical Observatory, United Kingdom Participation Group, Universidad Nacional Aut\'{o}noma de M\'{e}xico, University of Arizona, University of Colorado Boulder, University of Oxford, University of Portsmouth, University of Utah, University of Virginia, University of Washington, University of Wisconsin, Vanderbilt University, and Yale University.

WA acknowledges support from the MICINN projects PID2019-111317GB-C32, PID2022-141079NB-C32 as well as predoctoral program AGAUR-FI ajuts (2024 FI-1 00692) Joan Oró.

\section{Data Availability}
The DES Y6 data products used in this work, as well as the full ensemble of DES Y6 \maglim\ sample redshift distributions described by this work, will be publicly available at \citep{y6-data}. As cosmology likelihood sampling software we use \texttt{cosmosis}, available at \citep{cosmosis-github}.





\bibliography{references}


\appendix
\section{SOM}\label{app:soms}

\begin{figure*}
    \centering
    \includegraphics[width=0.49\linewidth]{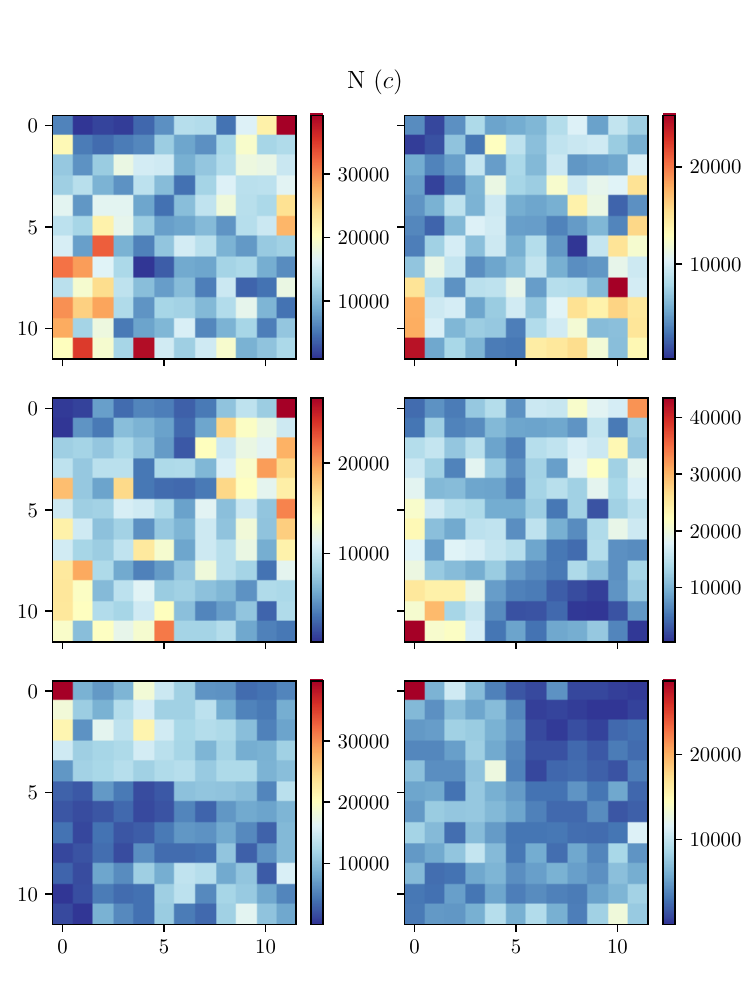}
    \includegraphics[width=0.49\linewidth]{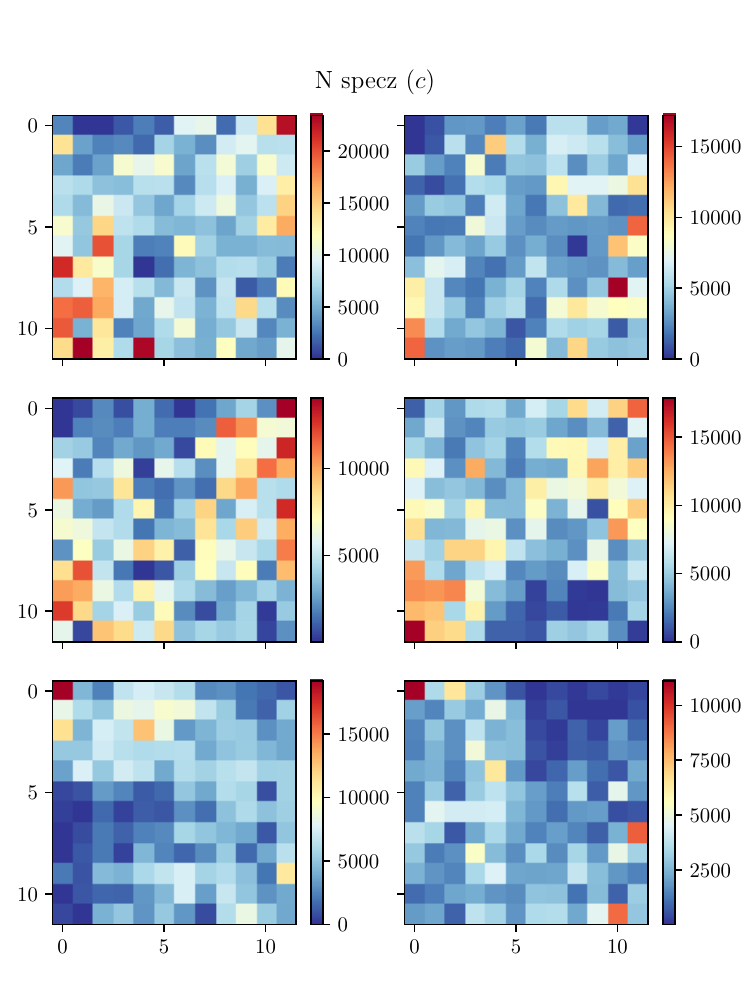}
    \caption{SOM cell occupancy and spectroscopic coverage for the deep sample. \textit{Left panel:} Number of galaxies in each deep SOM cell. \textit{Right panel:} Number of galaxies with secure spectroscopic redshifts in each deep SOM cell. These maps are used to assess the sampling of the SOM in both photometric and spectroscopic space. As discussed in the main text, high redshift scatter is not systematically correlated with low occupancy or poor spec-$z$ coverage.}
\label{fig:deep_som_occupancy}
\end{figure*}

\begin{figure*}
    \centering
    \includegraphics[width=0.49\linewidth]{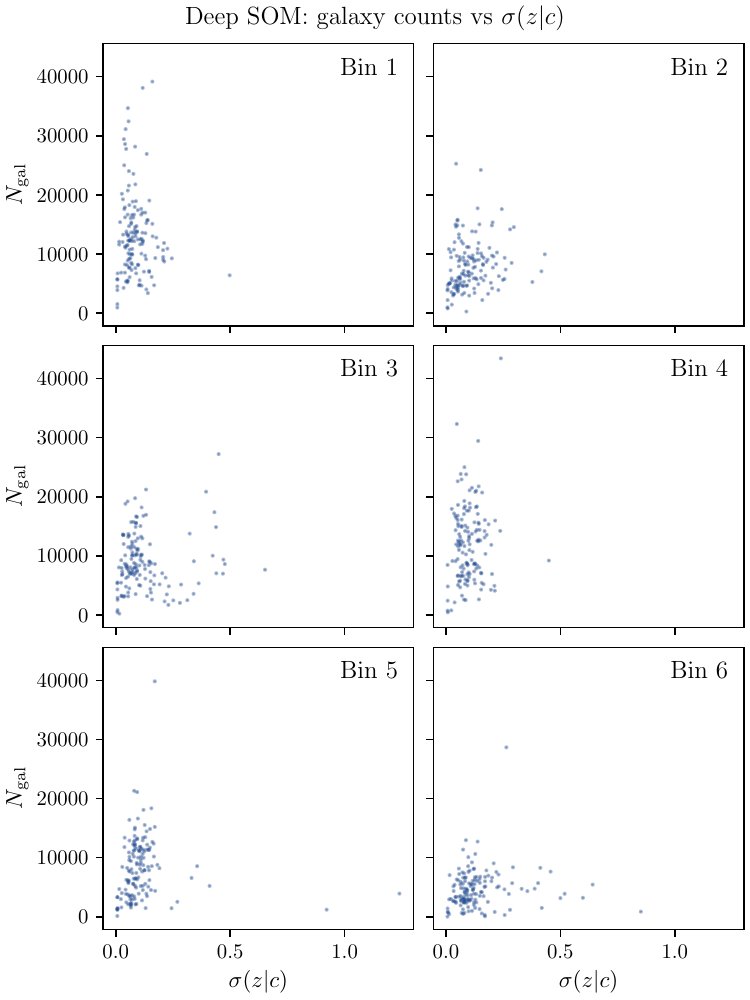}
    \includegraphics[width=0.49\linewidth]{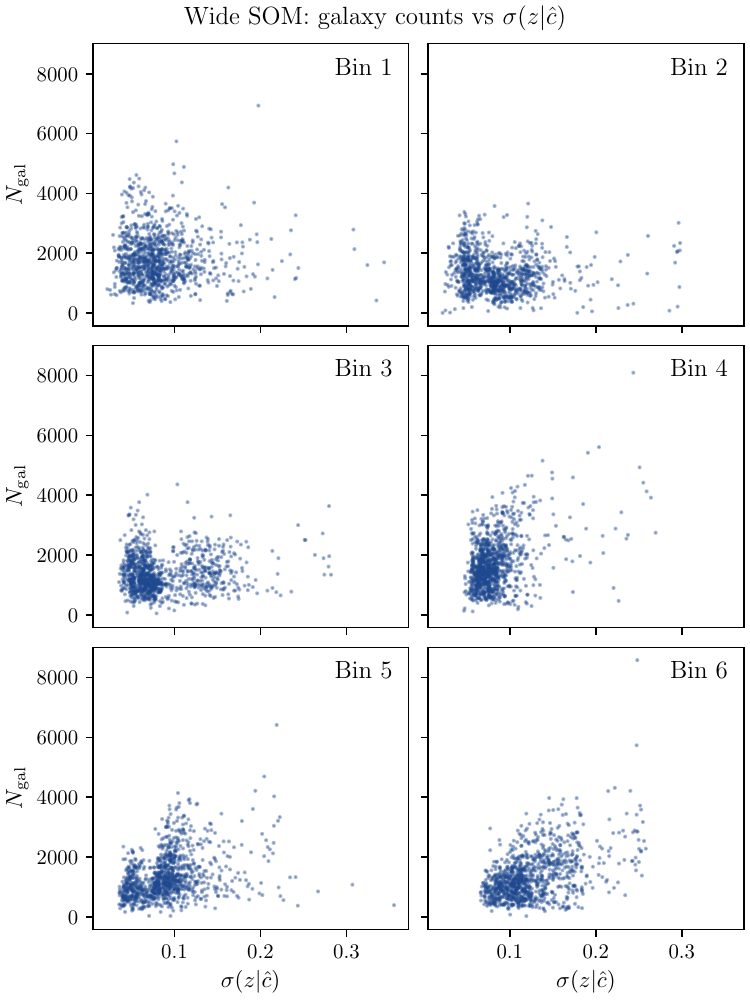}
    \caption{Number of galaxies per value of redshift scatter $\sigma(z|c)$ for each tomographic bin.  The left panel shows the deep SOM, and the right panel shows the wide SOM. Each point corresponds to a SOM cell; the $x$-axis indicates the scatter $\sigma(z|c)$ associated with that cell, while the $y$-axis shows the number of galaxies assigned to it. The deep Balrog sample displays a broader range of scatter values, but high-scatter cells contain very few galaxies. In the wide sample (right panel), the distribution is narrower and peaks at lower values of $\sigma(z|c)$, reflecting the fact that most wide-field galaxies are assigned to well-calibrated cells in the deep SOM. This confirms that redshift uncertainties are well controlled for the majority of galaxies in both samples.}
\label{fig:som_occupancy_vs_sigmaz}
\end{figure*}

Figure~\ref{fig:deep_som_occupancy} shows the number of galaxies (left) and the number of secure spectroscopic redshifts (right) assigned to each deep SOM cell. These maps illustrate how the deep SOM is populated by the full photometric sample and how well it is sampled by the redshift calibration set. While the spectroscopic coverage is somewhat heterogeneous across the SOM, we find no systematic correlation between $\sigma(z|c)$ and either the photometric or spectroscopic cell occupancy, consistent with the interpretation of redshift scatter provided in Section~\ref{sec:soms}. 
We note that the occupancy of the deep SOM cells is highly uneven, with a small number of cells containing a large fraction of the photometric sample. While this might appear concerning, these high-occupancy cells are typically located in regions of color--magnitude space with well-constrained redshifts and abundant spectroscopic coverage. To further assess the impact of high-$\sigma(z|c)$ cells, Figure~\ref{fig:som_occupancy_vs_sigmaz} shows the number of galaxies per value of $\sigma(z|c)$, for both the deep and wide SOMs. These histograms allow us to quantify the fraction of galaxies that fall into poorly calibrated SOM regions, defined here as cells with large redshift scatter. As shown in Figure~\ref{fig:som_occupancy_vs_sigmaz}, the vast majority of galaxies in these cells have low $\sigma(z|c)$ values, confirming that their contribution to the redshift distribution is well calibrated. Moreover, the SOMPZ framework naturally accounts for this imbalance by weighting each cell’s contribution according to its population, ensuring that sparsely populated or poorly calibrated regions have minimal impact on the final $n(z)$ estimate.

In the deep sample, we find that the vast majority of galaxies occupy SOM cells with $\sigma(z|c) < 0.1$, and fewer than 1\% of galaxies lie in cells with $\sigma(z|c) > 0.2$. This demonstrates that, despite the presence of a small number of cells with high redshift uncertainty, these regions contribute negligibly to the overall redshift distribution. The result reflects the effectiveness of the deep SOM in organizing galaxies according to photometric properties that are predictive of redshift.

In the wide sample, the distribution of $\sigma(z|c)$ is narrower, with a sharper drop-off beyond $\sim0.15$. This may seem counterintuitive given the increased photometric noise in the wide field; however, it is a natural outcome of the fact that redshift estimates for wide galaxies are inherited from their deep counterparts. In particular, the wide sample is dominated by cells where the deep-redshift calibration is precise, and relatively few wide galaxies fall into high-scatter regions of the deep SOM.

Overall, this analysis confirms that both the deep and wide samples are predominantly associated with SOM regions where redshift uncertainties are well controlled. High-$\sigma(z|c)$ cells are rare and contain only a small fraction of galaxies, indicating that the global $n(z)$ calibration is robust against localized failures in redshift estimation.



\section{Modes}\label{app:modes}

\begin{figure*}
    \centering
    \includegraphics[width=0.49\linewidth]{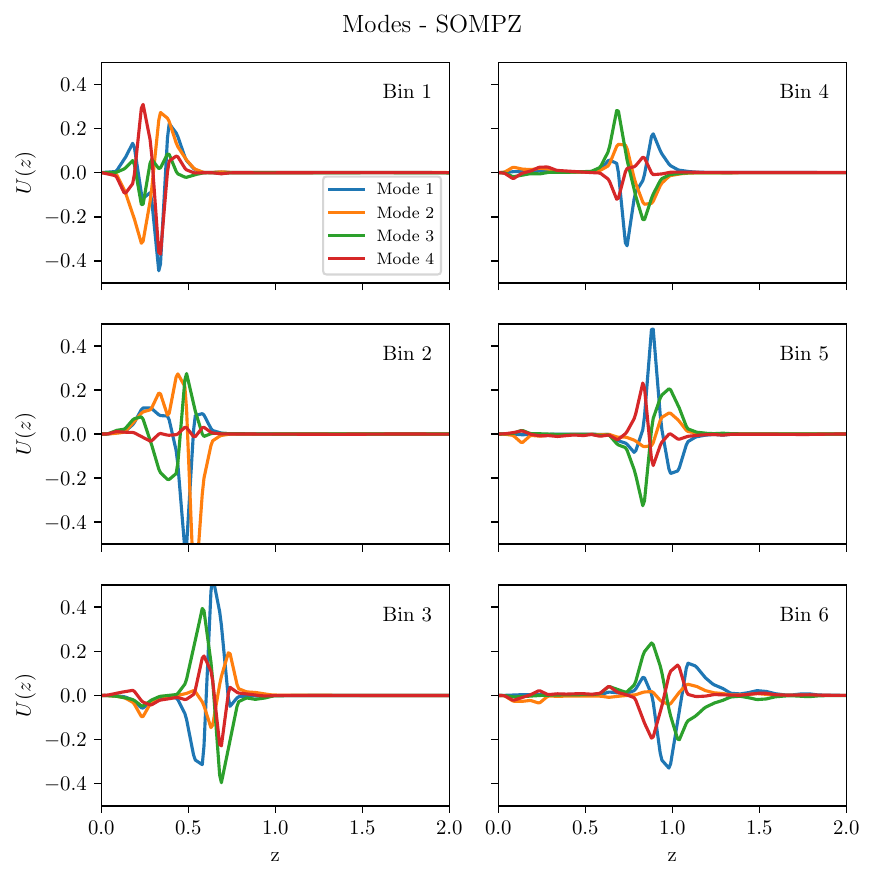}
    \includegraphics[width=0.49\linewidth]{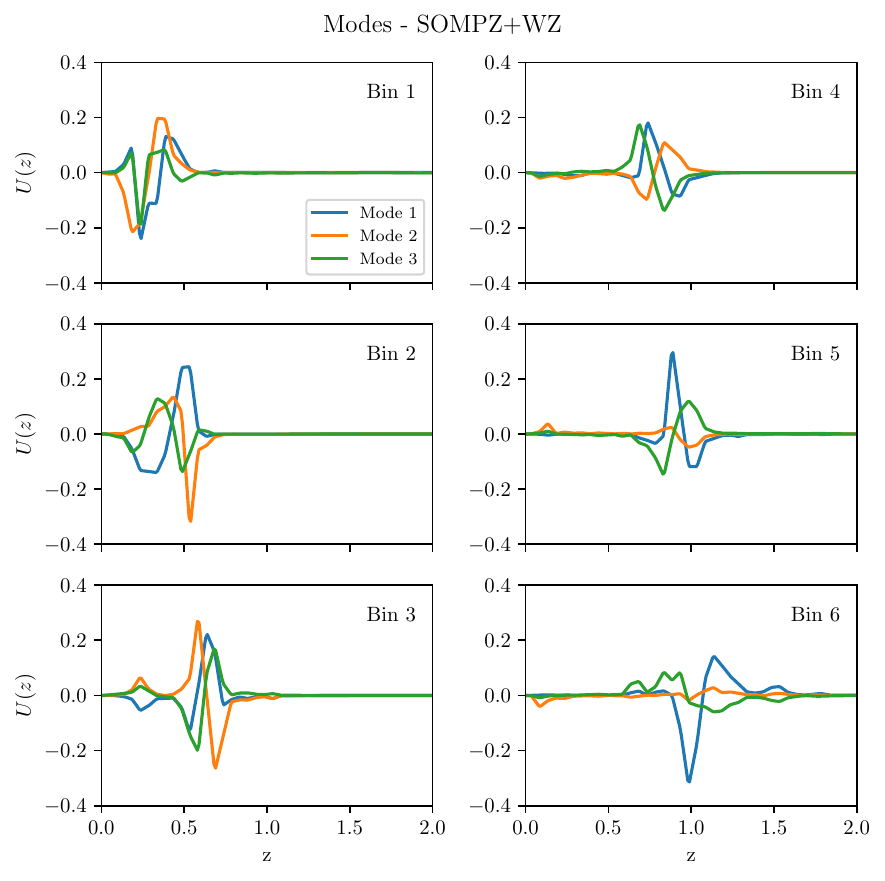}
    \caption{Modes \( U_i(z) \) retained in the compression. These modes define the primary directions of variation in \( n(z) \), capturing shifts in peak position, width, and other structural changes. Higher-order modes, which contribute minimally to the observable impact, are discarded. Left: SOMPZ only, right: SOMPZ+WZ.}
\label{fig:modes}
\end{figure*}

As described in Section~\ref{sec:modes}, we model redshift uncertainties using a basis of orthogonal perturbation modes applied to the fiducial redshift distributions \(n(z)\). Each coefficient  \(u_{i,j}\) modulates the amplitude of the \(j\)-th mode in tomographic bin \(i\) and is treated as a free parameter in the 3×2pt likelihood.

Figure~\ref{fig:modes} presents the leading modes \( U_i(z) \), which define the dominant directions of variation in \( n(z) \) retained by the compression. These modes correspond to cosmologically relevant deformations of the redshift distribution. Since they are derived from the Fisher information matrix of the observables, they are ordered by their impact on the likelihood: the first mode produces the largest change in observables, the second the next largest, and so on. In practice, the leading modes often resemble intuitive variations such as shifts in the mean redshift, changes in the width, or distortions in the tails, although this interpretation depends on the specific sensitivity of the observables used. Their smooth structure and orthogonality under the Fisher metric make them well-suited for parameterizing redshift uncertainty in a basis aligned with cosmological sensitivity.

\begin{figure*}
    \centering
    \includegraphics[width=\linewidth]{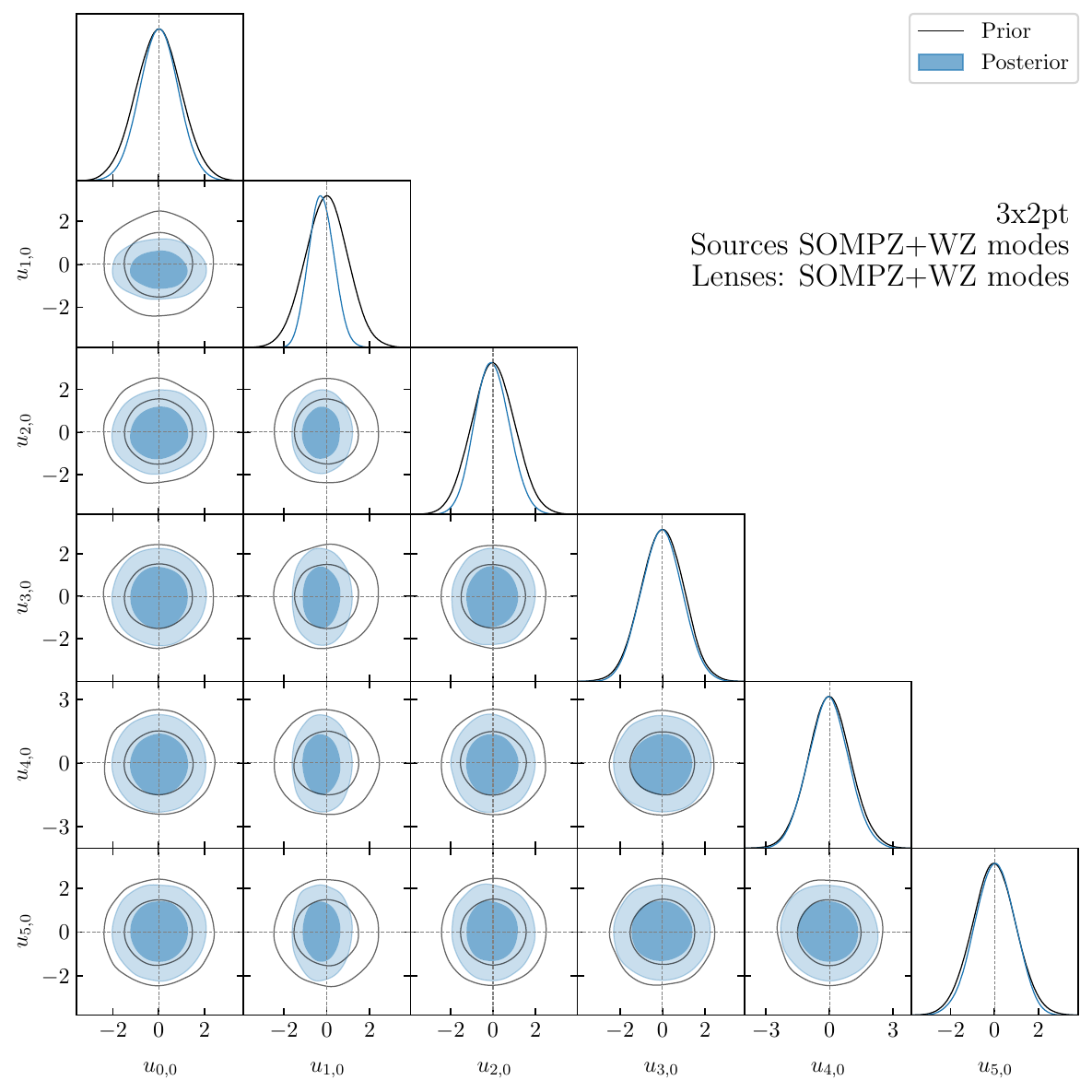}
    \caption{Prior (grey) and posterior (blue) distributions for the amplitudes of the main redshift modes, $u_{i,0}$, for all six tomographic bins. Each $u_{i,0}$ parameter controls the amplitude of the dominant mode of redshift uncertainty in bin $i$, as derived from the SOMPZ+WZ calibration. The priors reflect this propagated uncertainty, while the posteriors are constrained by the 3$\times$2pt data. The plot shows that the $u_{1,0}$ parameter (corresponding to bin 2) is noticeably more constrained by the 3$\times$2pt data compared to the other bins, likely because this bin overlaps more strongly with the regions where the lensing and clustering signals have the highest sensitivity to redshift calibration.}
    \label{fig:u_posteriors}
\end{figure*}

Figure~\ref{fig:u_posteriors} shows the prior and posterior distributions of \(u_{i,0}\), meaning the leading mode, for the 6 tomographic bins. The priors, centered at zero, are derived from the variance of redshift distributions estimated through the SOMPZ+WZ calibration and encode the level of uncertainty in each mode. The posteriors are obtained by sampling over the redshift, cosmological, and nuisance parameters in the likelihood analysis.

We find that the leading mode of bin 2 is well constrained by the simulated data, reflecting sensitivity to large-scale shifts in the redshift distribution.
We are not showing higher-order modes, since they remain completely unconstrained and are dominated by the prior in all bins. This supports our decision to truncate the mode expansion to a small number of dominant modes per bin.

The independent treatment of redshift uncertainties for the MagLim lens bins is well-justified by both the binning strategy and empirical tests. The MagLim sample was designed to optimize 2x2pt constraints, with each tomographic bin carefully selected to maximize cosmological information while maintaining well-separated redshift distributions. Since the binning is explicitly based on galaxy properties such as magnitude and color, any systematic redshift uncertainties are expected to primarily affect individual bins rather than introduce strong correlations across them. This is in contrast to the source redshift uncertainties, where modes are applied globally to capture correlated shifts across tomographic bins.

A key test to validate this assumption is to examine the behavior of the mode parameters $u_i$ which parameterize the redshift uncertainty within each bin. If the uncertainties were significantly correlated between bins, we would expect to see strong posterior correlations between these parameters. However, as shown in Figure \ref{fig:u_posteriors}, the posterior distributions of the $u_i$ parameters remain uncorrelated across different tomographic bins, confirming that the redshift uncertainties in each bin behave independently. This result is consistent with the calibration approach used in \maglimpp, where redshift uncertainties are sampled independently across three deep fields using a Latin Hypercube and then propagated to the final redshift distributions. While this process does not strictly enforce full independence, it ensures that any correlations arising from the calibration procedure remain small. The lack of significant correlations in the posteriors further supports the use of independent nuisance parameters for the \maglimpp lens bins, as incorporating artificial correlations would only add unnecessary complexity without improving the accuracy of the uncertainty marginalization.

\section{Impact of clipping redshift distributions to enforce positivity}\label{app:clipped}

\label{appendix:clipping}

As discussed in Section~\ref{sec:modes}, our lens redshift distributions are modeled as perturbations around a fiducial \(n(z)\) using a set of orthogonal modes. 
Because the reconstructed distributions result from a linear combination of these modes, they can occasionally take on slightly negative values in low-density regions, particularly at the tails of the distribution.

To test whether these small negative excursions affect cosmological inference, we run a version of the 2×2pt likelihood in which the reconstructed \(n(z)\) is clipped to zero wherever it is negative. 
We then compare the resulting constraints to those obtained using the original, un-clipped distributions.

\begin{figure}
    \centering
    \includegraphics[width=\linewidth]{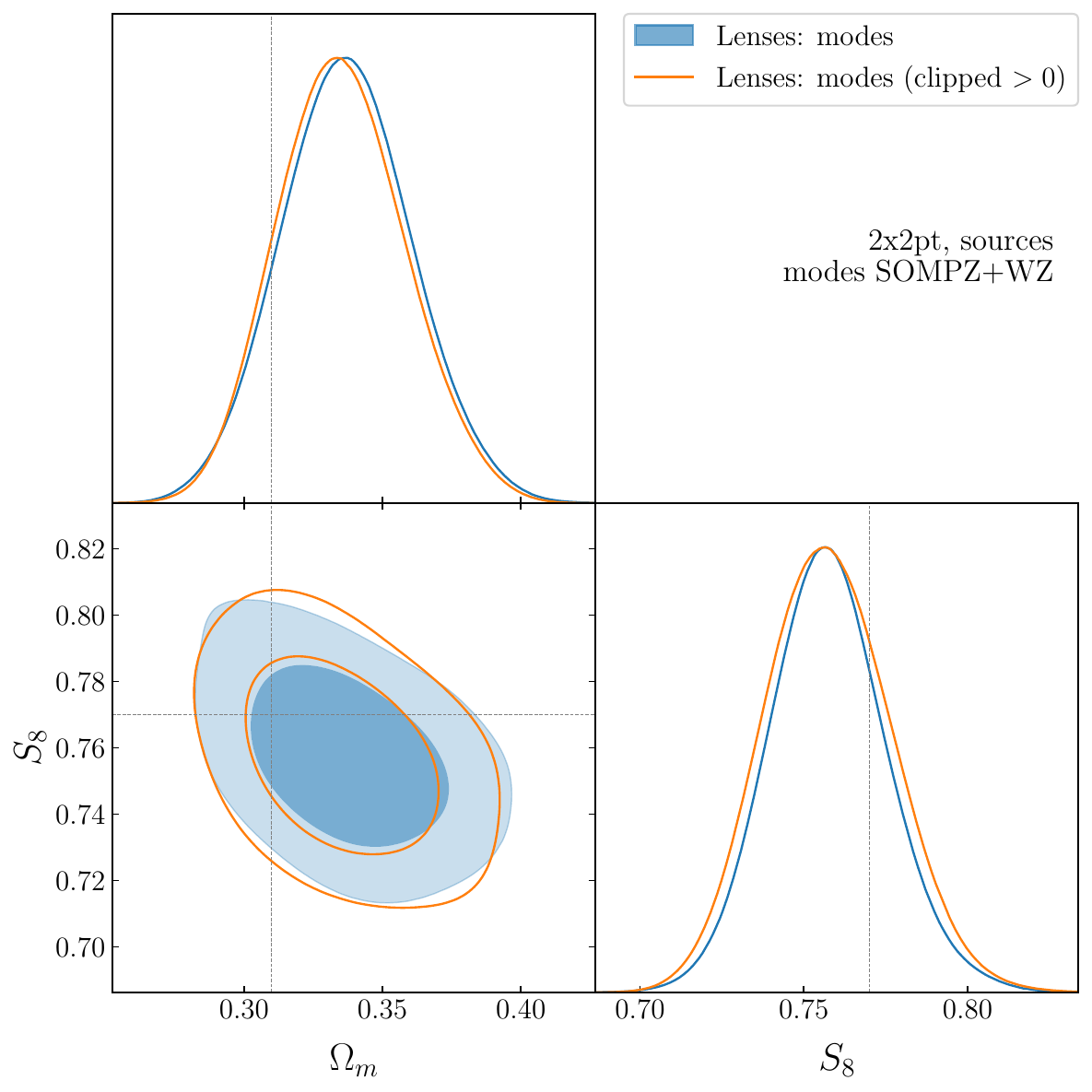}
    \caption{Impact of clipping the reconstructed lens redshift distributions at $n(z) > 0$ when using a mode-based representation. 
    We compare cosmological constraints from 2×2pt measurements using lens $n(z)$ reconstructed from modes with and without enforcing positivity. 
    The blue contours correspond to the default case where no clipping is applied, and the orange contours correspond to the case where the reconstructed $n(z)$ is clipped at zero. 
     The constraints are very similar, indicating that the small negative values introduced by the mode-based reconstruction do not impact the final results.}
    \label{fig:nz_clipped}
\end{figure}

Figure~\ref{fig:nz_clipped} shows that the contours are virtually unchanged between the clipped and un-clipped cases, confirming that these small artifacts do not impact the final results.

\section{Impact on galaxy bias and lens magnification parameters}
\label{app:bias_mag}

In Section~\ref{sec:cosmology_calib_method} we examined the robustness of cosmological constraints to mismatches between the redshift distributions used to construct the compressed data vector and those used to define the compression basis. 
Since both galaxy bias and magnification slope parameters are coupled to the lens $n(z)$, we extend that analysis here by inspecting the corresponding posteriors. 

\begin{figure*}
    \centering
    \includegraphics[width=0.8\linewidth]{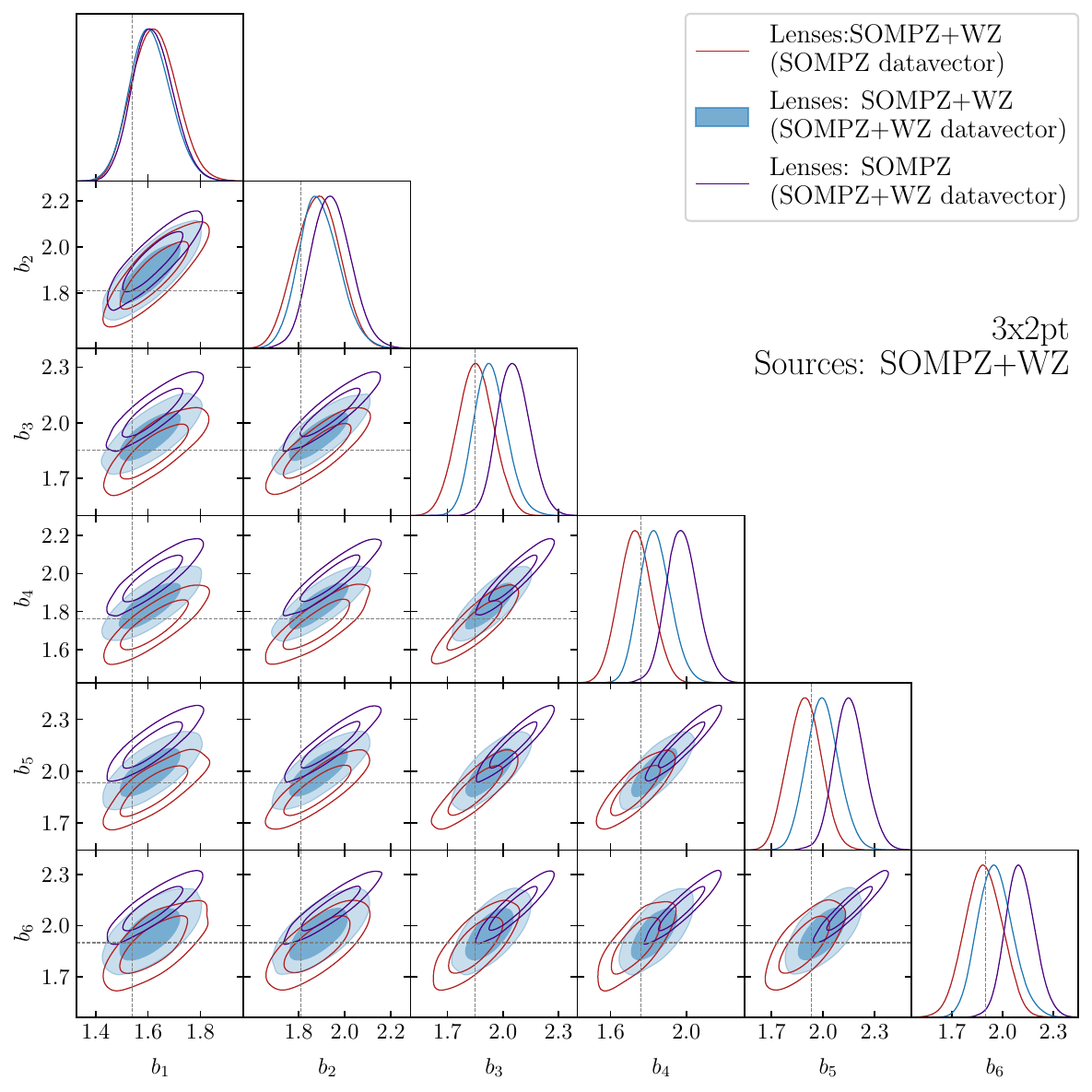}
    \caption{Posterior constraints on the linear galaxy bias parameters for the three data-vector/compression-basis combinations tested in Section~\ref{sec:cosmology_calib_method}. 
    Results are consistent across cases, indicating that moderate mismatches in the lens $n(z)$ treatment do not significantly affect the inference of galaxy bias.}
    \label{fig:bias_constraints}
\end{figure*}

Figure~\ref{fig:bias_constraints} shows the marginalized constraints on the six linear galaxy bias parameters for the three configurations considered in Section~\ref{sec:cosmology_calib_method}: 
DV(\textsc{SOMPZ})–Basis(\textsc{SOMPZ+WZ}), 
DV(\textsc{SOMPZ+WZ})–Basis(\textsc{SOMPZ}), 
and the fiducial DV(\textsc{SOMPZ+WZ})–Basis(\textsc{SOMPZ+WZ}). 
We find that the inferred biases are stable across the three cases, with shifts always well within the prior widths, in line with the values reported in Section~\ref{sec:cosmology_calib_method}.

For the magnification parameters, the posteriors in the three configurations exhibit only minimal shifts—considerably smaller than those seen for the galaxy bias parameters. Since the results are nearly identical across cases, we do not include a separate figure.

\label{lastpage}

\end{document}